\documentclass{aastex631}

\usepackage{amsmath}
\usepackage{graphicx}	

\begin{document}

\title[Predictions]{Physical Correlations and Predictions Emerging from Modern Core-Collapse Supernova Theory}

\author[0000-0002-3099-5024]{Adam Burrows}
\affiliation{Department of Astrophysical Sciences, Princeton University, Princeton NJ 08544 and the Institute for Advanced Study, 1 Einstein Drive, Princeton NJ  08540}

\correspondingauthor{Adam Burrows}
\email{burrows@astro.princeton.edu}

\author[0000-0002-0042-9873]{Tianshu Wang}
\affiliation{Department of Astrophysical Sciences, Princeton University, Princeton, NJ 08544}

\author[0000-0003-1938-9282]{David Vartanyan}
\affiliation{Carnegie Observatories, 813 Santa Barbara St., Pasadena, CA 91101}



\date{Accepted XXX. Received YYY}

\begin{abstract}
 In this paper, we derive correlations between core-collapse supernova observables and progenitor core structures that emerge from our suite of twenty state-of-the-art 3D core-collapse supernova simulations carried to late times. This is the largest such collection of 3D supernova models ever generated and allows one to witness and derive testable patterns that might otherwise be obscured when studying one or a few models in isolation. From this panoramic perspective, we have discovered correlations between explosion energy, neutron star gravitational birth masses, $^{56}$Ni and $\alpha$-rich freeze-out yields, and pulsar kicks and theoretically important correlations with the compactness parameter of progenitor structure. We find a correlation between explosion energy and progenitor mantle binding energy, suggesting that such explosions are self-regulating. We also find a testable correlation between explosion energy and measures of explosion asymmetry, such as the ejecta energy and mass dipoles. While the correlations between two observables are roughly independent of the progenitor ZAMS mass, the many correlations we derive with compactness can not unambiguously be tied to a particular progenitor ZAMS mass. This relationship depends upon the compactness/ZAMS mass mapping associated with the massive star progenitor models employed.  Therefore, our derived correlations between compactness and observables may be more robust than with ZAMS mass, but can nevertheless be used in the future once massive star modeling has converged.   
\end{abstract}





\section{Introduction}
\label{intro}









Building on the early pioneering work of \citet{colgate1966}, \citet{arnett1967}, \citet{wilson1971}, \citet{bowers1982}, and \citet{Bethe1985}, with the recognition of the importance of neutrino-driven turbulence and convection \citep{Herant1994,Burrows1995,janka1996}, and paralleling the developments in neutrino physics and supercomputers over the decades \citep{burrows_paris_2019}, core-collapse supernova (CCSN) theory has emerged after $\sim$60 years of progress to become a mature, if complicated and multi-physics, international enterprise.
Neutrino heating, aided by the stress of neutrino-driven turbulence, is {commonly} accepted as the key mechanism of explosion. However, turbulence is chaotic and this implies that a range of outcomes and observable values are likely, even for the same massive star progenitor. The explosive expansion of the initially stalled bounce shock wave seems to be a critical phenomenon \citep{goshy} and is associated with the shock's pulsational instability, but a robust analytic model has proven elusive. This has necessitated detailed multi-dimensional numerical simulations, whose complexity and expense had slowed conceptual and quantitative progress. The accretion before explosion of the density jump associated with the Si/O interface in the massive star progenitor, itself correlated with ``compactness" (see \S\ref{compact}), oftimes inaugurates explosion \citep{vartanyan2018,wang2022}. However, this is not always the case, and its invocation as an exclusive trigger for explosion should be nuanced. Curiously, a quick explosion does not in general lead to an energetic explosion (see \S\ref{time}).

Rotation, though likely always present at some level, seems critically important in only a small subset of core collapses (hypernovae and Type Ic broad-line CCSNe), and its role in general may be subdominant, but has yet to be determined\footnote{We note that for rotation to be of central importance generically and energetically, the proto-neutron star (PNS) birthed in collapse would need a spin period near $\sim$5 milliseconds (ms). The pulsar data would seem to mitigate against this possibility \citep{Chevalier1986,Lyne1994,Manchester2005,Faucher2006,Popov2012,Igoshev2013,Noutsos2013}.}. Moreover, the role and evolution of magnetic fields, likely tightly coupled with  
rotation, is a subject of current intense research \citep{Muller2020,varma21,Obergaulinger2020,Kuroda2020,Obergaulinger2021}, but is thought to be only rarely of central importance.

Interestingly, in recent late-time 3D simulations, \citet{wang2023} observe aspherical winds \citep{duncan1986,burrows1987,Burrows1995,qian1996} emerging from the PNS after a delay of seconds following the onset of explosion; these produce secondary shock waves upon encountering the primary ejecta. As much as $\sim$10\% of the explosion energy can be due to these winds. However, other groups have yet to simulate 
to late enough times after bounce to verify the details of this phenomenon, itself also associated with the important nucleosynthetic phase of $\alpha$-rich freeze-out \citep{woosley2002,nomoto2013,arcones2023,wang2023b}. Furthermore, which stars leave neutron stars and which black holes is still not resolved, though it is known that previous mappings were too simplistic \citep{Burrows_PT,heger2003b,Sukhbold2016,Ertl2016}.  There are new ideas on this partitioning \citep{Burrows2023}, but these have not been thoroughly vetted and depend a great deal on the still-evolving progenitor models inherited.

\begin{figure}
    \centering
    \includegraphics[width=0.48\textwidth]{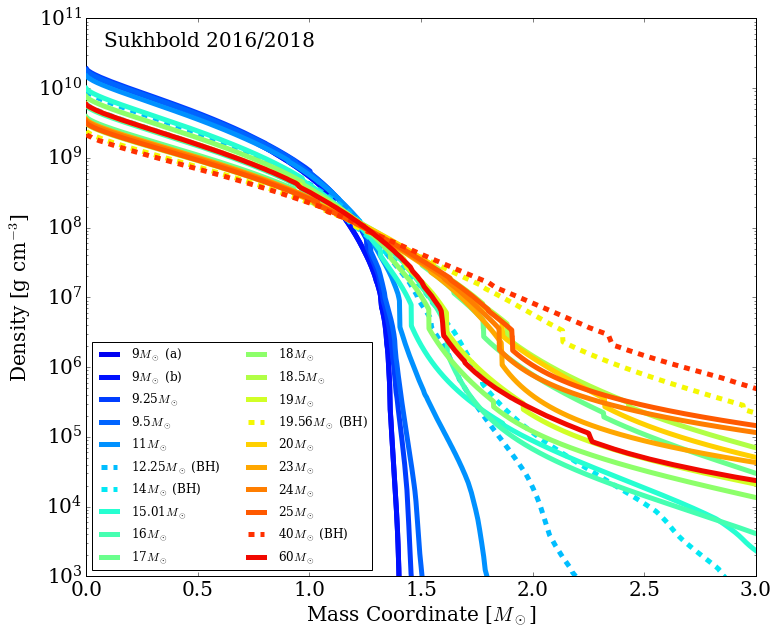}
    \caption{Progenitor mass density profiles as a function of interior mass coordinate for the twenty models we employ for this 3D supernova explosion investigation from the \citet{Sukhbold2016} and \citet{Sukhbold2018} collection of unstable massive star cores. The colors correlate with ZAMS mass and the dashed curves are for models we see leave black holes, either explosively (the 19.56- and 40-$M_{\odot}$ models) or quiescently.}
    \label{fig:M-rho}
\end{figure}

Despite these many remaining issues and the physical and numerical complexity of the phenomenon, and despite the fact that chaotic turbulence renders a clear one-to-one mapping of progenitor to outcome naturally blurred, there has emerged a broad consensus on the viability of the turbulence-aided neutrino-driven mechanism of core-collapse supernovae (CCSNe).  Modern sophisticated 3D simulations witness neutrino-driven explosions naturally and without artifice \citep{Lentz2015,roberts2016,Takiwaki2016,Muller2017,ott2018_rel,Glas2019,muller_low_kick2019,Burrows2019,Burrows2020,Muller2020,Powell2020,Stockinger2020,Kuroda2020,Bollig2021,Vartanyan2021,Vartanyan2023} and numerous reviews have articulated the mechanism's details \citep{Janka2012,Burrows2013,Muller2016,Muller2017,Burrows2018,Burrows2020,Mandel2020,Burrows2021,wang2022}. With this paper, we summarize the physical and observable correlations that have emerged from the panoramic view enabled by our new and unprecedented set of twenty state-of-the-art 3D core-collapse simulations taken to late times. 

In \S\ref{progenitors}, we discuss progenitor issues and their structures to provide context. In \S\ref{basics}, the basic explosion phenomenology is provided. This is followed by a graphical summary of the variety of immediate post-explosion morphologies and their general characteristics. Then, in \S\ref{correlations}, we provide for the first time using such a comprehensive long-term suite of 3D models the physical and observable correlations that define the explosion systematics and byproducts.  These include the neutron star birth masses, explosion energies, $^{56}$Ni yields, and recoil kicks. Finally, in \S\ref{conclusion}, we summarize our salient results and provide a set of caveats to consider going forward. To create the set of twenty sophisticated 3D late-time simulations that we use in this correlation study, we employed the CPU machines TACC/Frontera \citep{Stanzione2020} and ALCF/Theta and the GPU machines ALCF/Polaris and NERSC/Perlmutter. The specific setups and run parameters are described in \citet{Burrows2023} and \citet{Vartanyan2023} and in the Appendix.

\section{Initial Progenitor Structures}
\label{progenitors}

For this study we employ solar-metallicity, initially non-rotating, spherical massive star progenitor models from \citet{Sukhbold2016} and \citet{Sukhbold2018} and we do not {include} magnetic fields. The F{\sc{ornax}} code and calculational details are described in the Appendix and in \citet{Skinner2019}.  As discussed in \citet{wang2022}, it is predominantly the density profile at collapse that determines explodability and this profile is not rigorously monotonic with zero-age main-sequence (ZAMS) mass. 
Figure \ref{fig:M-rho} depicts these mass density profiles for the models we include in this paper. As one sees, while the steepness of the envelope in the Chandrasekhar-like core is roughly an increasing function of ZAMS mass, it is not rigorously so, nor is the ``compactness" parameter \citep{oconnor2011} (see \S\ref{compact}); we will find in this paper a tighter relation of the observables with compactness than with ZAMS mass (\S\ref{correlations}). 

Importantly, there remain many ambiguities in the initial
progenitor models and, as important as they are to core-collapse supernova theory, the progenitor models can not yet be considered definitive.  Issues such as the $^{12}$C($\alpha$,$\gamma$) rate, binarity, overshoot, wind mass loss, shell mixing, semi-convection, magnetic field effects, and rotation are still to be retired. Moreover, a new generation of 3D progenitor models \citep{2015ApJ...808L..21C,jones2016,Chatzopoulos2016,Muller2016b,jones2017,yoshida2019,Fields2020,Bollig2021,McNeill2021,Fields2021,varma21,Yoshida2021} is emerging. As a result of these many remaining questions and uncertainties, models for the stellar evolution of massive stars to the point of collapse can not be considered to have converged. Nevertheless, we use the \citet{Sukhbold2016} and \citet{Sukhbold2018} model suite
as the best and most comprehensive available. Moreover, we suggest that our derived correlations between observables and compactness may be usefully robust, providing ongoing insight as the progenitor model studies evolve into the future.

\begin{figure}
    \centering
    \includegraphics[width=0.48\textwidth]{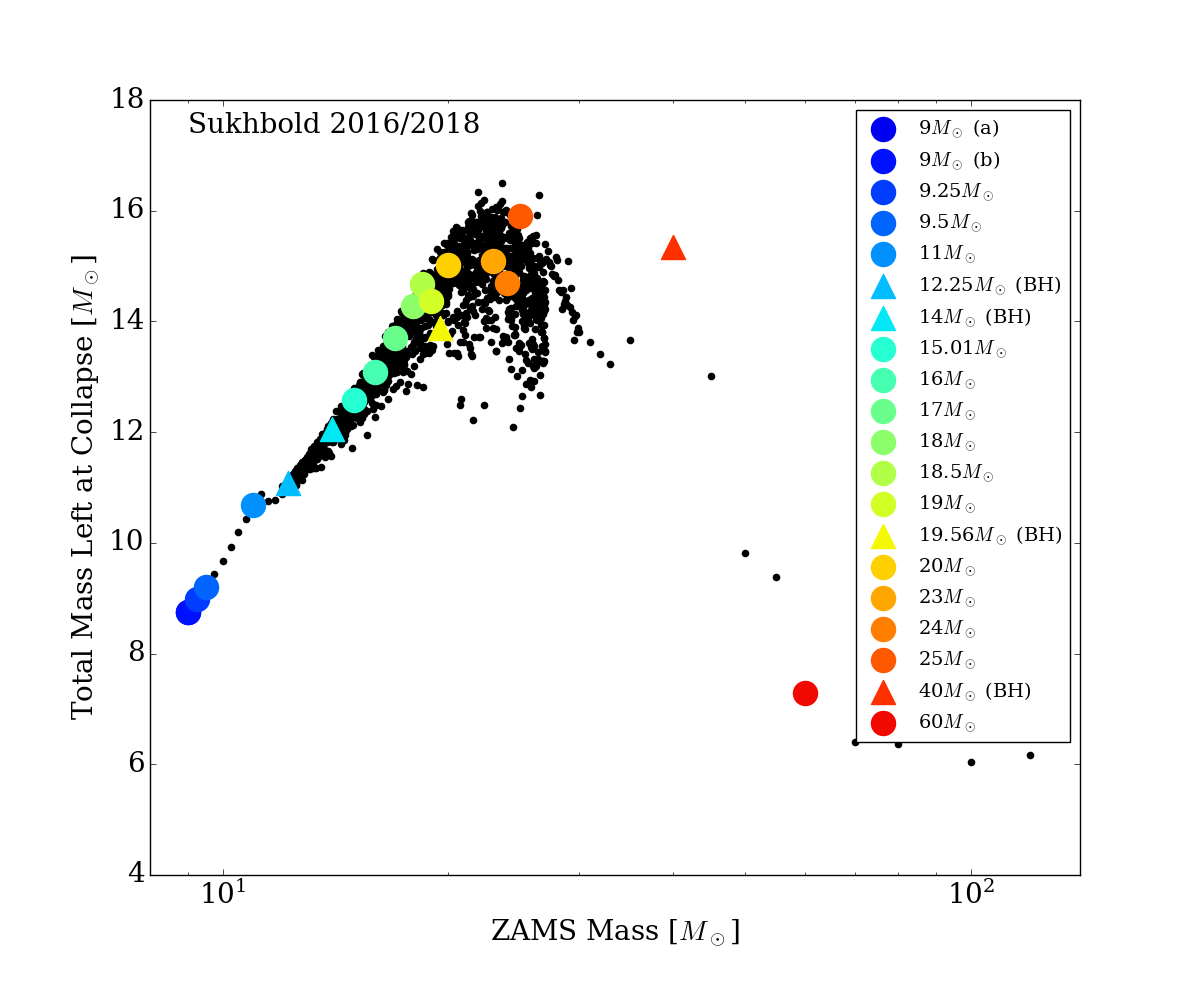}
    \includegraphics[width=0.48\textwidth]{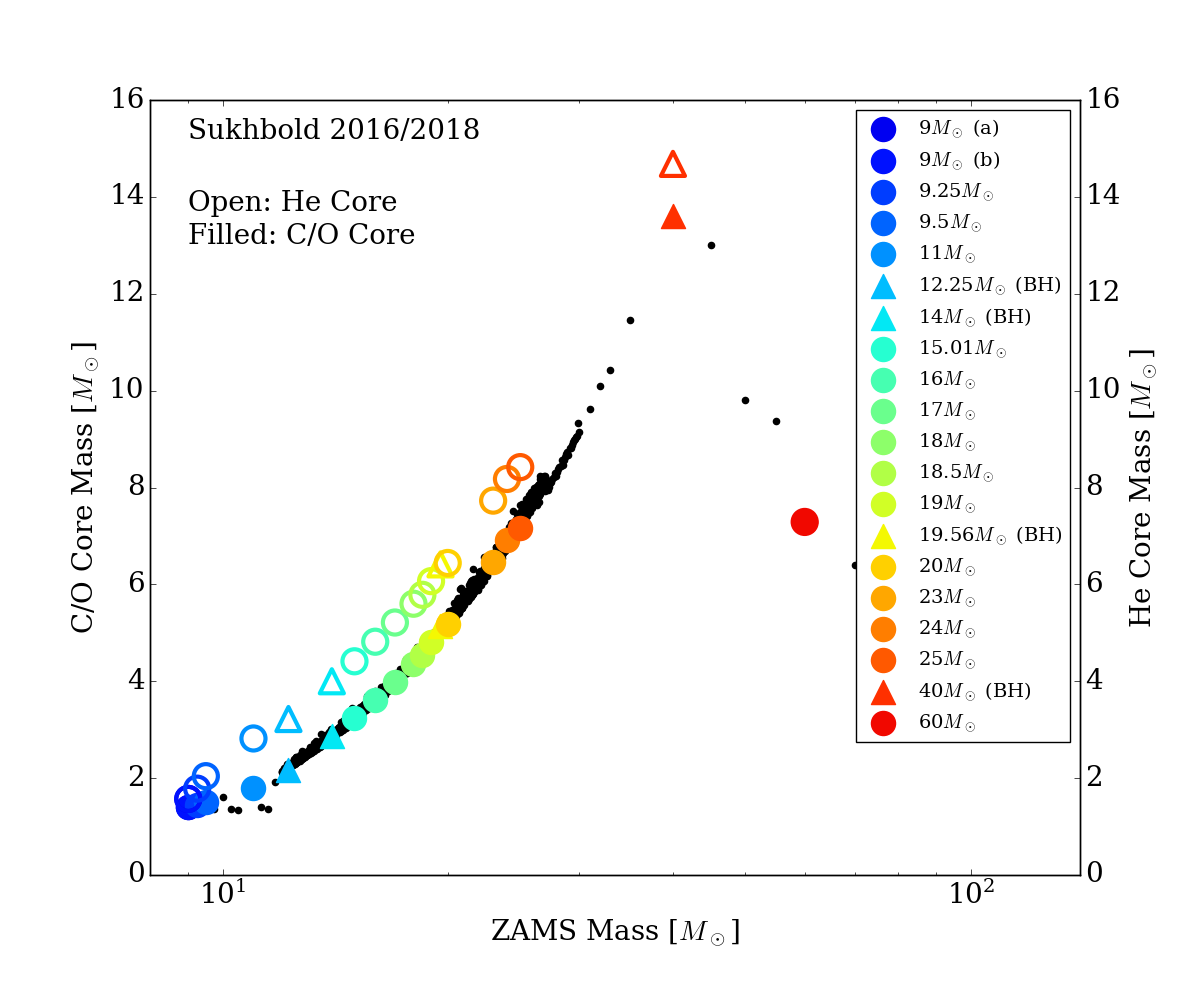}
    \caption{Total stellar mass (left) that remains at collapse and the corresponding C/O and He core masses (right) as a function of the {log$_{10}$} of the ZAMS mass. On the right panel, the filled symbols are the C/O core masses (left label), while the unfilled symbols are the He core masses (right label). The 60 $M_\odot$ progenitor has the same C/O and He core mass because it has ejected its outer envelope exterior to the C/O core in strong winds. The black dots in both panels are all the models in the \citet{Sukhbold2016} and \citet{Sukhbold2018} solar-metallicity set. The triangles are the black hole formers.}
    \label{fig:zams-MCO}
\end{figure}

Figure \ref{fig:zams-MCO} shows the total stellar mass (left) that remains at collapse and the corresponding C/O and He core masses (right) as a function of ZAMS mass for the models we have chosen for this comprehensive study. We note that despite the unprecedented number of models we have included in this study, there are many more models from the \citet{Sukhbold2016} and \citet{Sukhbold2018} left unsimulated.
Nevertheless, our goal was to well sample the collection, if not optimally.

\subsection{Compactness}
\label{compact}

\begin{figure}
    \centering
    \includegraphics[width=0.48\textwidth]{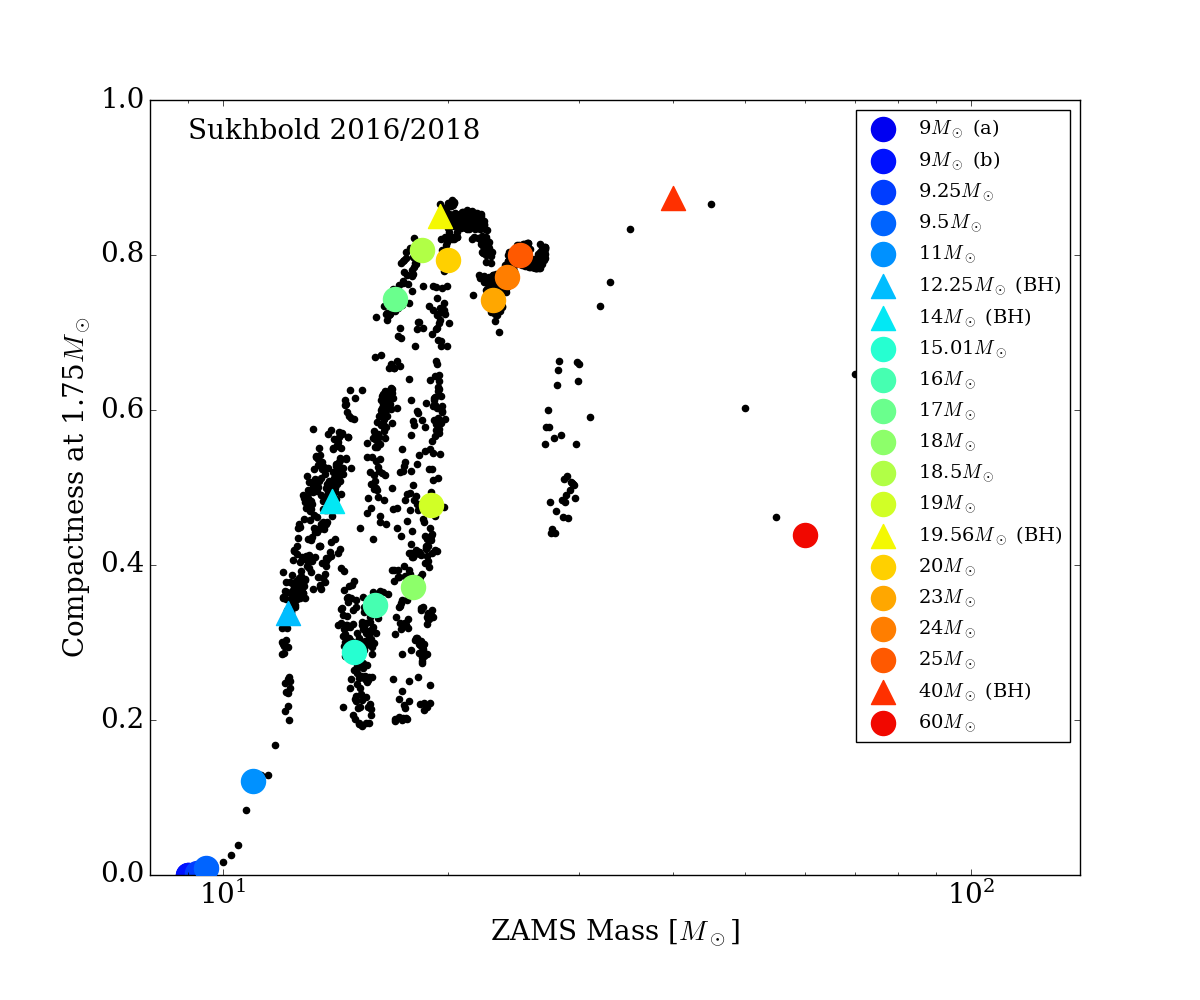}
    \includegraphics[width=0.48\textwidth]{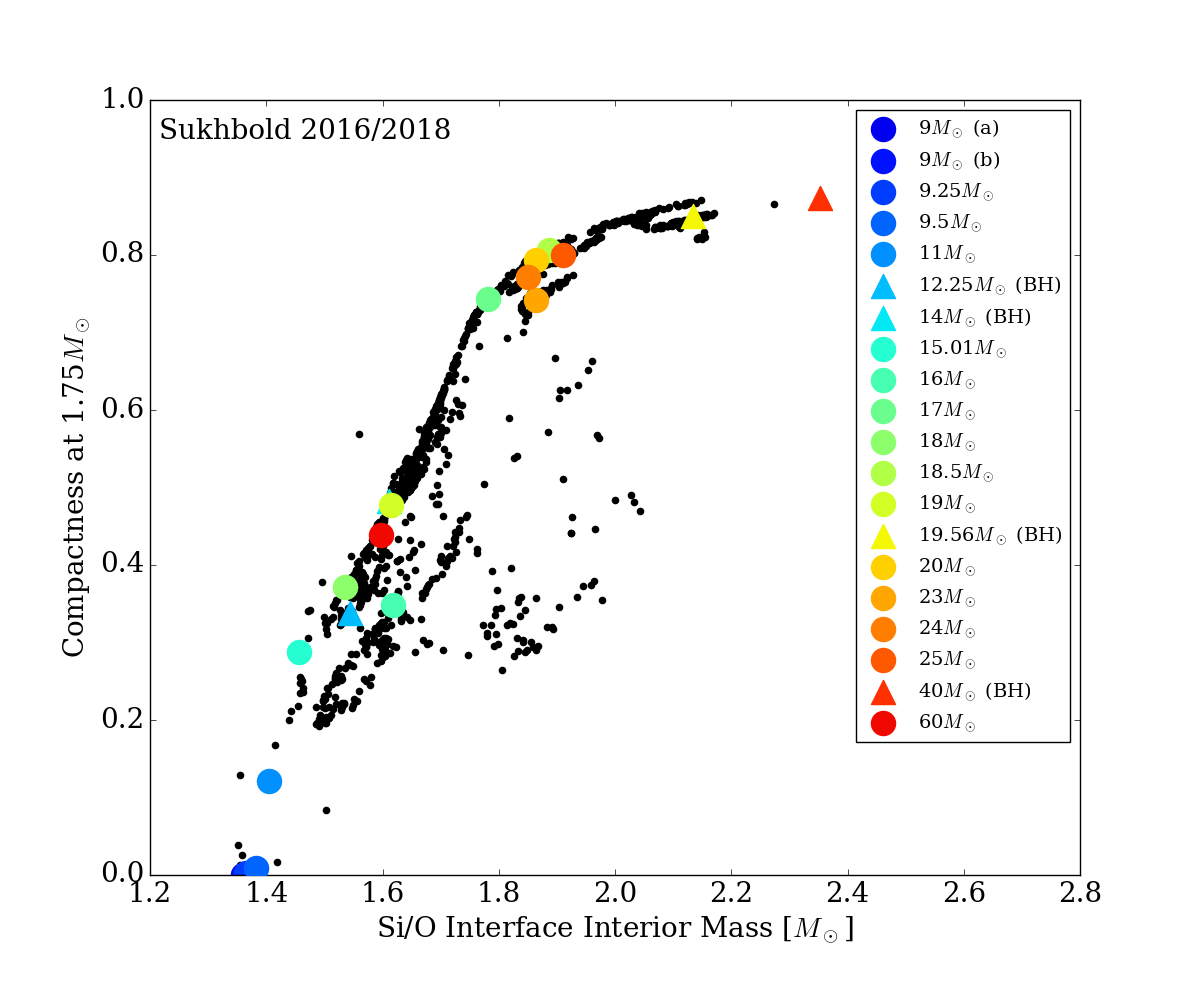}
    \caption{The dependence of the compactness on the {log$_{10}$} of the ZAMS mass (left) and the Si/O interface interior mass (right).  The black dots identify the full \citet{Sukhbold2016} and \citet{Sukhbold2018} collection of solar-metallicity models, while the colored symbols indicate the models in this study. The colored triangles identify the models that we see form black holes.}
    \label{fig:cpt-zams-MSi}
\end{figure}

The shallowness of the mass density profiles depicted in  Figure \ref{fig:M-rho} is related to the post-bounce mass accretion rate history and the accretion component of the neutrino luminosities.  It is through these quantities that the complicated dynamics of the supernova mainly rests. One crude, but useful, metric of this shallowness is the compactness. Compactness \citep{oconnor2011}\footnote{Introduced to explore black hole formation, it is defined as $\frac{M/M_\odot}{R(M)/1000\text{km}}$, for which we prefer to set $M$ equal to 1.75 $M_{\odot}$ when studying supernova explosions, not $2.5 M_{\odot}$, as did \citet{oconnor2011}.} is not a predictor of explodability \citep{wang2022} {in our simulations}. Nevertheless, it is a useful, if gross, measure of progenitor structure and correlates well with numerous observables and physical characteristics of supernova dynamics.

Figure \ref{fig:cpt-zams-MSi} shows the relation between the compactness parameters, ZAMS masses, and Si/O interface interior masses. Echoic in some sense of the plots in Figure \ref{fig:zams-MCO}, the differences are important. The accretion of the Si/O interface frequently kicks the core's mantle into explosion \citep{wang2022} and the near monotonicity of compactness with Si/O mass is suggestive. 
Figures \ref{fig:M-rho}, \ref{fig:zams-MCO}, and \ref{fig:cpt-zams-MSi} set the initial structural context of our results.

\section{Basic Explosion Results}
\label{basics}

\begin{figure}
    \centering
    \includegraphics[width=0.48\textwidth]{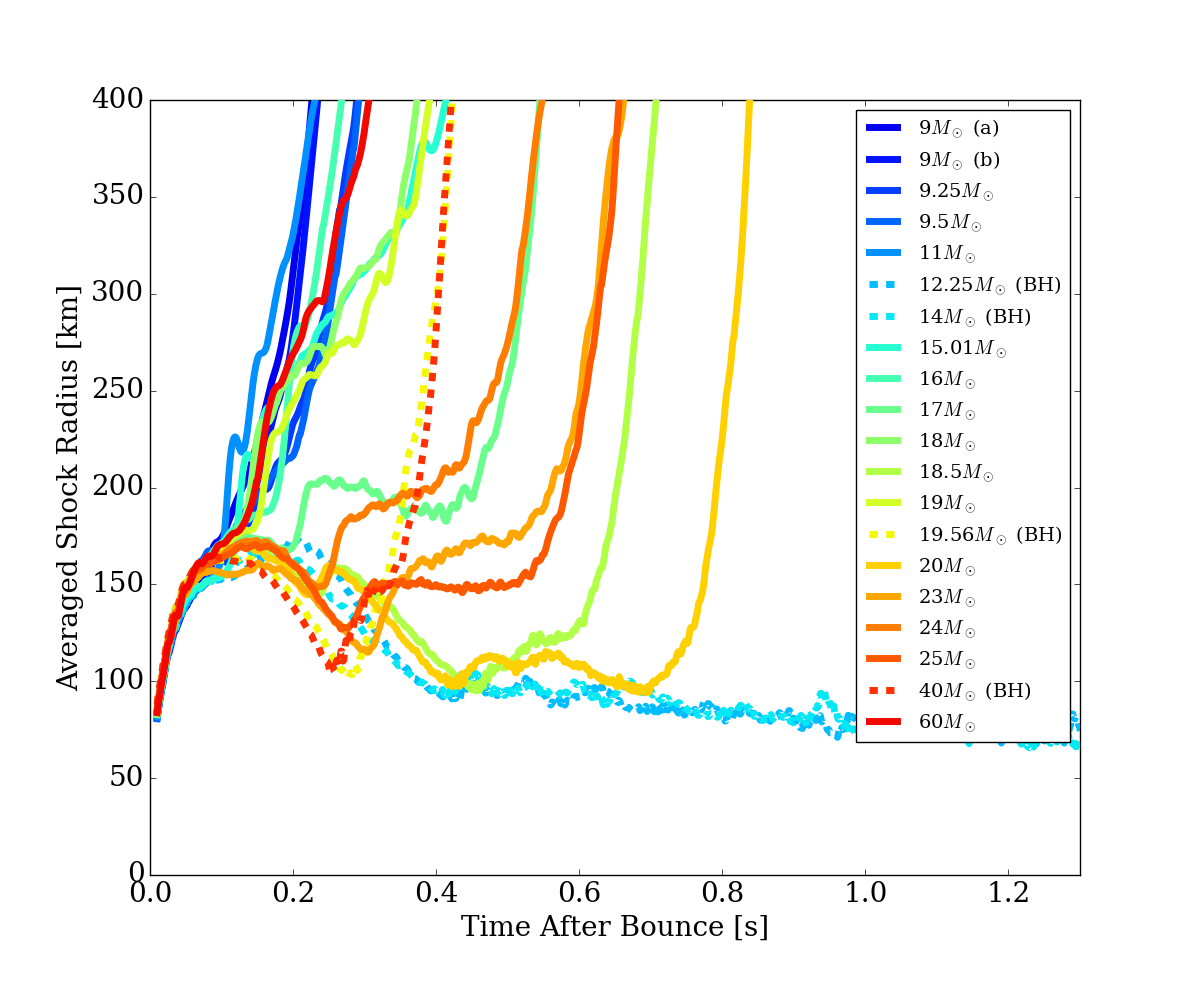}
    \includegraphics[width=0.48\textwidth]{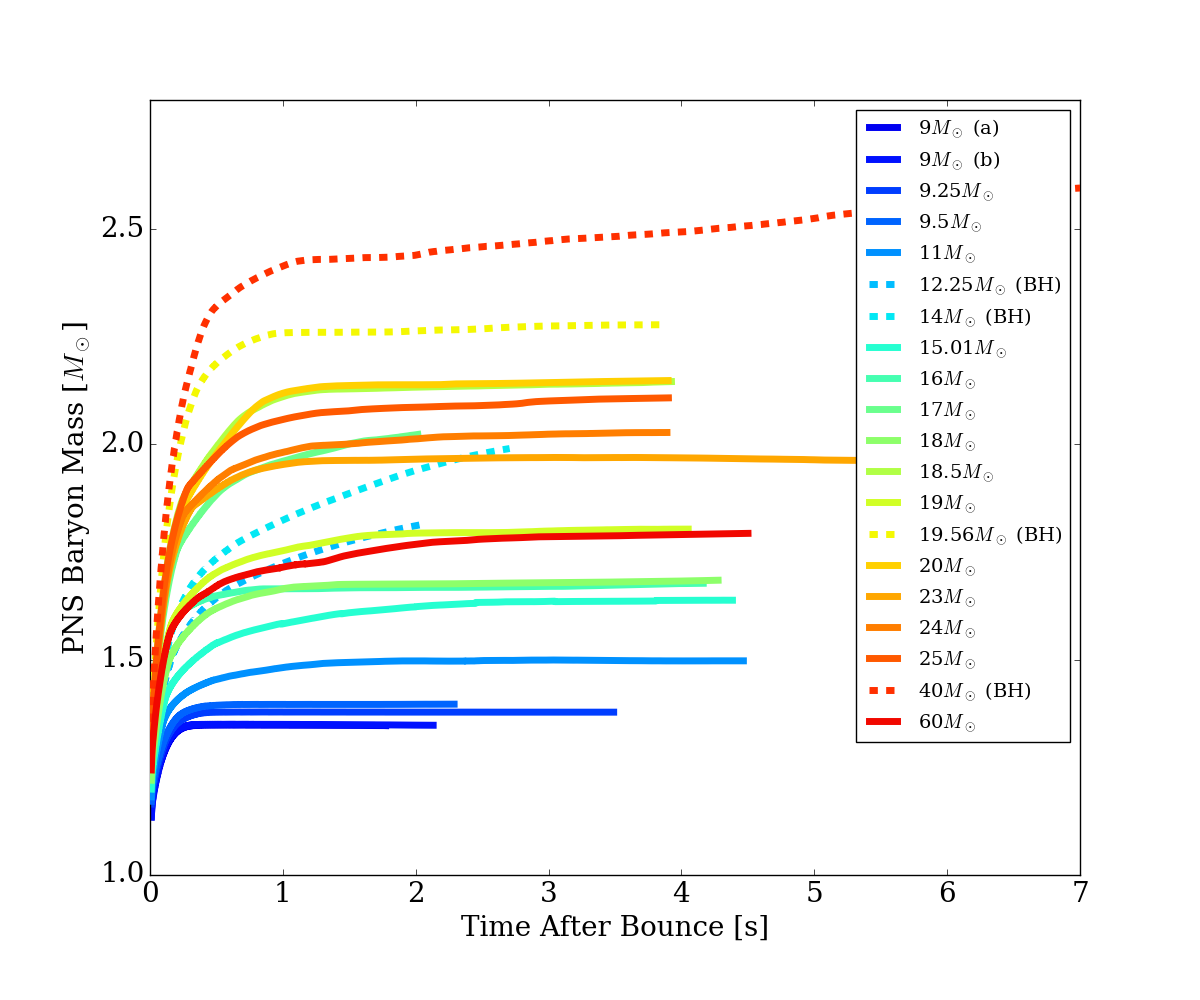}
    \includegraphics[width=0.48\textwidth]{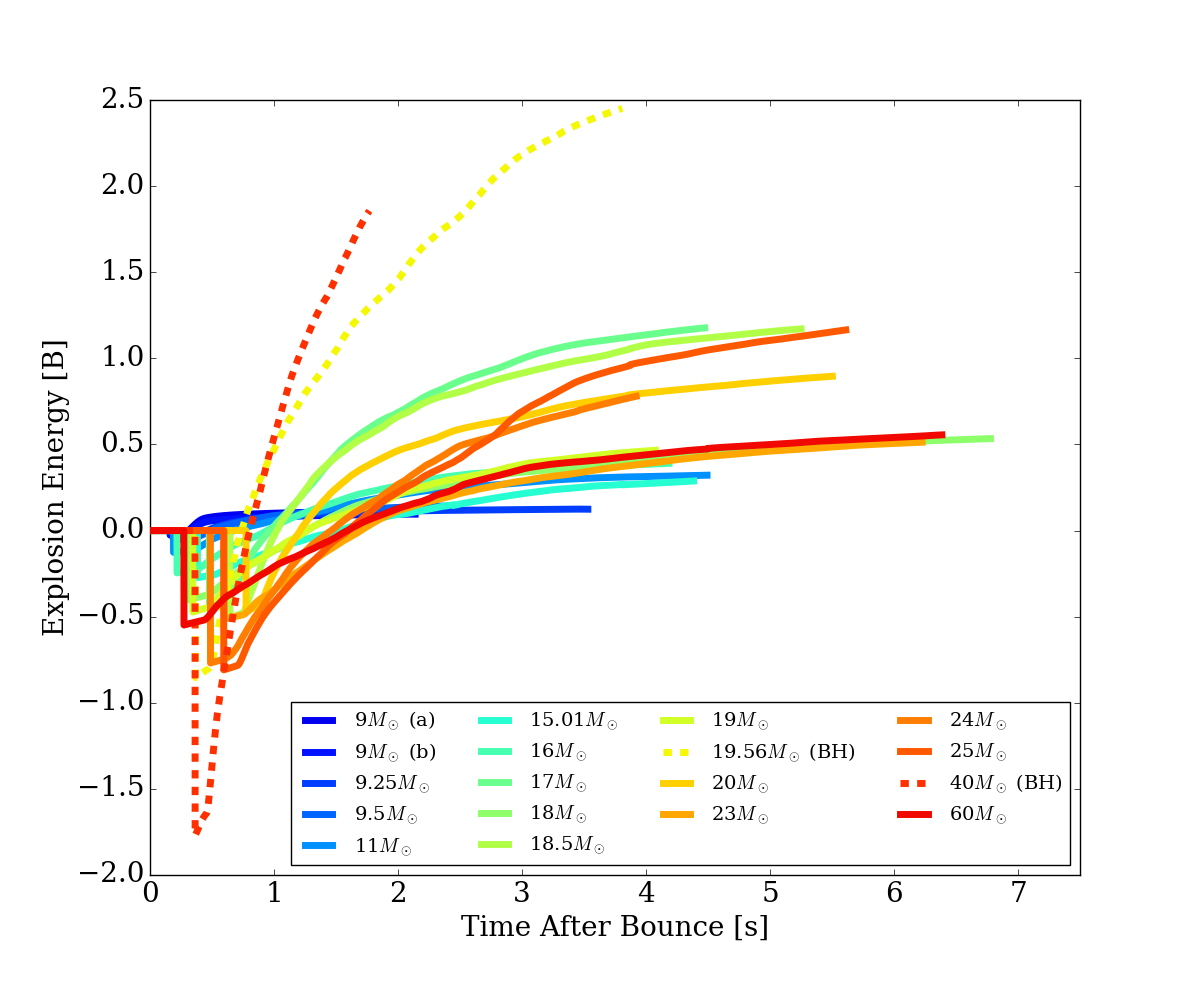}    
    \includegraphics[width=0.48\textwidth]{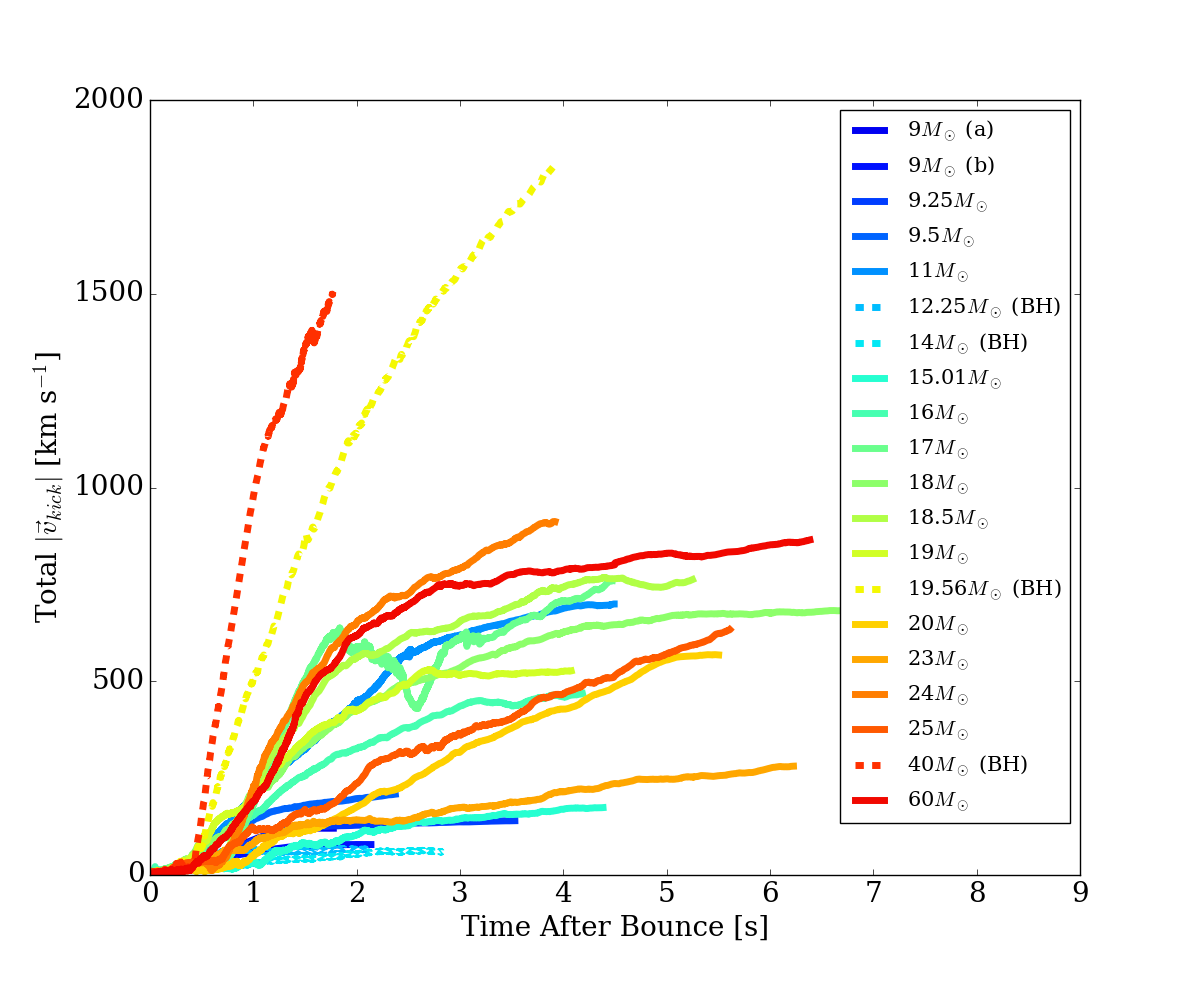}
    \caption{The temporal evolution of averaged shock radius (top left), PNS baryon mass (top right), the minimum explosion energy (bottom left, in Bethes $\equiv$10$^{51}$ ergs), and kick speed (bottom right). {Before shock revival, we set the ``explosion energy" arbitrarily to zero.  Upon shock revival, the matter behind the shock is still bound and early on most of the deposited neutrino energy goes into overcoming this negative gravitational binding energy and the (negative) contribution of the overburden exterior to the simulation grid (which is included in the accounting).  This is why the explosion energy starts negative and only gradually grows to become positive over time.} See text in \S\ref{time} for a discussion.}
    \label{fig:t-rs-Mpns}
\end{figure}

Table $1$ provides various final-state quantities for each of the twenty 3D progenitor evolutions that comprise this study.  Included are the baryon mass, gravitational mass, explosion energy, $^{56}$Ni mass, and kick speed at the end of each simulation (at post-bounce time $t_{max}$)\footnote{We use the formalism in \citet{burrows_kick_2023} and \citet{coleman} to obtain the kick speed.}. Also given is whether we witness the formation of a neutron star or black hole, the ZAMS mass, and the compactness at 1.75 $M_{\odot}$.  It is this collection of numbers that informs our discussions to follow of the systematics found between stellar property, explosion characteristics, and observables.

\begin{table}[]
\centering
\begin{tabular}{c|cccccccc}
\hline
$M_\text{ZAMS}$ & $\xi_{1.75}$       & $t_\text{max}$ & Type   & $M_\text{baryonic}$ & $M_\text{grav}$  & Explosion Energy & $^{56}$Ni              & $v_\text{kick}^\text{total}$  \\
{[}$M_\odot${]} &                    & {[}s{]}        &        & {[}$M_\odot${]}     & {[}$M_\odot${]}  & {[}B{]}          & {[}$10^{-2}M_\odot${]} & {[}km s$^{-1}${]}                  \\ \hline
9(a)            & $6.7\times10^{-5}$ & 1.775          & NS     & 1.347               & 1.237            & 0.111--0.111            & 0.168                  & 120.7                                  \\
9(b)            & $6.7\times10^{-5}$ & 1.950          & NS     & 1.348               & 1.238            & 0.094--0.095            & 0.612                  & 78.6                                  \\
9.25            & $2.5\times10^{-3}$ & 3.532          & NS     & 1.378               & 1.263            & 0.124--0.124            & 1.04                   & 140.1                                    \\
9.5             & $8.5\times10^{-3}$ & 2.375          & NS     & 1.397               & 1.278            & 0.142--0.143            & 1.47                   & 208.6                                   \\
11              & 0.12               & 4.492          & NS     & 1.497               & 1.361            & 0.321--0.326            & 2.92                   & 699.4                                  \\
15.01           & 0.29               & 4.384          & NS     & 1.638               & 1.474            & 0.288--0.352            & 5.42                   & 173.9                                  \\
16              & 0.35               & 4.184          & NS     & 1.678               & 1.505            & 0.390--0.463            & 6.06                   & 468.0                                  \\
17              & 0.74               & 4.473          & NS     & 2.044               & 1.785            & 1.123--1.173            & 9.99                   & 759.8                                  \\
18              & 0.37               & 6.778          & NS     & 1.684               & 1.510            & 0.516--0.587            & 10.3                   & 686.0                                  \\
18.5            & 0.80               & 5.250          & NS     & 2.139               & 1.854            & 1.155--1.207            & 13.8                   & 762.4                                  \\
19              & 0.48               & 4.075          & NS     & 1.803               & 1.603            & 0.466--0.642            & 7.73                   & 526.7                                  \\
20              & 0.79               & 5.503          & NS     & 2.143               & 1.857            & 0.881--0.988            & 9.94                   & 567.9                                 \\
23              & 0.74               & 6.228          & NS     & 1.959               & 1.722            & 0.513--0.619            & 8.77                   & 280.8                                  \\
24              & 0.77               & 3.919          & NS     & 2.028               & 1.773            & 0.779--1.228            & 12.5                   & 912.1                                 \\
25              & 0.80               & 5.611         & NS      & 2.106               & 1.830            & 1.101--1.401            & 16.8                   & 633.0                                  \\
60              & 0.44               & 6.386          & NS     & 1.791               & 1.594            & 0.541--0.722            & 10.6                   & 864.8                                 \\ \hline
$M_\text{ZAMS}$ & $\xi_{1.75}$       & $t_\text{max}$ & Type   & $M_\text{baryonic}$ & $M_\text{final}$ & Explosion Energy & $^{56}$Ni              & $v_\text{kick}^\text{total}$             \\
{[}$M_\odot${]} &                    & {[}s{]}        &        & {[}$M_\odot${]}     & {[}$M_\odot${]}  & {[}B{]}          & {[}$M_\odot${]}        & {[}km s$^{-1}${]}                             \\ \hline
12.25           & 0.34               & 2.090          & BH (F) & 1.815               & $\sim$11.1       & N/A              & N/A                    & 39.9 (10.7)                    \\
14              & 0.48               & 2.824          & BH (F) & 1.990               & $\sim$12.1       & N/A              & N/A                    & 42.5 (9.59)                   \\
19.56           & 0.85               & 3.890 (7.5)          & BH (S) & 2.278               & ?               & 2.451--2.542            & 25.6                   & 1826 (?)                        \\
40              & 0.87               & 1.760 (21)          & BH (S) & 2.434               & $\sim$3.5        & 1.804--1.851            & 16.5                   & 1504(1034)                  
\end{tabular}
\caption{Summary of simulation outcomes and characteristics.  The range of explosion energies provided is from minimum to projected (likely) and reflects the fact that not all models have completely asymptoted. See text for details.}
\end{table}
\label{table1}

To set the stage for the subsequent comparisons, we display in Figure \ref{fig:t-rs-Mpns} the temporal evolution of the averaged shock radius (top left), PNS baryon mass (top right), explosion energy (bottom left), and kick speed (bottom right).  Note that explosion occurs over a range of post-bounce times from $\sim$100 milliseconds (ms) to $\sim$800 ms, with the lower mass progenitors generally exploding more quickly and the more massive progenitors taking longer. In fact, the shock waves for the 9 to 9.5 $M_{\odot}$ progenitors barely stall, while that for the 20 $M_{\odot}$ progenitor recedes significantly before exploding. We note that among those models for which the shock initially recedes it is only those for which the compactness and associated mass accretion rate and accretion luminosity are high that explode.  The 12.25 and 14 $M_{\odot}$ models have too low a compactness to reverse the recession of their stalled accretion shocks. As Figure \ref{fig:t-rs-Mpns} indicates, the baryon mass asymptotes earlier than the energy and kick speeds.  The latter both take many seconds to ``flatten," {except in the case of the lowest-mass and lowest compactness progenitors}. We note that explosion energies can ramp up quickly in some MHD explosions \citep{Obergaulinger2018,Muller2018_ultra,Obergaulinger2020}. The explosion energy (here including the stellar mantle overburden binding energy) starts negative and can take more than a second to become positive. Most of the neutrino energy deposited early on by absorption goes into overcoming the gravitational term and only later is the positive value necessary for explosion achieved.  Those numerical simulations in papers that stop within $\sim$1 second of bounce can not possibly provide an informed estimate of the explosion energy. The same can be said for the kick speed \citep{coleman,burrows_kick_2023} $-$ as Figure \ref{fig:t-rs-Mpns} reveals, models halted within a second of bounce can not provide a useful final kick magnitude.

\begin{figure}
    \centering
    \includegraphics[width=0.48\textwidth]{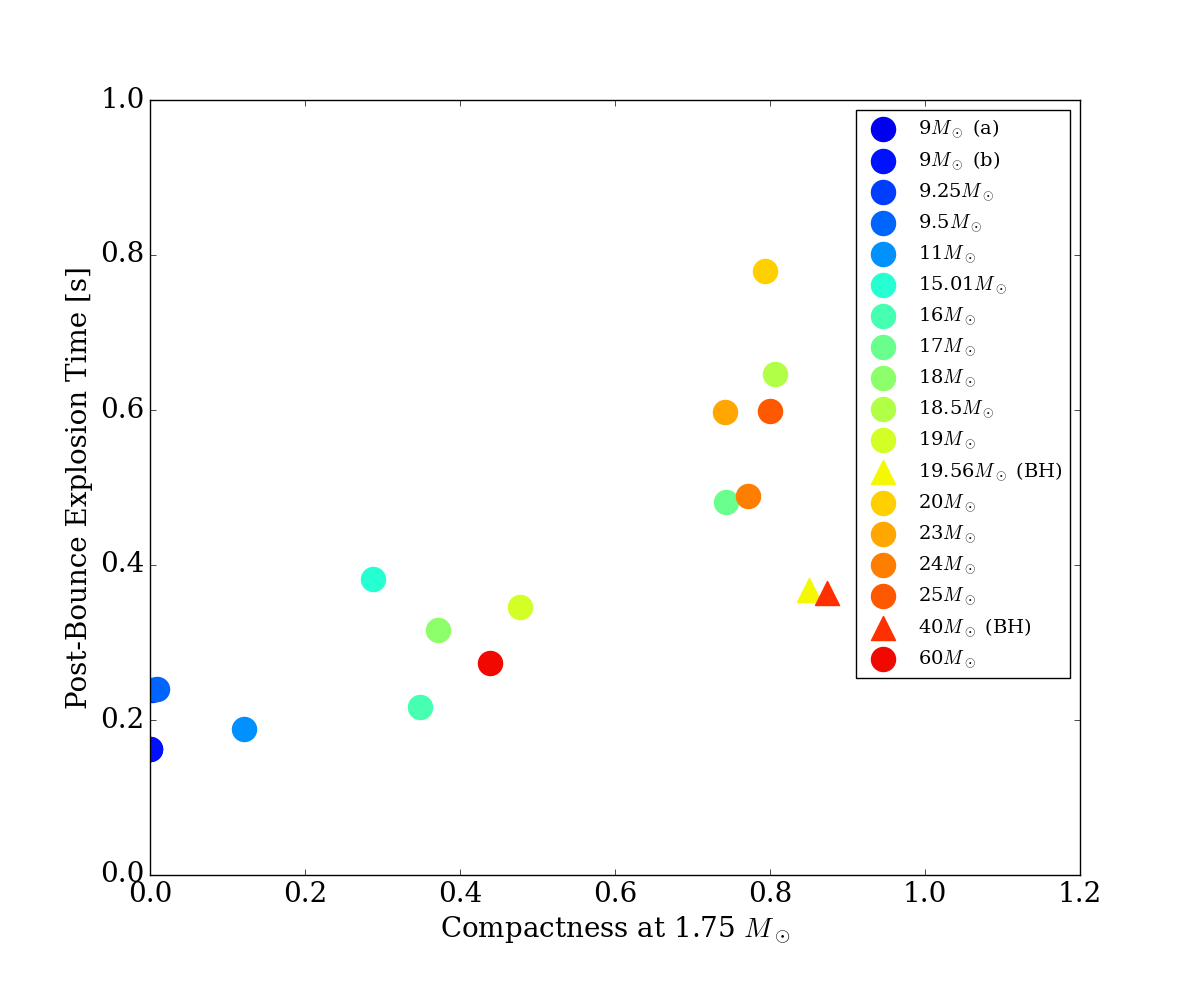}
    \caption{Relation between the explosion time and compactness parameter for the set of twenty models in this study. Note the roughly monotonic behavior, with lower ZAMS mass and lower compactness models exploding more quickly, though with some scatter. See the text in \S\ref{time} for a discussion.}
    \label{fig:cpt-texp}
\end{figure}

\begin{figure}
    \centering
    \includegraphics[width=0.48\textwidth]{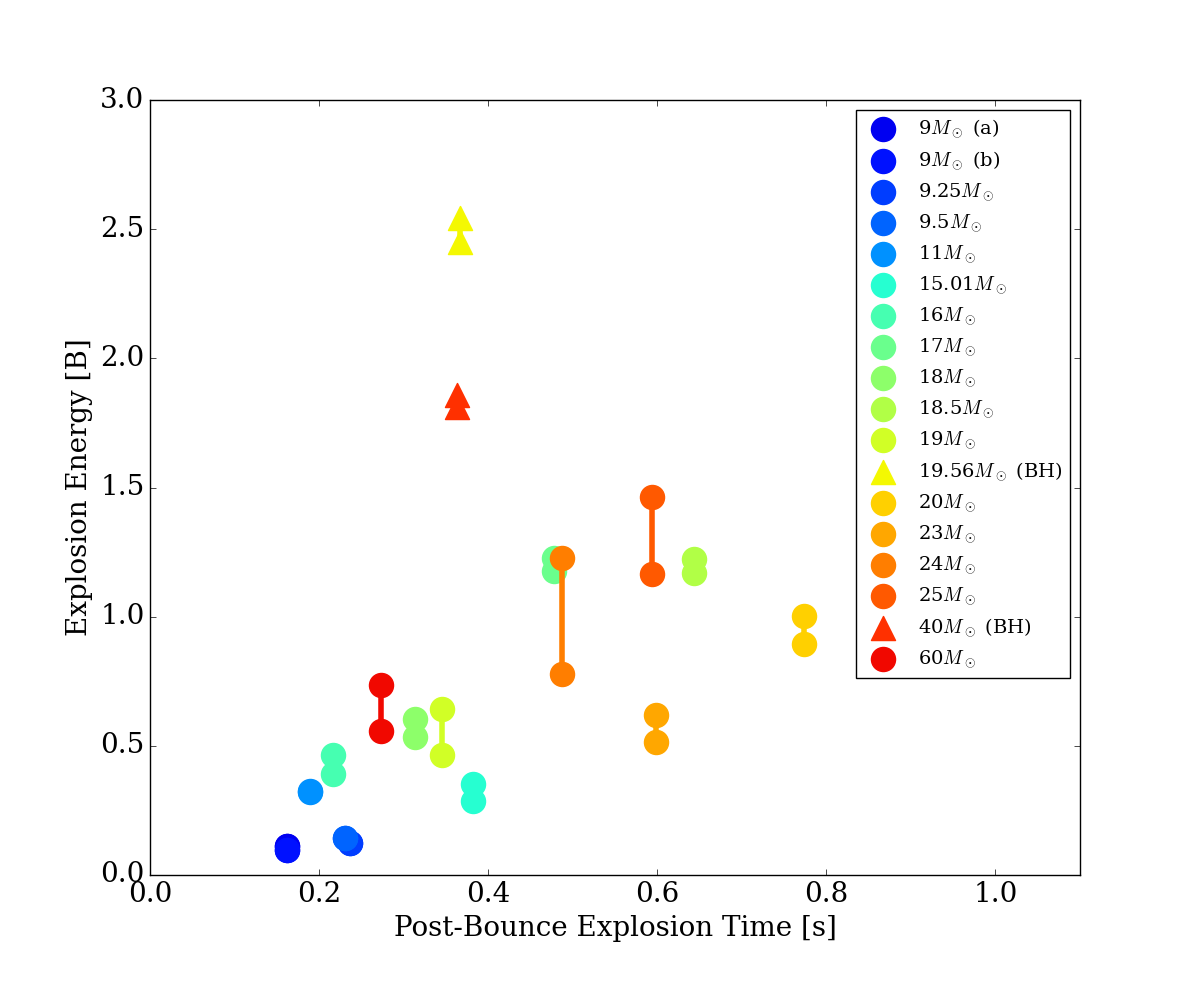}
    \includegraphics[width=0.48\textwidth]{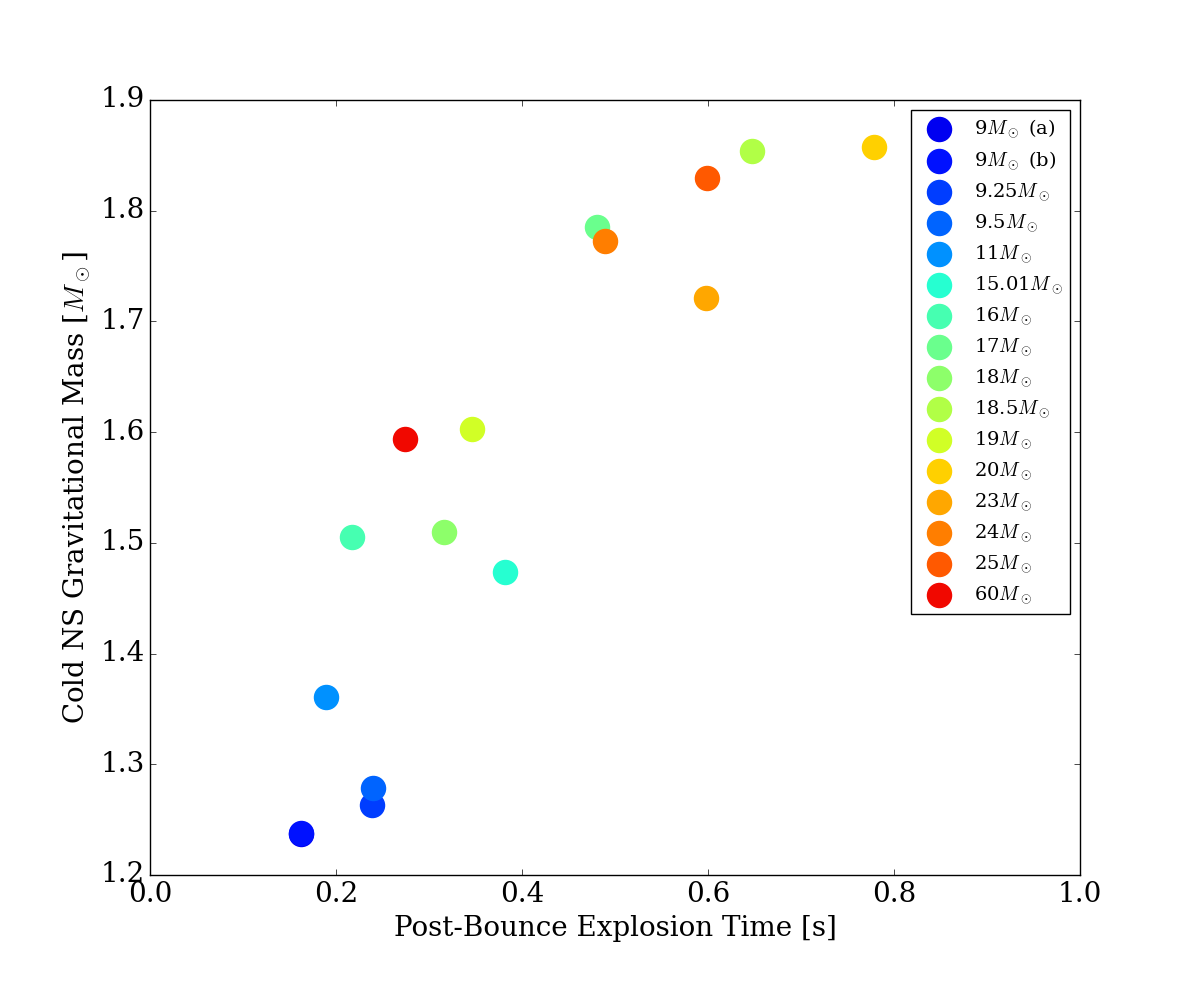}
    \caption{The behavior of the explosion energy and final-state neutron star mass with explosion time.  For some models that haven't quite asymptoted, a range of explosion energies is given, from minimum to projected (likely). Note the roughly monotonic behavior of both the energy and neutron star mass with explosion time. See the discussion in \S\ref{time}.}
    \label{fig:texp-E-mgrav}
\end{figure}

\subsection{Time of Explosion}
\label{time}

Figure \ref{fig:cpt-texp} displays the relation between the compactness parameter and the explosion time. {The explosion time in this work is defined as the time after which the shock never
again recedes in radius.} We see that the relation between the two is very roughly monotonic 
$-$ models with larger compactness take longer to explode, but there is a lot of scatter.  This scatter may in part reflect the chaos of the dynamics, but the degree to which this is true has yet to be determined.  We emphasize that this theme of chaos, stochasticity, and scatter in the correlations suffuses our results and may reflect the very real scatter in the outcomes, even for the same progenitor.  In fact, one expects a distribution function in the observables for the same progenitor, arising from the turbulent chaos in the dynamics.  The width and form of such distribution functions
has yet to be determined and is an important topic for future investigation. One of the motivations for this study of many models, as opposed to only a few, is the idea that in so doing one can glimpse the various correlations that would otherwise be obscured. We think we have achieved this in this paper, but clearly an even more expanded (and expensive!) investigation would be better.    

A more interesting plot is Figure \ref{fig:texp-E-mgrav}, which depicts the relationship between explosion energy (left) and cold, final-state neutron star gravitational mass (right) and time to explosion.  The behavior of the neutron star mass with post-explosion time is roughly monotonic and expected. The longer the delay to explosion, the more mass can be accumulated for a given accretion rate. Moreover, as Figure \ref{fig:cpt-texp} reflects, the explosion time is greater for greater accretion rate (compactness), making the trend all the more robust. However, the roughly increasing explosion energy with explosion time might seem surprising and this behavior is opposite to naive expectation. Nevertheless, the greater delay to explosion is generally (if not universally) associated with higher accretion-powered luminosities, which after the onset of explosion drive it more vigorously.  This trend is hinted at in the bottom left panel of Figure \ref{fig:t-rs-Mpns}.

\section{Explosion and Debris Morphology}
\label{morph}

\begin{figure}
    \centering
    \includegraphics[width=0.32\textwidth]{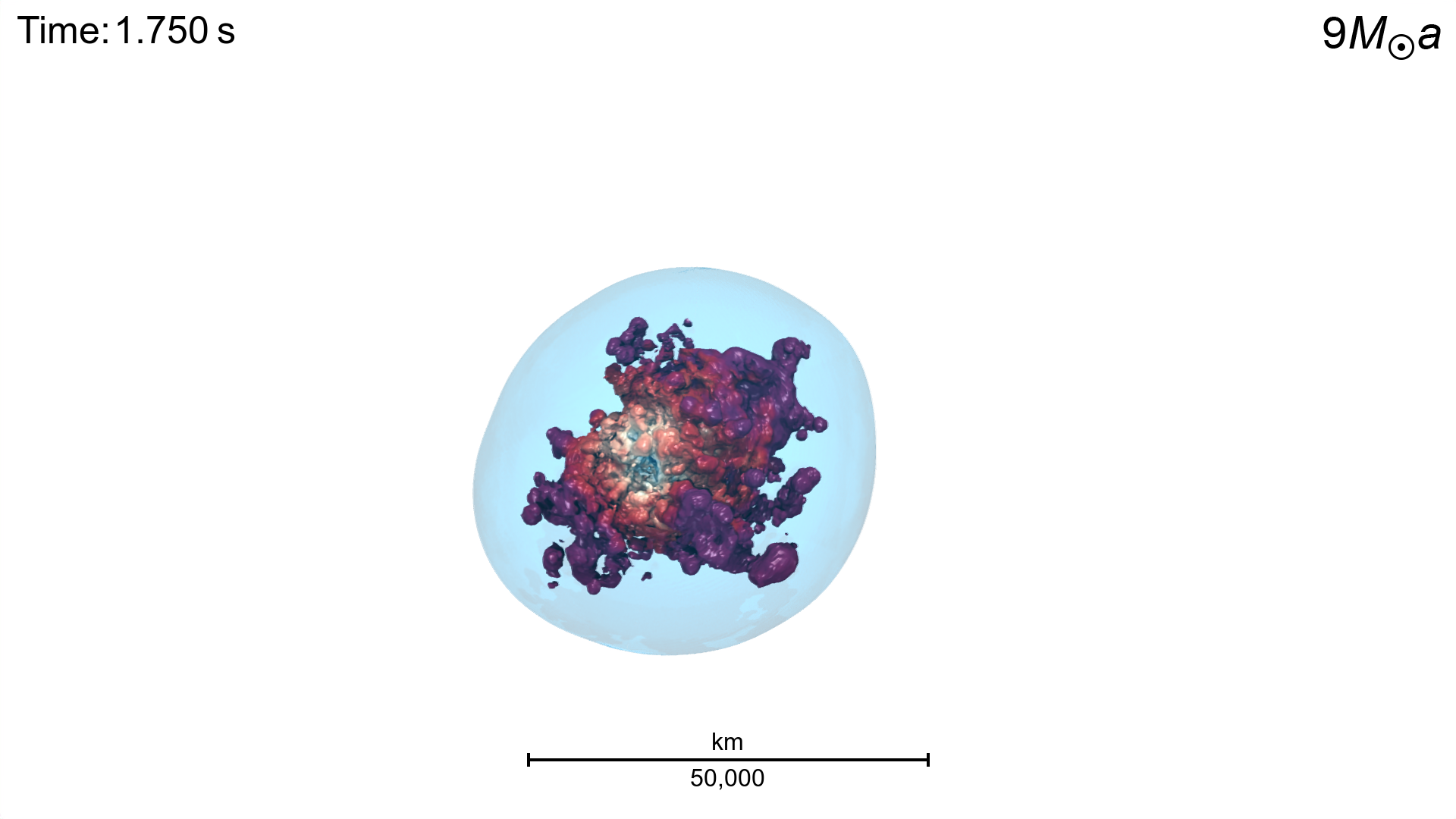}
    \includegraphics[width=0.32\textwidth]{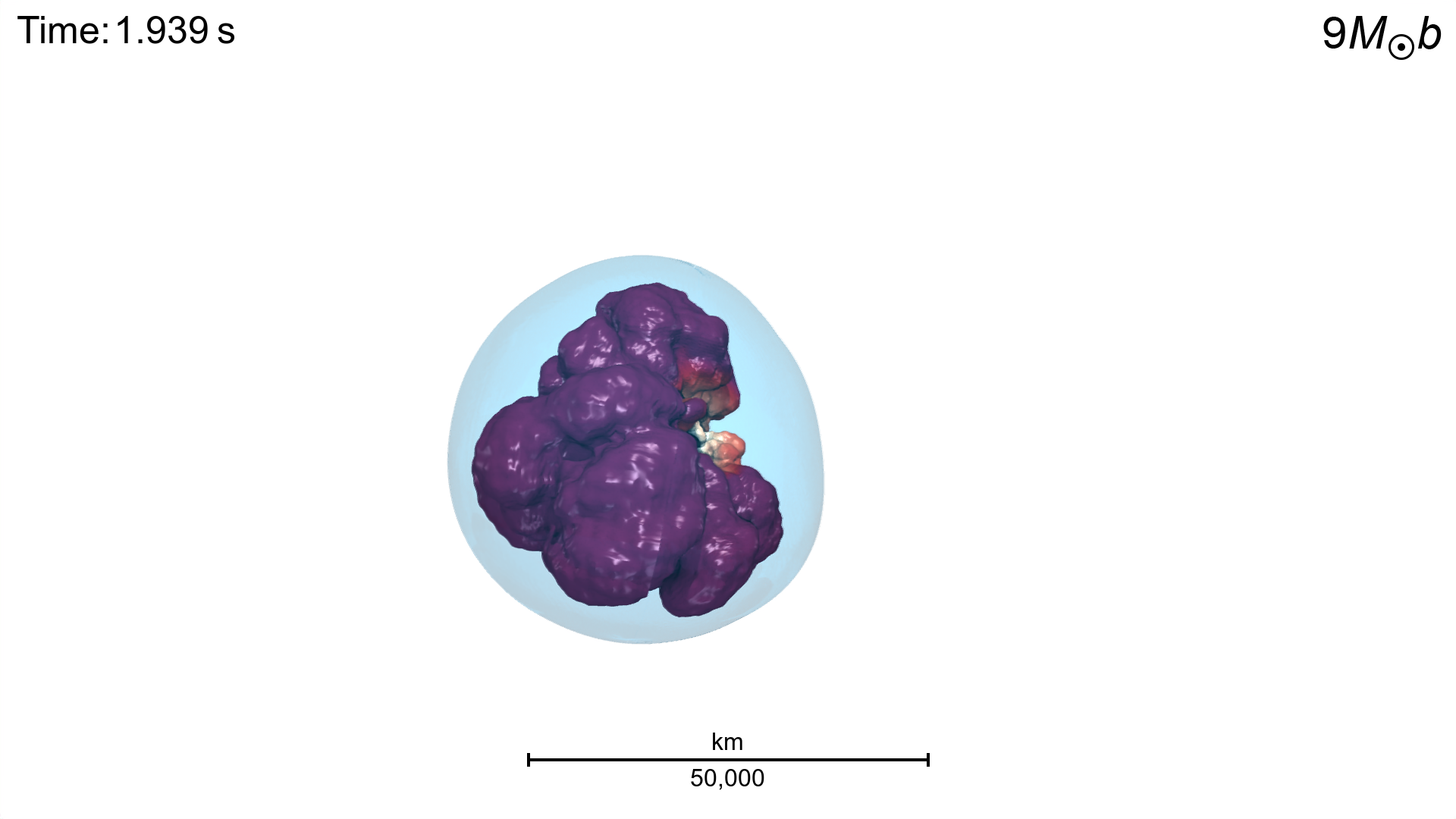}
    \includegraphics[width=0.32\textwidth]{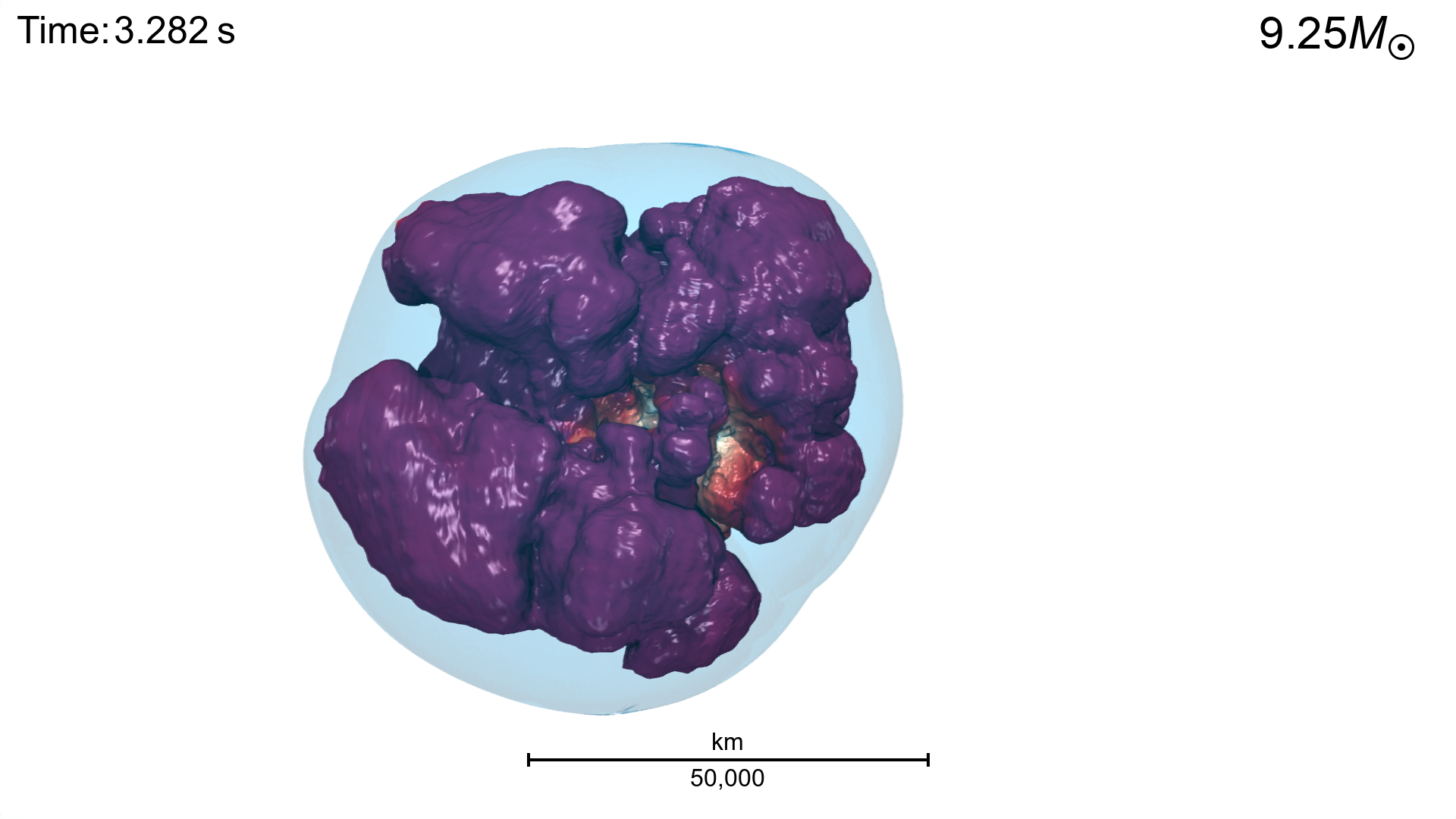}
    \includegraphics[width=0.32\textwidth]{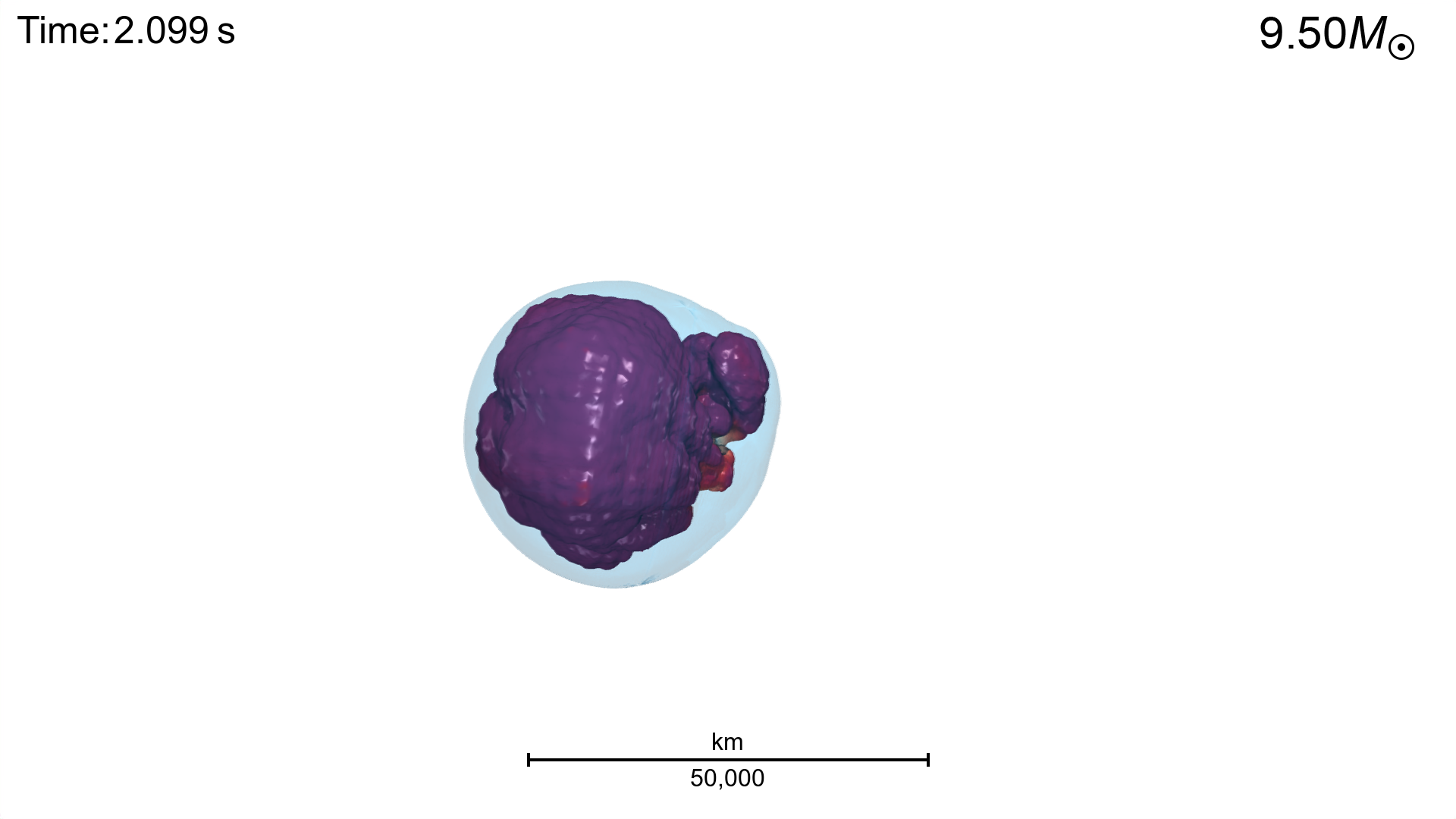}
    \includegraphics[width=0.32\textwidth]{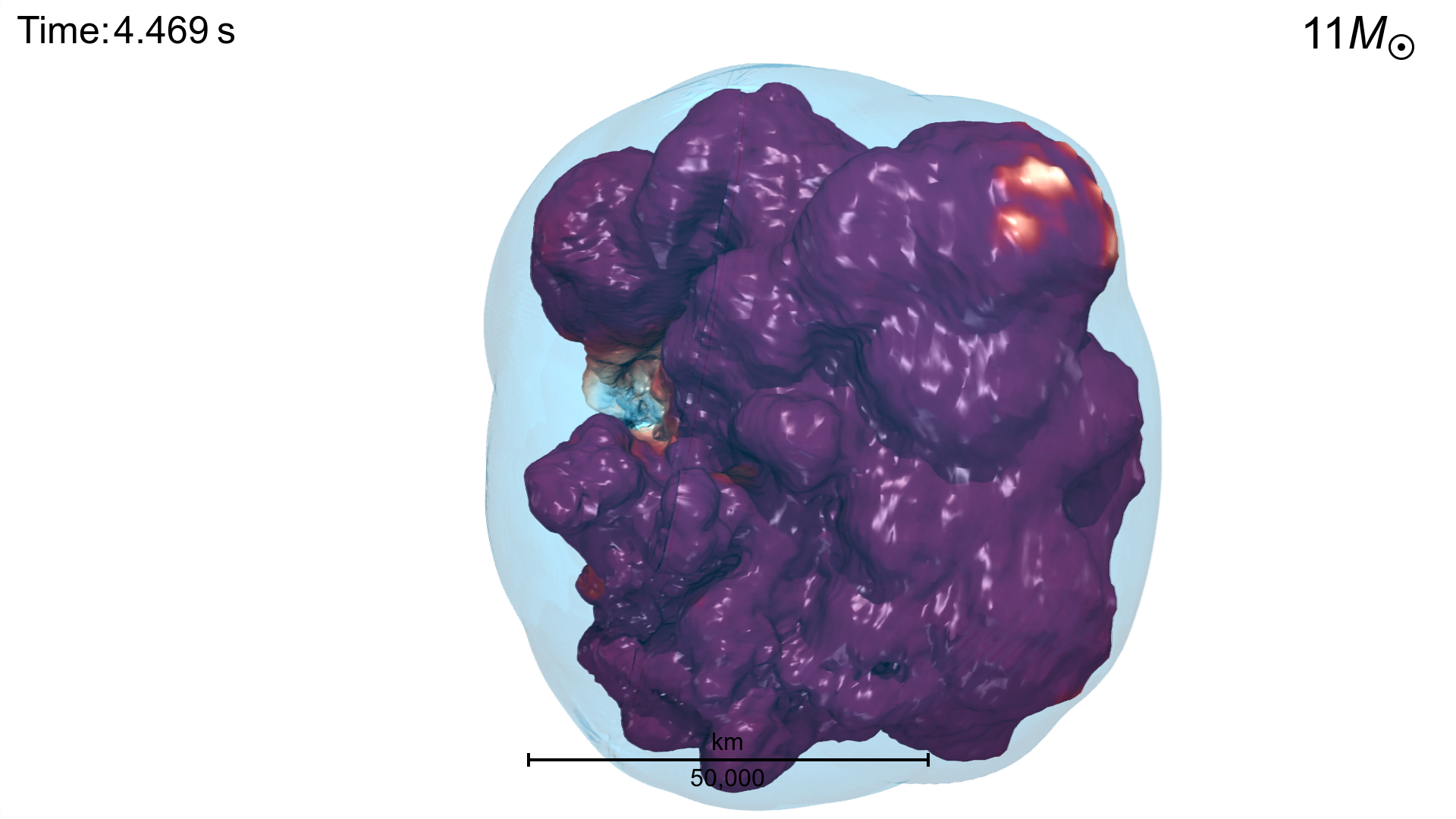}
    \includegraphics[width=0.32\textwidth]{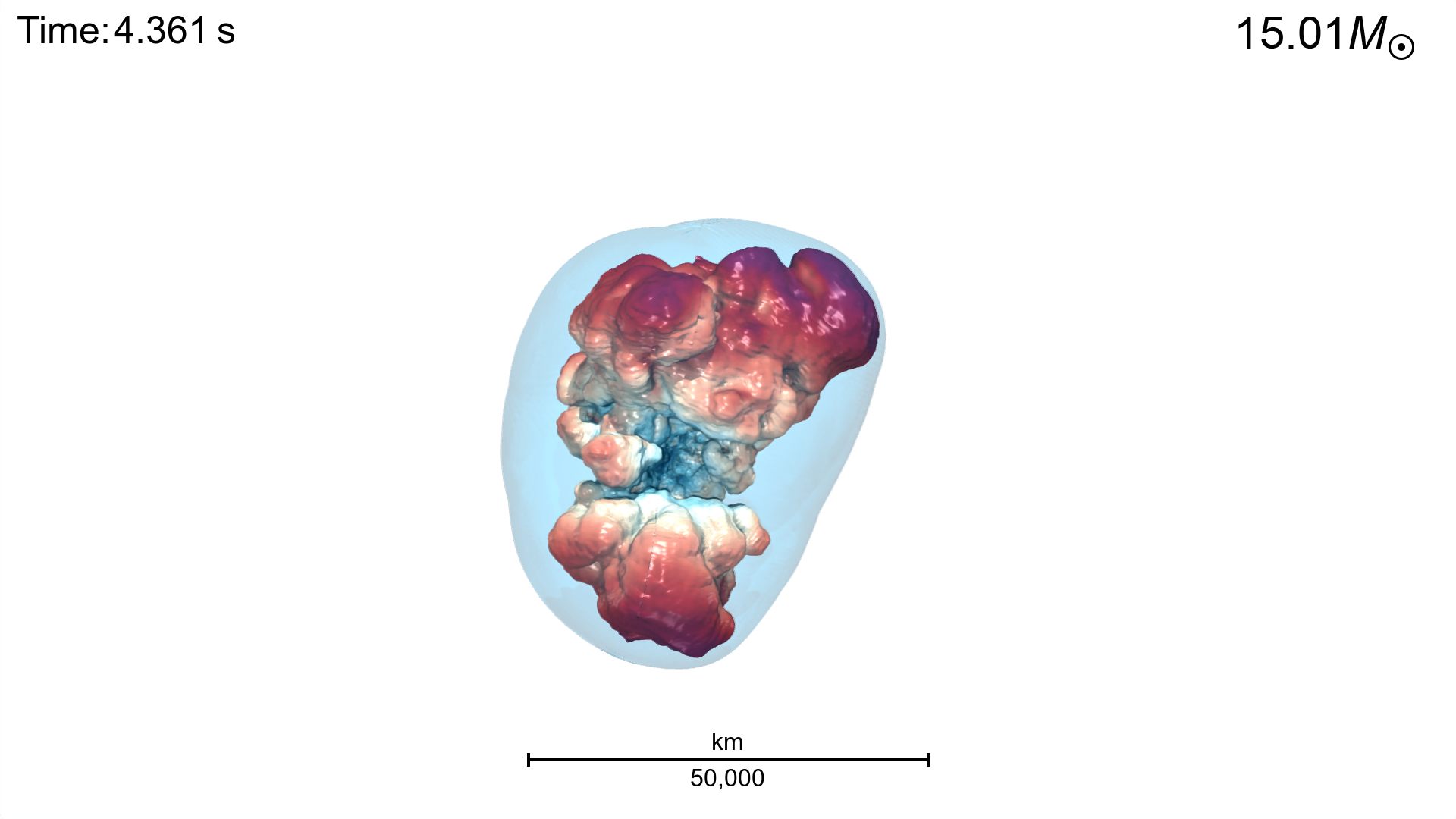}
    \includegraphics[width=0.32\textwidth]{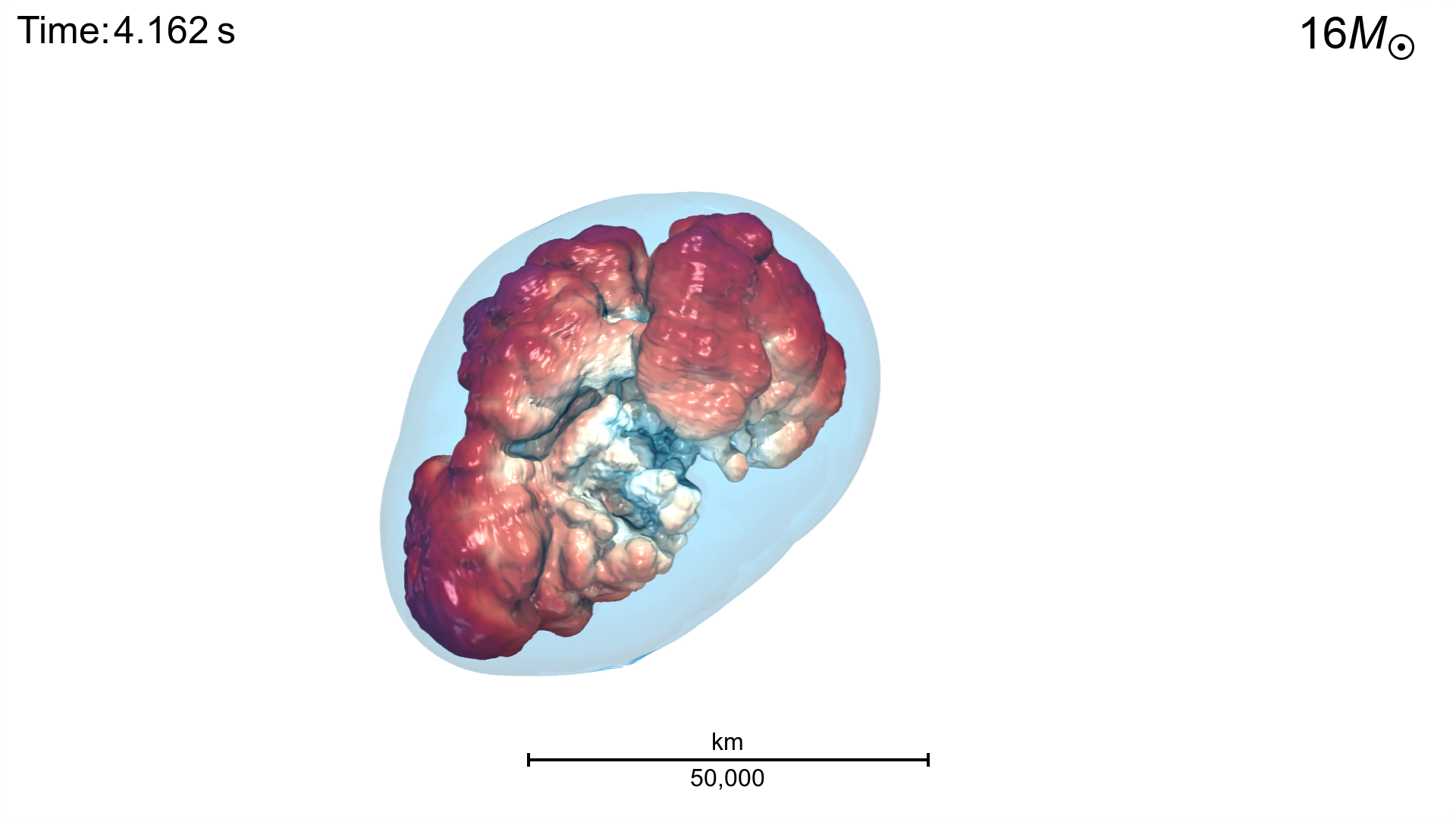}
    \includegraphics[width=0.32\textwidth]{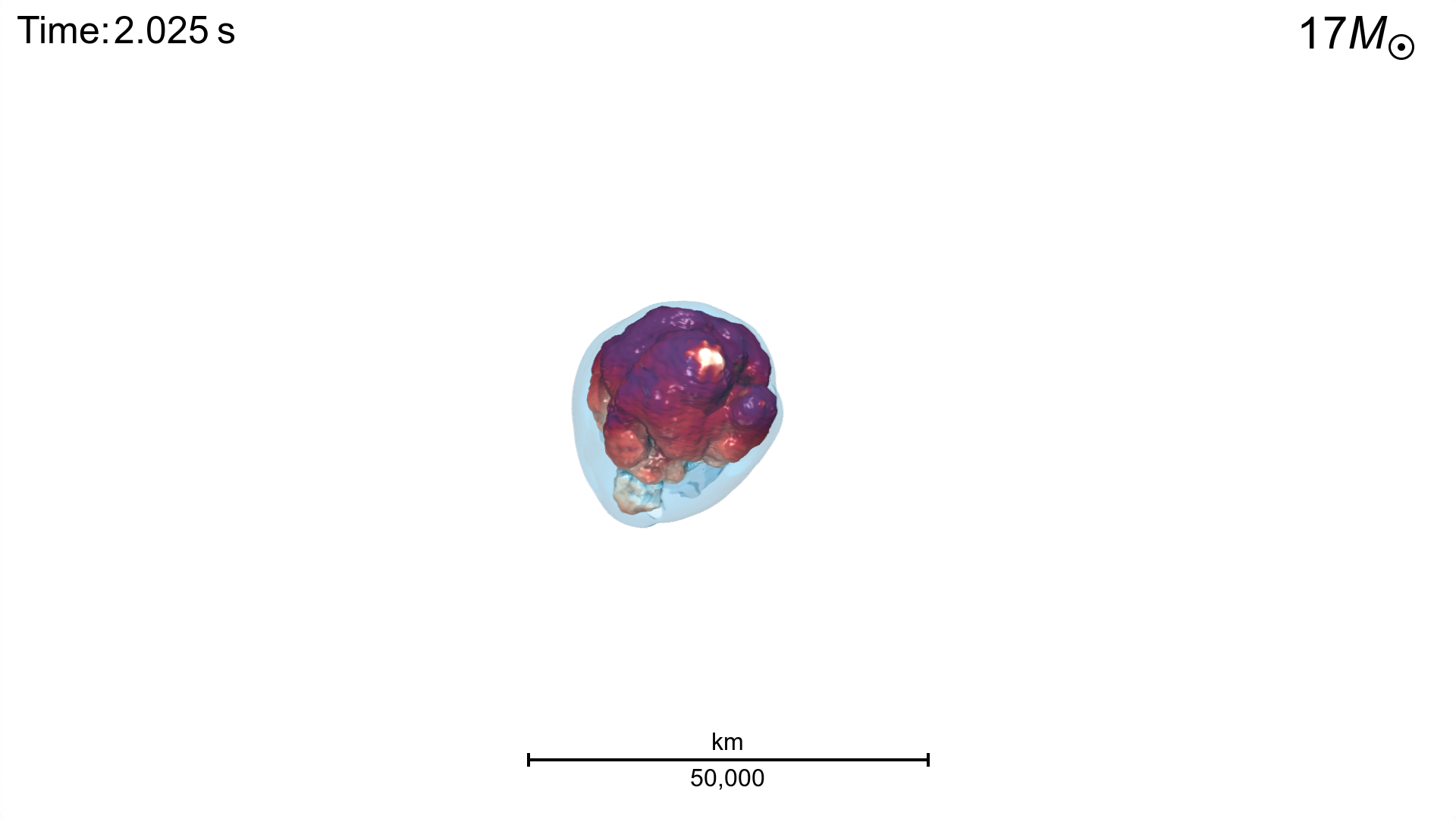}
    \includegraphics[width=0.32\textwidth]{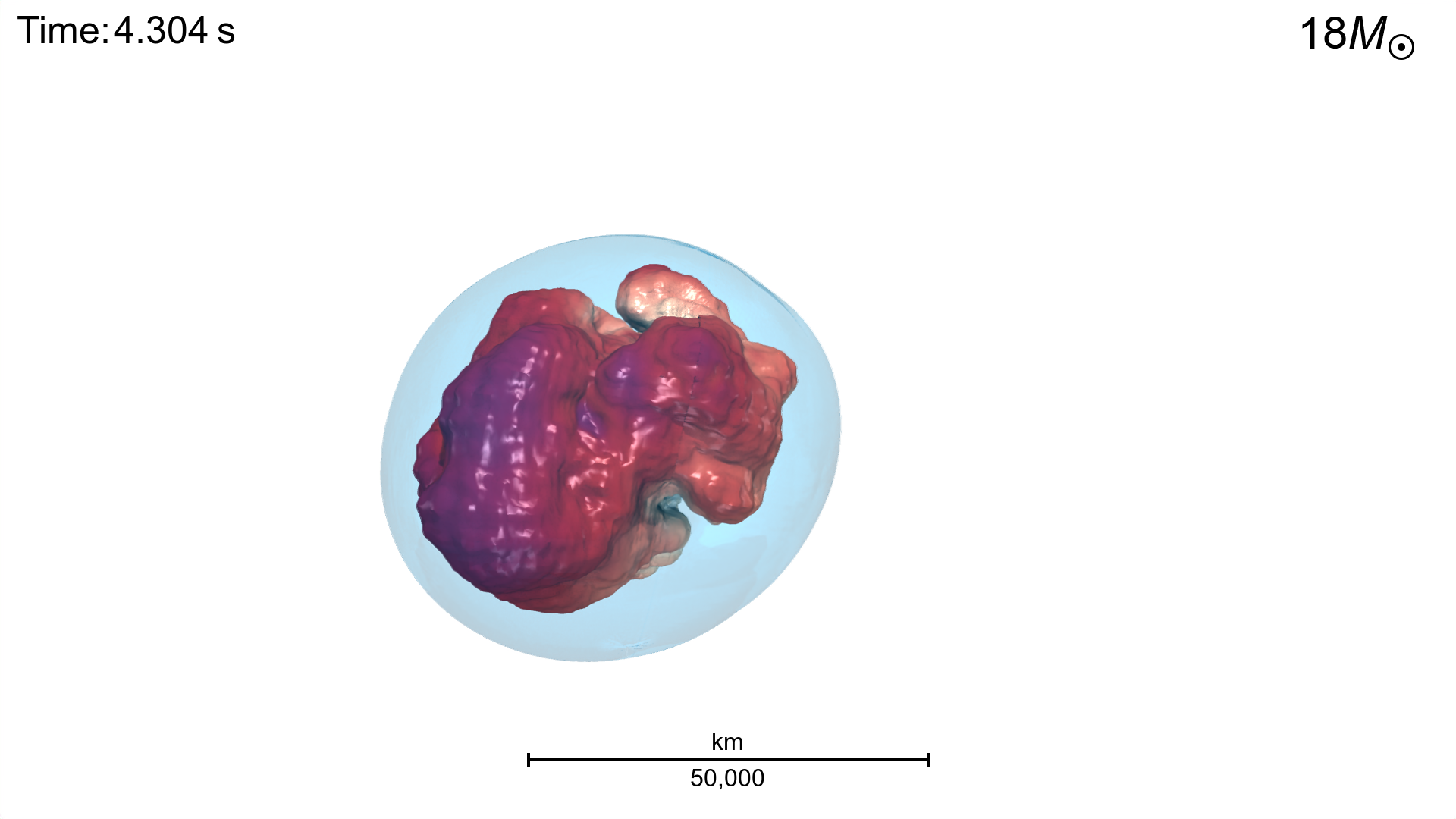}
    \caption{Ejecta morphology of the 9(a), 9(b), 9.25, 9.5, 11, 15.01, 16, 17, and 18 $M_\odot$ models. For most models, the snapshot is taken at the end of the simulation. The 10\% $^{56}$Ni isosurface is shown, colored by the Mach number of the radial velocity. Red means positive Mach number (outflow), while blue means negative Mach number (inflow). The light blue envelope shows the shock position. The shock is rendered as a bluish bounding veil. Unbound $^{16}$O is distributed between the nickel isosurface and the shock. The only difference between models 9(a) and 9(b) is the imposition of slight velocity perturbations upon infall in the former \citep{wang2023}.}
    \label{fig:snapshot}
\end{figure}

\begin{figure}
    \centering
    \includegraphics[width=0.32\textwidth]{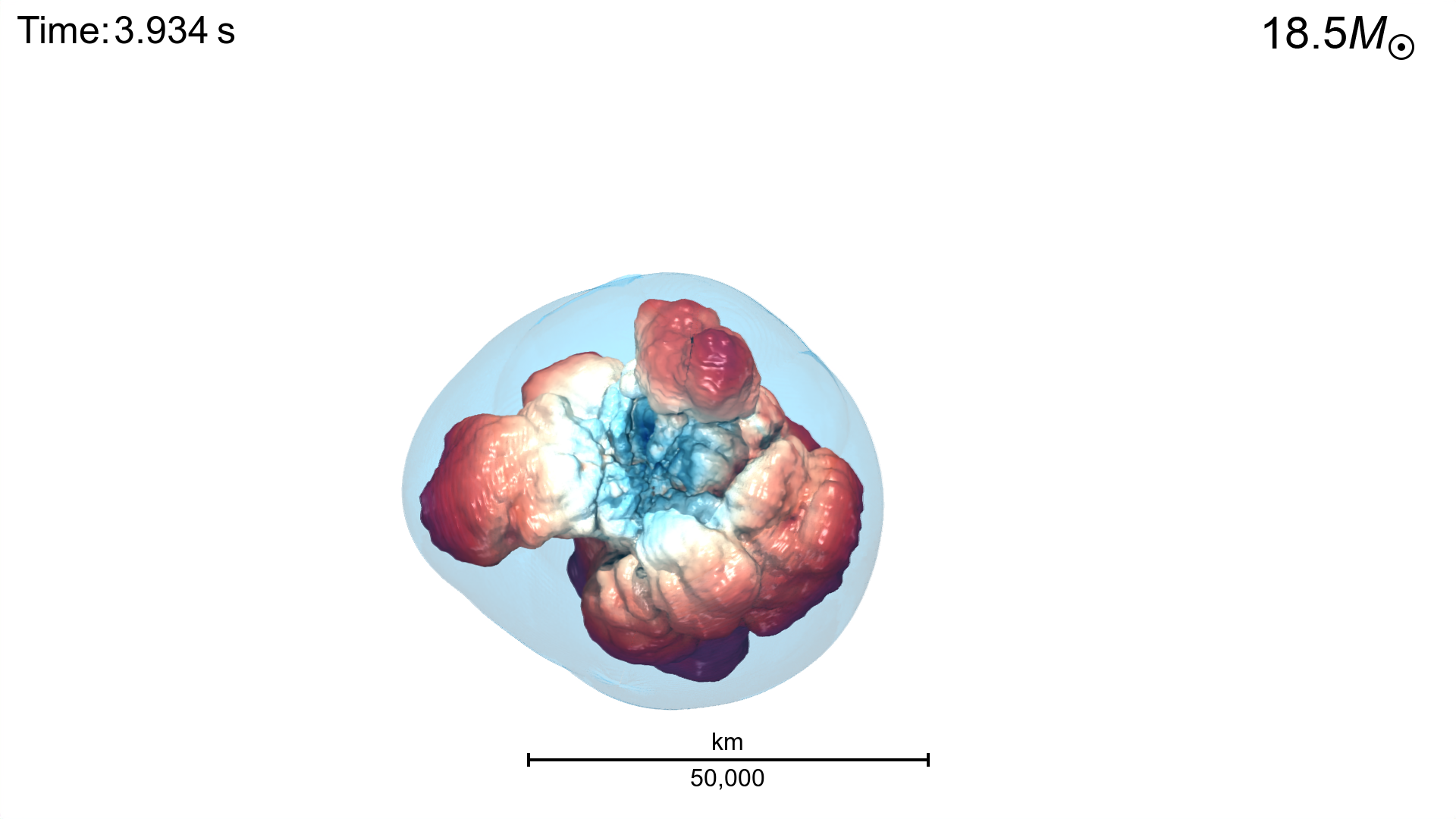}
    \includegraphics[width=0.32\textwidth]{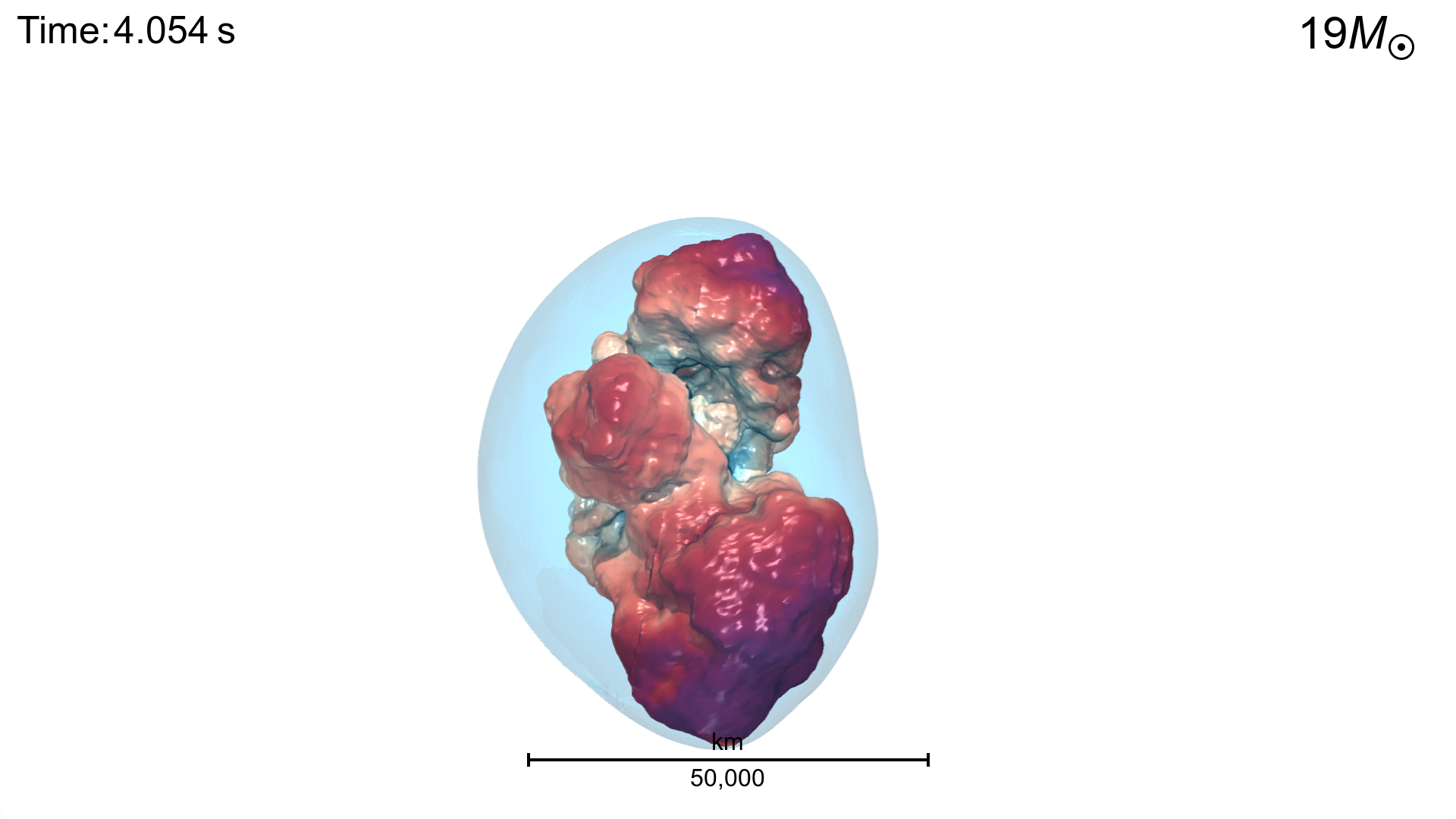}
    \includegraphics[width=0.32\textwidth]{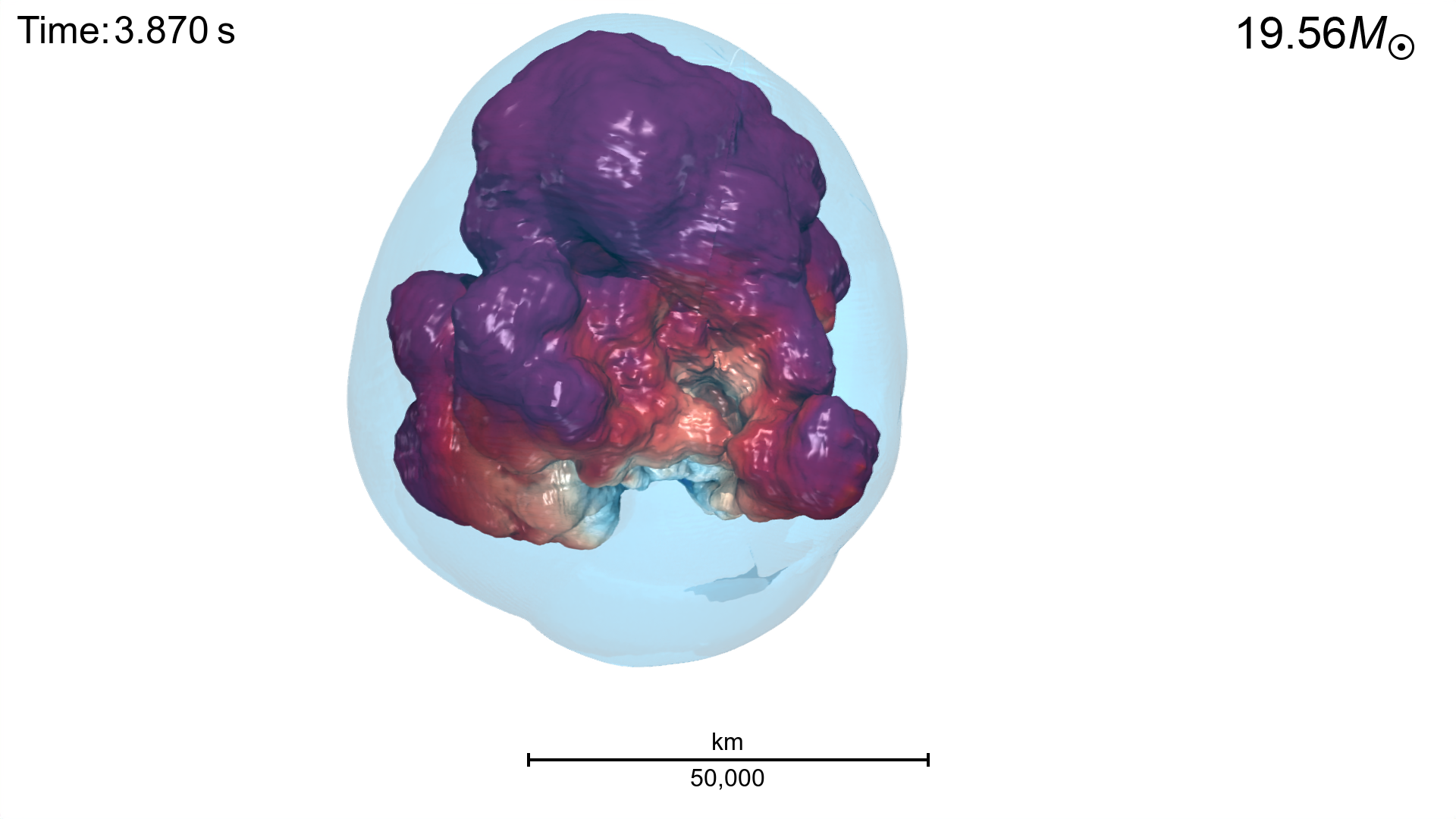}
    \includegraphics[width=0.32\textwidth]{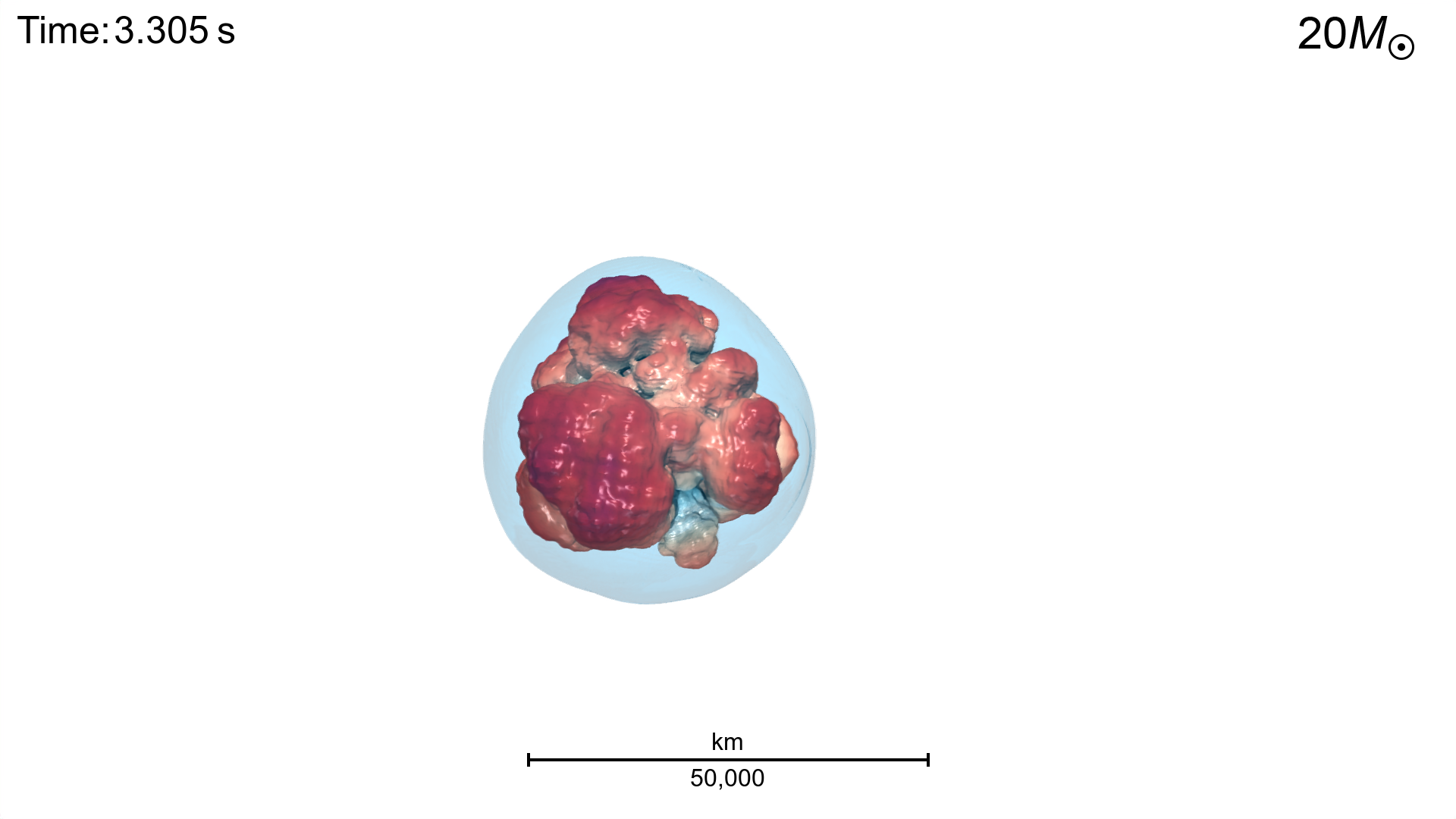}
    \includegraphics[width=0.32\textwidth]{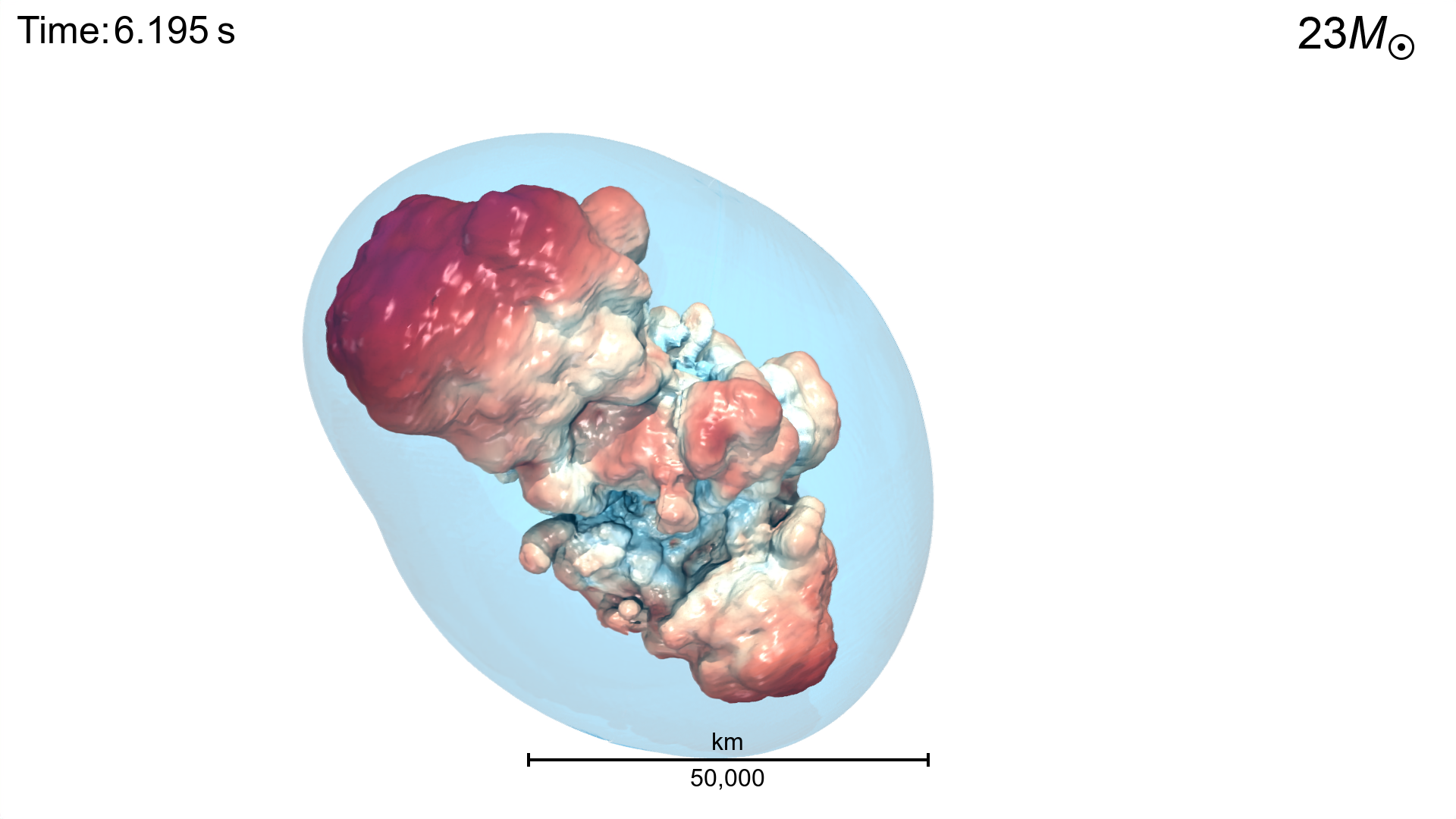}
    \includegraphics[width=0.32\textwidth]{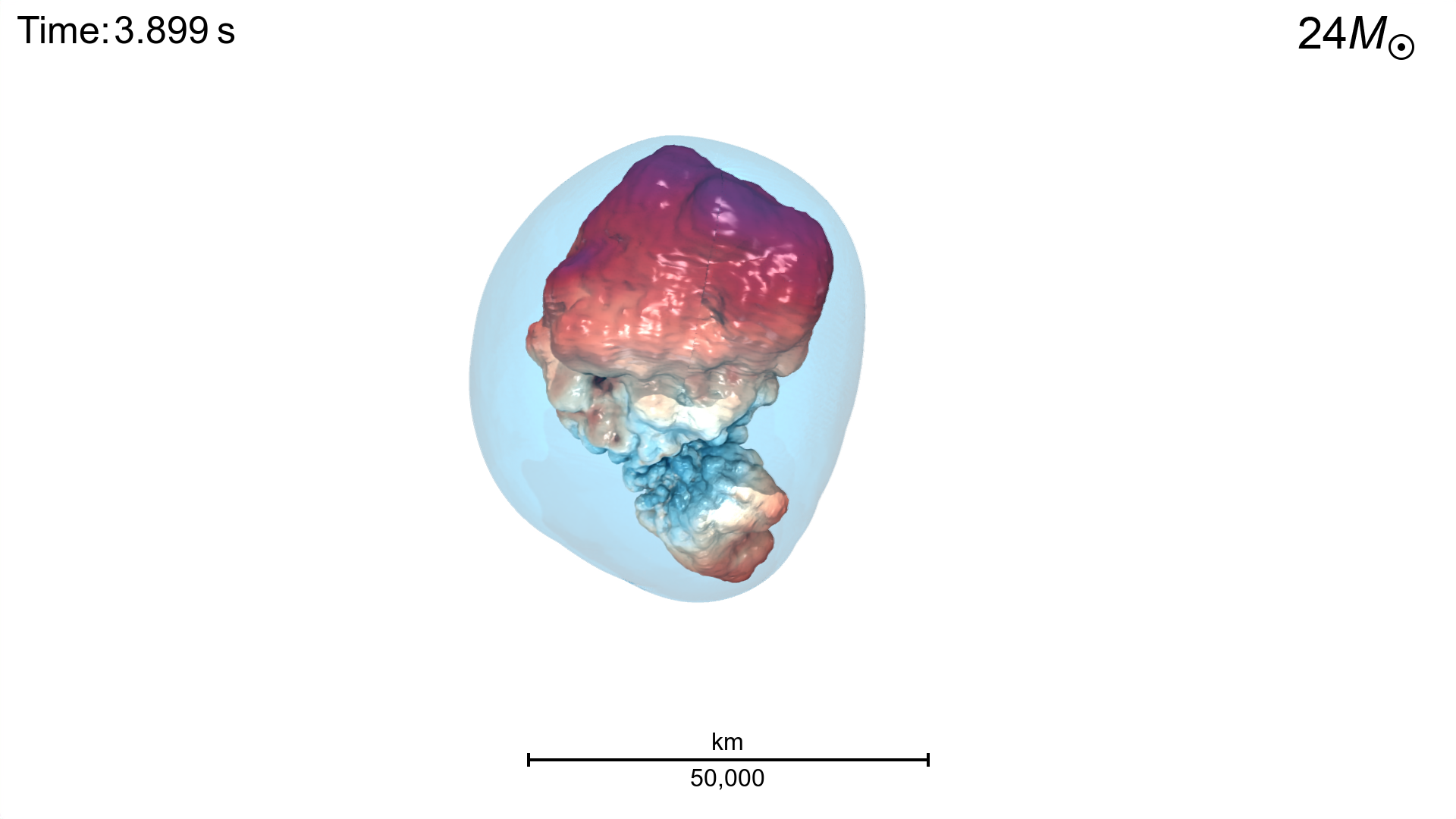}
    \includegraphics[width=0.32\textwidth]{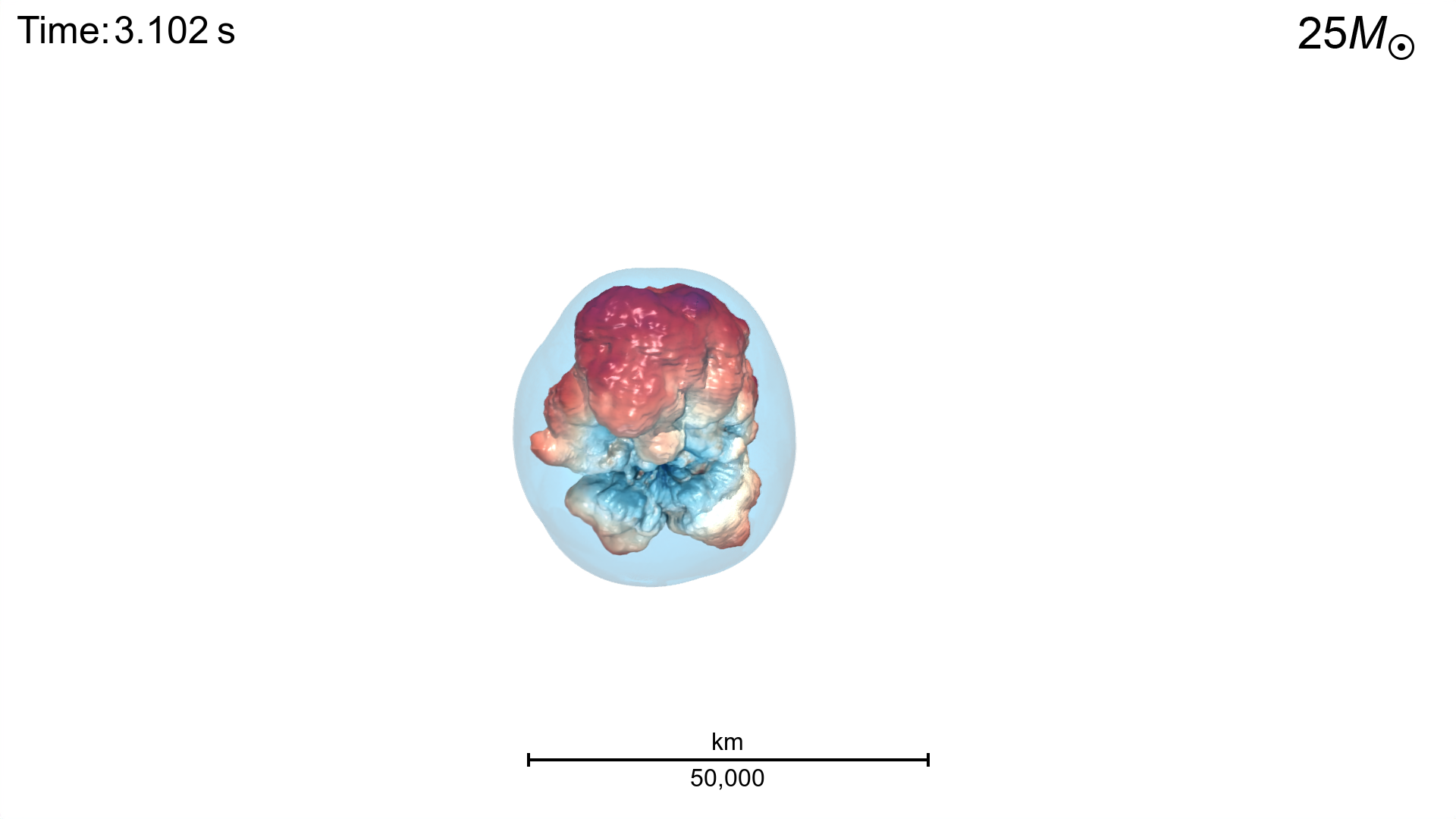}
    \includegraphics[width=0.32\textwidth]{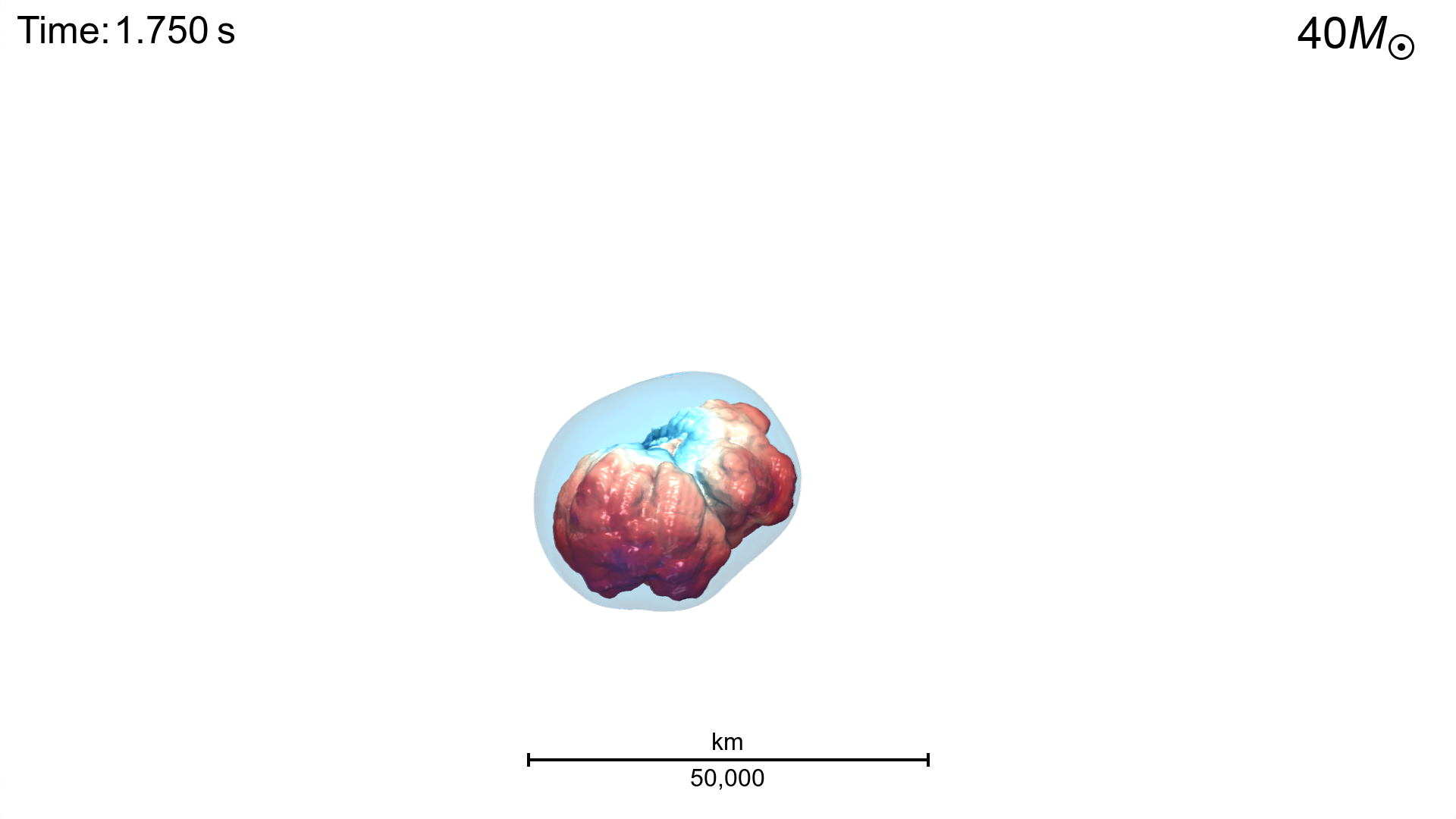}
    \includegraphics[width=0.32\textwidth]{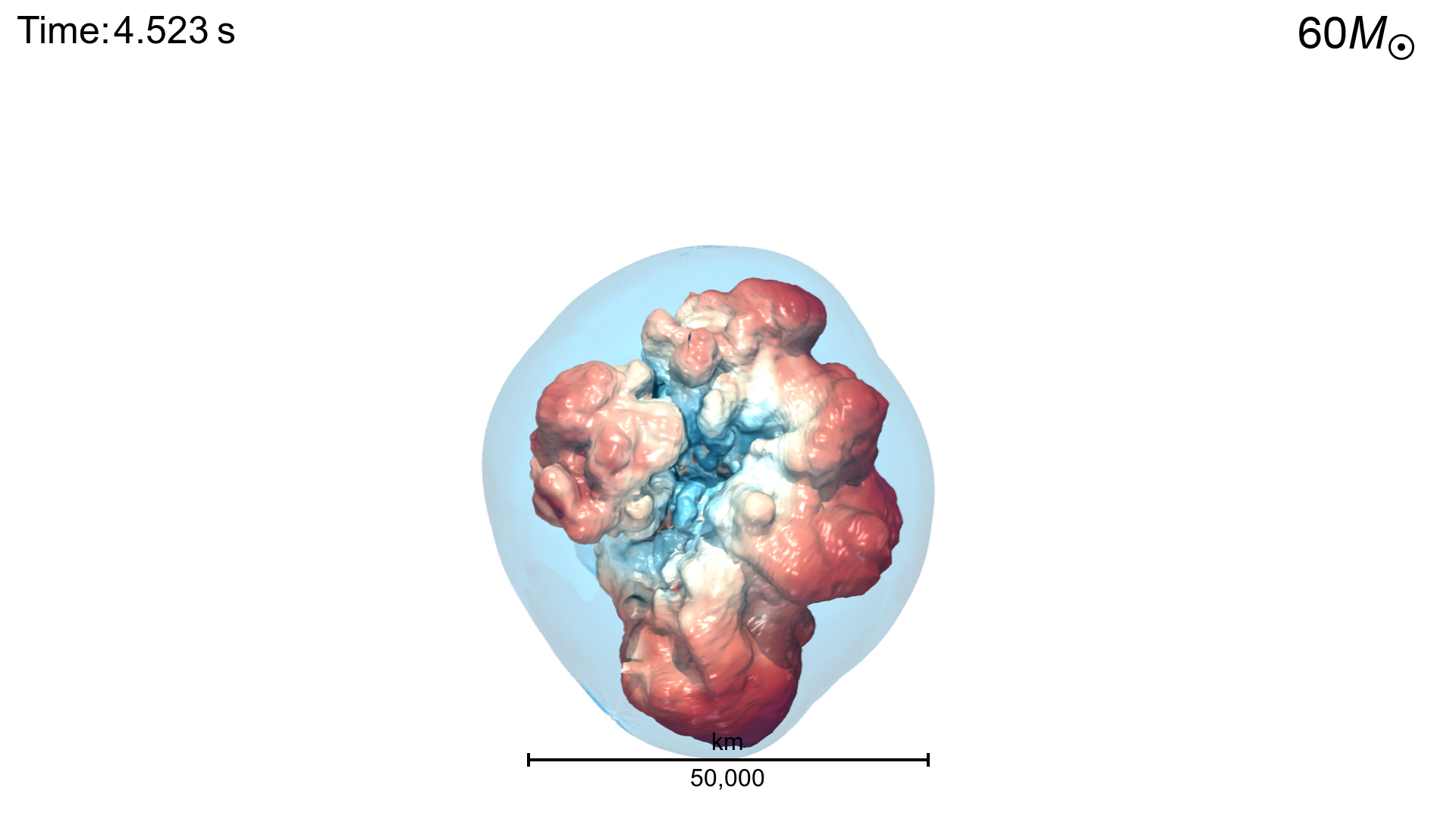}
    \caption{Same as Figure \ref{fig:snapshot-continue}, but for the 18.5, 19, 19.56, 20, 23, 24, 25, 40, and 60 $M_\odot$ models.}
    \label{fig:snapshot-continue}
\end{figure}

The aspherical morphology of the explosion debris reflects the turbulence of the pre- and post-explosion dynamics that has been a central feature of CCSN theory for decades \citep{Burrows1995}\footnote{The Standing Accretion Shock Instability (SASI) \citep{blondin2003}, a vortical-acoustic instability \citep{foglizzo2007}, once seemed manifest in most CCSN simulations. However, the associated up and down ``sloshing" was {in most models} an artifact of the axial effect in 2D simulations; the turbulence witnessed was in fact neutrino-driven convection. However, \citet{foglizzo2006} has shown that when the stalled shock recedes a SASI can emerge, but current 3D modeling \citep{Burrows2020,Burrows2023} reveals it likely to be the spiral SASI \citep{Blondin2007b} {(However, see \citet{Melson2015} for an exception)}. Its emergence can presage a late-time explosion \citep{Andresen2019}, an explosion followed by the formation of a black hole \citep{Burrows2023}, or (as we see quite often \citep{Burrows2023}), a fizzled collapse that will over time lead to the ``quiescent" collapse to a black hole.}.  Bubble structures are universally in evidence (see \citet{sato_hughes_bubbles_2021}), with larger bubbles generally driving the explosions. Importantly, the longer the delay to explosion, the more vigorous the turbulence achieved, with the Mach number of the convective overturning motions prior to explosion getting as high as $\sim$0.5.  However, if the explosion is more prompt, the post-shock turbulence does not have sufficient time to grow to such vigor and large convective structures are not produced.  The result in such cases is a more spherical explosion, albeit with many small bubble structures.  Given the trends of compactness and progenitor ZAMS mass with explosion time exposed in Figure \ref{fig:cpt-texp}, we would expect lower compactness, lower ZAMS mass progenitors would explode more spherically, while those with larger compactness would explode more aspherically. 

\begin{figure}
    \centering
    \includegraphics[width=0.48\textwidth]{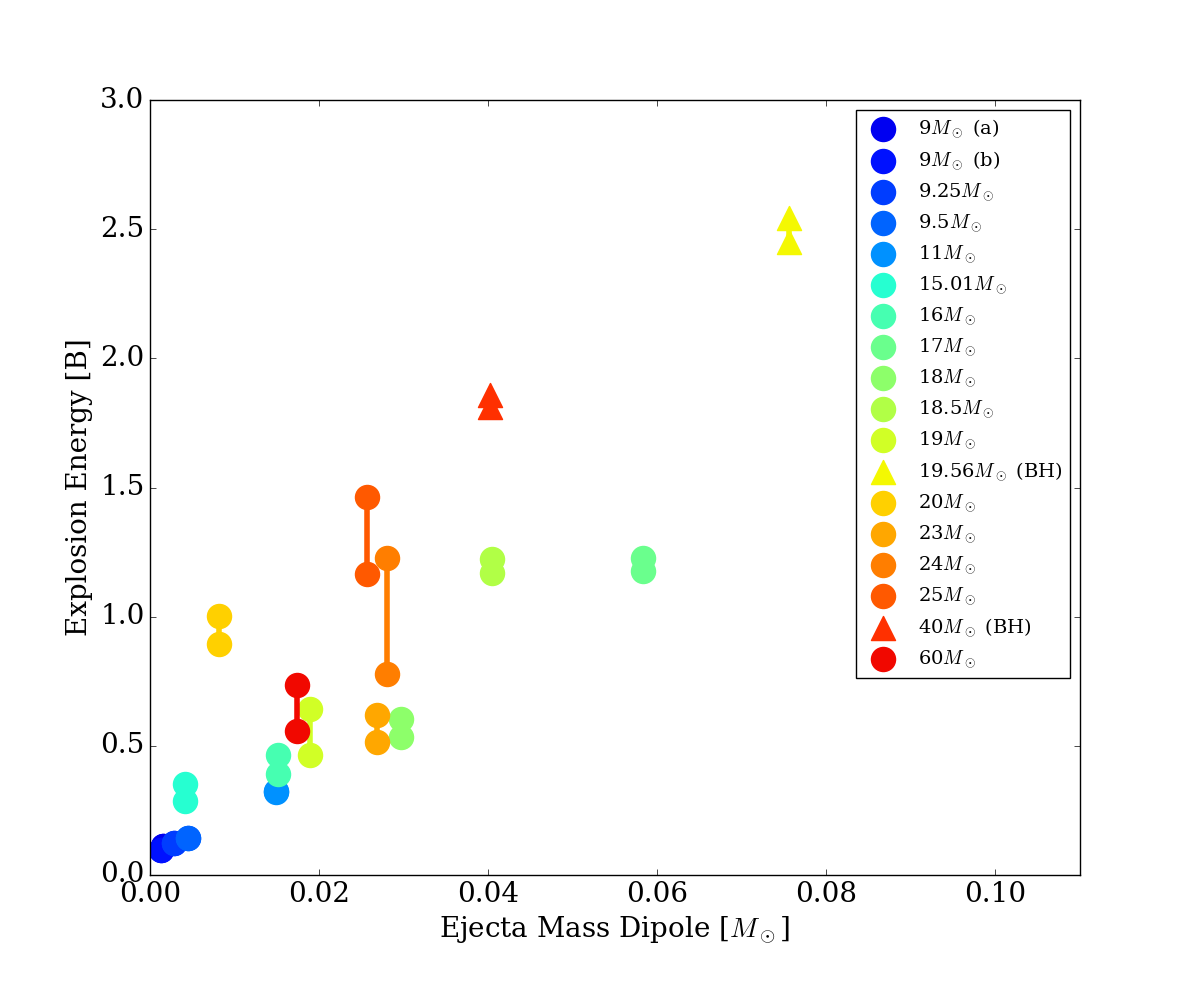}
    \includegraphics[width=0.48\textwidth]{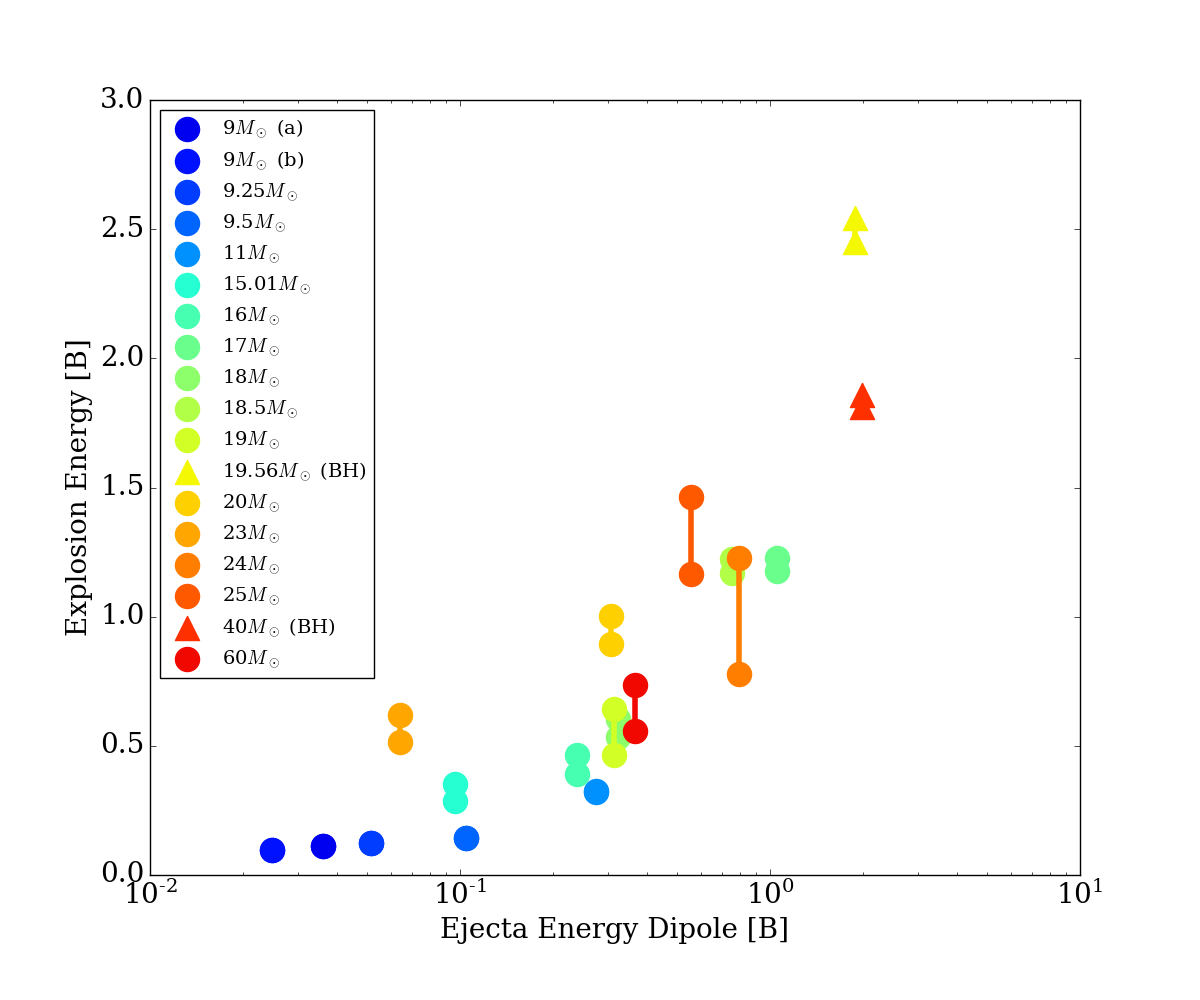}
    \caption{Relation between explosion energy and 1) the ejecta mass dipole (left) and 2) explosion energy dipole (right) for the exploding models in this compilation. {The mass and energy dipoles are the amplitudes of the $l=1$ spherical harmonic components of their angular distributions.}  The roughly monotonic behavior of these indices of asymmetry with explosion energy is a prediction of the neutrino-driven model of core-collapse supernovae.}
    \label{fig:E-dipoleM}
\end{figure}

This is qualitatively what we see. Figures \ref{fig:snapshot} and \ref{fig:snapshot-continue} depict $^{56}$Ni isosurfaces, colored by Mach number (red positive and blue negative), rendered in order of ZAMS mass.  We see that the lower compactness progenitors explode more spherically, while those with higher compactness and longer explosion times explode more axially, with larger driving bubbles.  In fact, the latter generally explode along an axis with a net dipolar and/or quadrupolar structure. Interestingly, we generically see simultaneous explosion in one set of directions, accompanied by accretion in the others \citep{Burrows2020}. This breaking of spherical symmetry is a core behavior of 3D CCSN models. The accretion in one set of directions continues to feed neutrino emissions that drive the explosion in the others, something that can't happen in 1D. Moreover, explosive nucleosynthesis is predominantly along the directions of the driving bubbles.  Complementary to those directions, the shock is weaker and nucleosynthesis by the primary shock is reduced. The upshot is the expectation that unburned oxygen will be concentrated very roughly perpendicular to the regions where most of the $^{56}$Ni and explosive burning products reside.  

Such structures have a direct bearing on the relative line profiles and the spatial distribution of the elements in supernova debris \citep{taubenberger_profiles_2009,sato_hughes_bubbles_2021,qiliang2023} and the associated (anti)correlation of $^{56}$Ni and progenitor oxygen shell material is a qualitative prediction of these 3D models. Moreover, one would expect a correlation with the polarization of the inner debris \citep{leonard_polar_2000,leonard2006,tanaka_polar_2017,nagao_2023_polar}.
Figure \ref{fig:E-dipoleM} portrays the relationship between the explosion energy and the dipole anisotropies of the ejecta mass (left) and the ejecta energies (right) among our set of twenty models. {The dipoles are the amplitudes of the $l=1$ spherical harmonic components of their angular distributions.} These panels suggest a roughly monotonic correlation of two indices of anisotropy with explosion energy that is a new prediction that emerges from our study.  We expect that, though with the common caveat of some scatter, more energetic explosions to evince greater anisotropy, but this needs to be tested.

\section{Correlations}
\label{correlations}

\subsection{Neutron Star Mass}
\label{mass}

\begin{figure}
    \centering
    \includegraphics[width=0.48\textwidth]{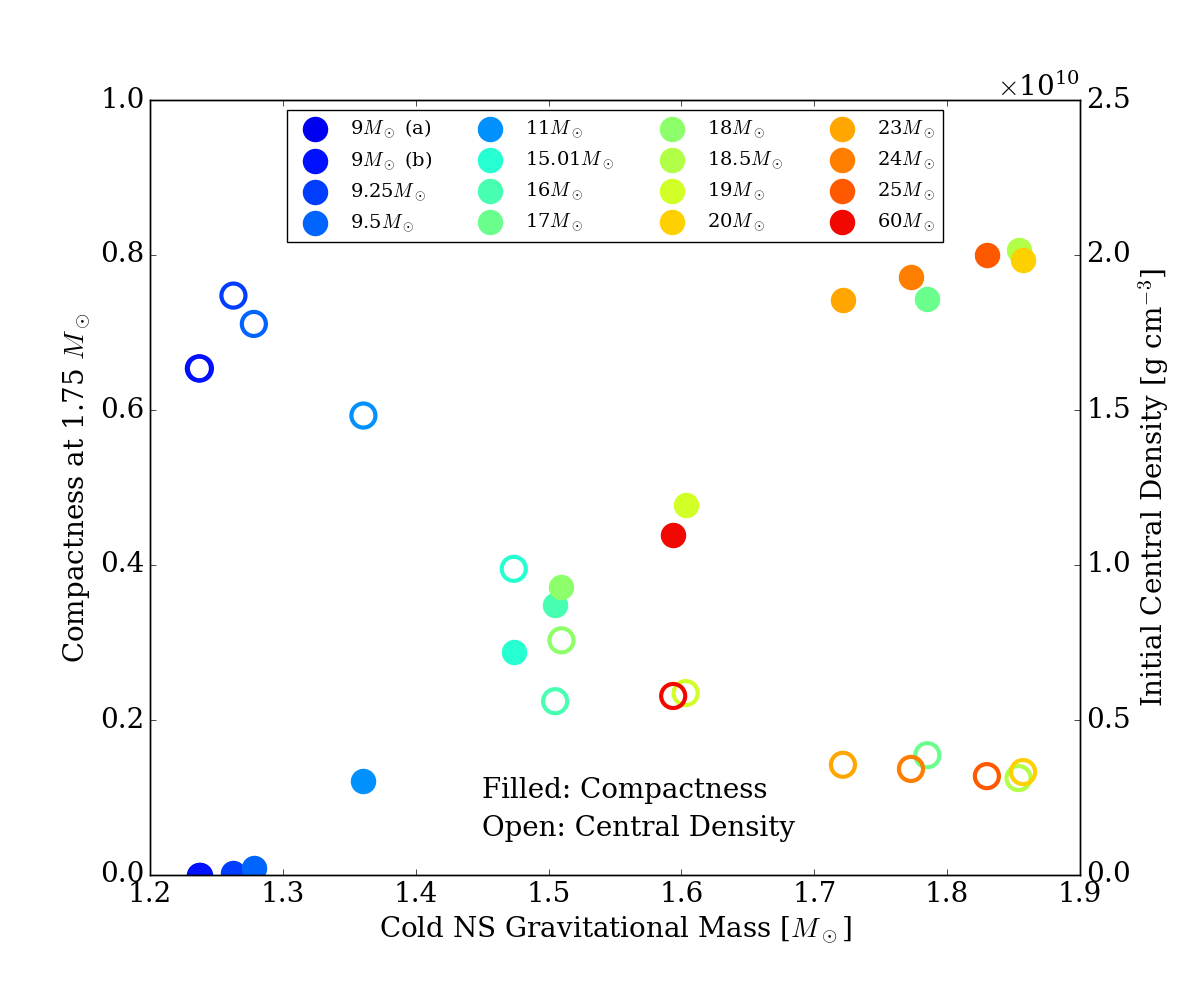}
    \includegraphics[width=0.48\textwidth]{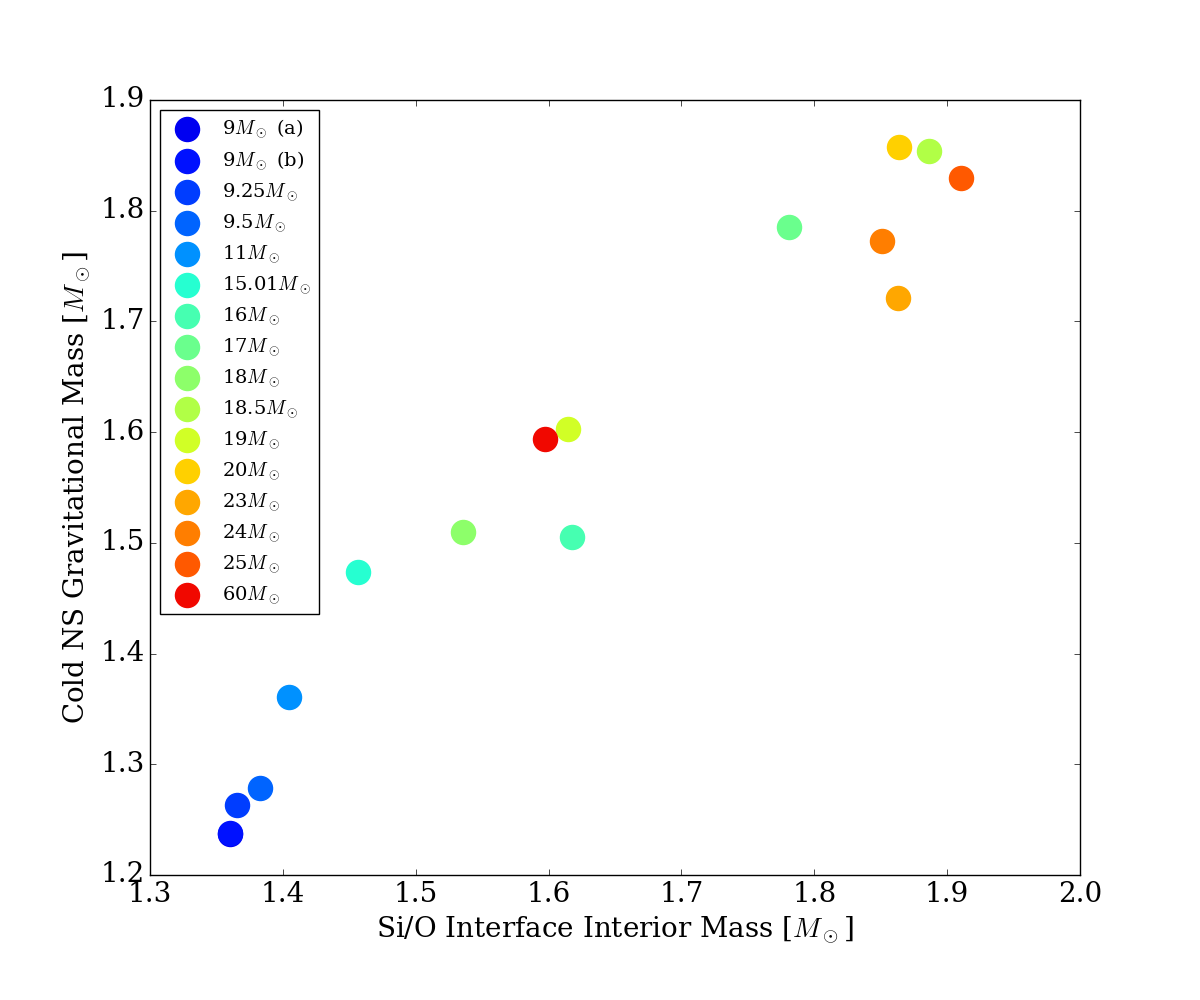}
    \caption{{\bf Left:} Relation between the final-state neutron star gravitational mass and both the compactness parameter (filled symbol) and the initial central density (open symbol), and {\bf Right:} the relation between the final-state neutron star gravitational mass and the Si/O interface mass. The approximately monotonic behavior of the neutron star mass with all three of these quantities is a robust prediction of the theory. {We note that \citet{nakamura2015} derived a similar correlation between neutron star mass and compactness, but calculated in 2D, used the all-too-approximate ``IDSA" neutrino scheme, and did not calculate beyond $\sim$1 second after bounce. 
    A related set of correlations was anticipated by \citet{Ertl2016}.}}
    \label{fig:mgrav-cpt-rhoc-texp}
\end{figure}

The panels in Figure \ref{fig:mgrav-cpt-rhoc-texp} show the dependence of the cold neutron star gravitational mass upon compactness parameter and upon initial central density (left) and the dependence of the neutron star gravitational mass upon the interior mass of the Si/O interface (right). The correlation between the compactness and the neutron star mass succinctly expresses the direct mapping between initial progenitor structure and an observable.

This correlation is stronger than with ZAMS mass, which reflects the non-monotonicity of ZAMS mass with collapsing structure in the \citet{Sukhbold2016} and \citet{Sukhbold2018} progenitor compilation.
Since compactness and mass accretion rate are tightly aligned, and the mass accretion rate is the primary factor in the accumulated mass of the residue of collapse, this relationship is a natural prediction of the neutrino-driven mechanism of CCSNe. With this derived relationship between compactness and gravitational mass, one can use the compactness/ZAMS mass correlation (itself a function of the provenance of the progenitor models) to derive a birth mass function for neutron stars \citep{feryal_masses_2016,fan2023,Vartanyan2023}.  This should be a program for future work.

\subsection{Explosion Energy}
\label{energy}

\begin{figure}
    \centering
    \includegraphics[width=0.80\textwidth]{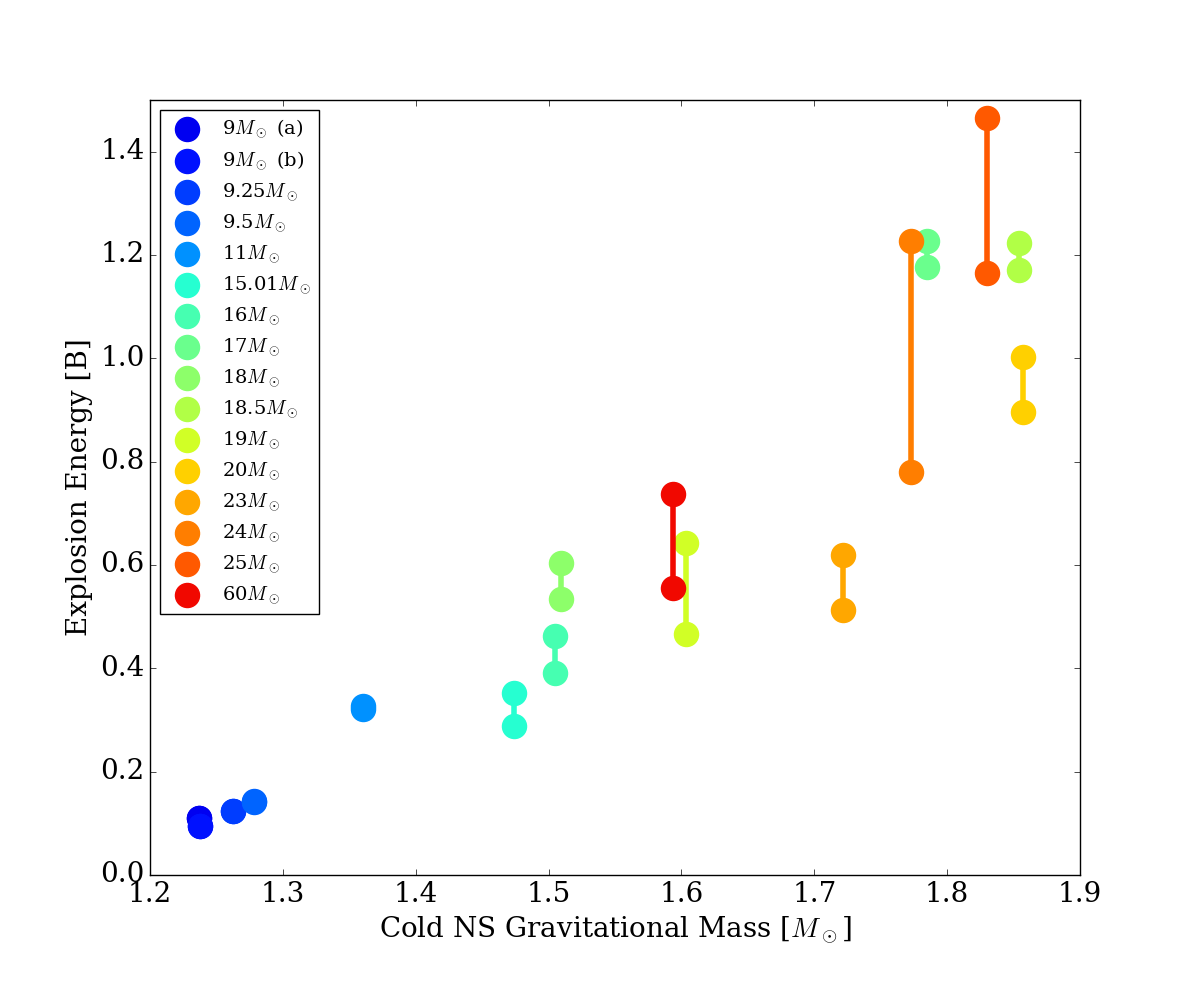}
    \caption{The explosion energy versus the cold, catalyzed neutron star gravitational mass. The range for some models reflects the remaining uncertainty of the final state energies. The monotonic, roughly linear, relationship emerges as one of the most interesting predictions of the neutrino-driven core-collapse model of supernova explosions.  See \S\ref{energy} for a discussion.}
    \label{fig:E-observables}
\end{figure}

Two of the most important observables of CCSN theory are the explosion energy and the gravitational mass of the residual neutron star.  Our theoretical results reveal that the two are tightly coupled. Figure \ref{fig:E-observables} portrays the relation between the two and indicates that, with the caveat that some of our models have not quite asymptoted to their final explosion energies (but are generally within $\sim$10\% or better), the explosion energy is {an almost} monotonic function of neutron star gravitational mass.  This relationship between two observables is one of the central results of this paper and speaks to the importance of our philosophy to explore many long-term 3D state-of-the-art simulations to extract meaningful predictions. This strong mapping between observables would not have been clearly discerned without a 3D investigation of many progenitors to late times and should inform future observational campaigns to extract physically useful parameters of core-collapse supernova explosions.
The study of but one single model would not be so illuminating.

\begin{figure}
    \centering
    \includegraphics[width=0.48\textwidth]{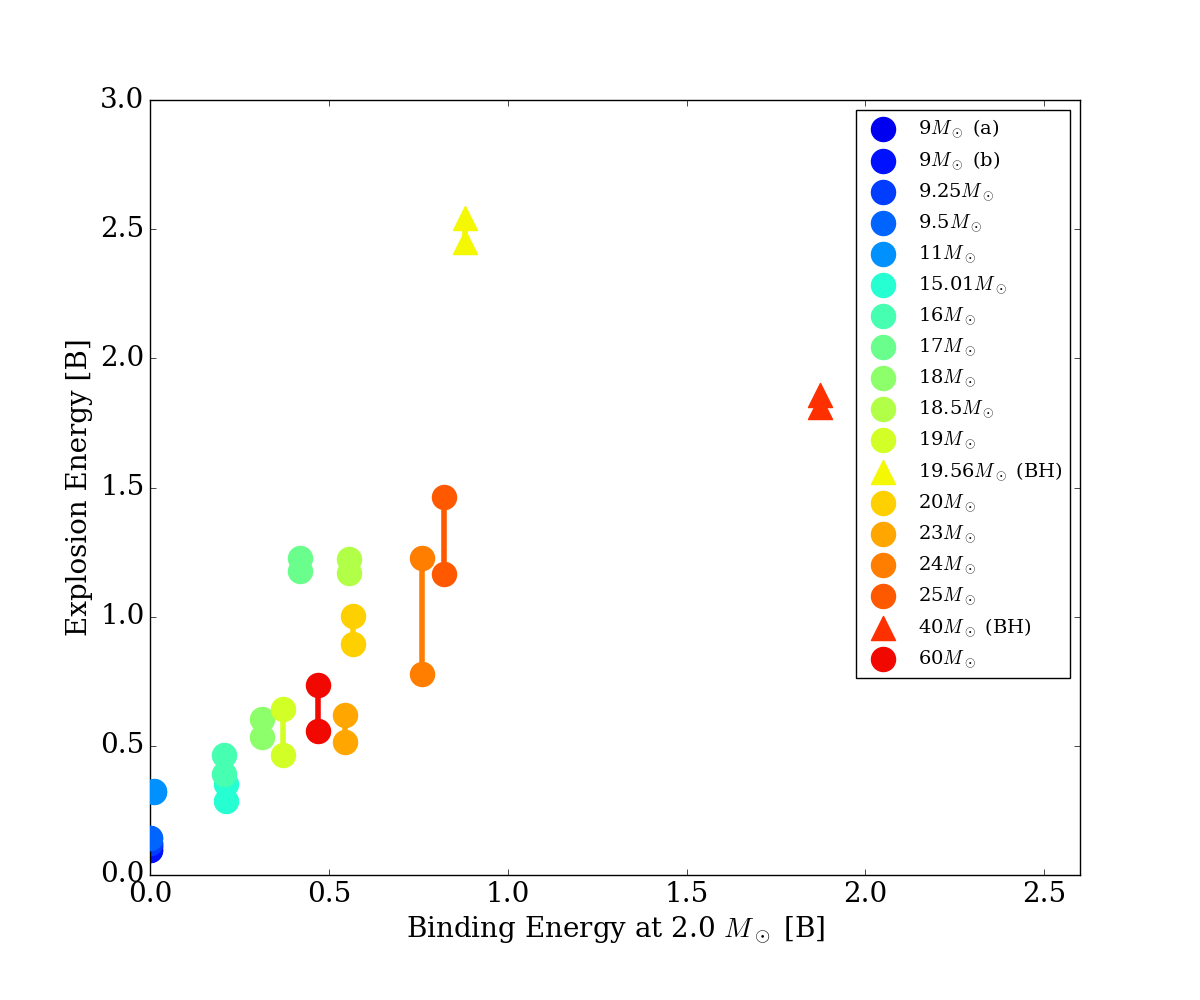}
    \includegraphics[width=0.48\textwidth]{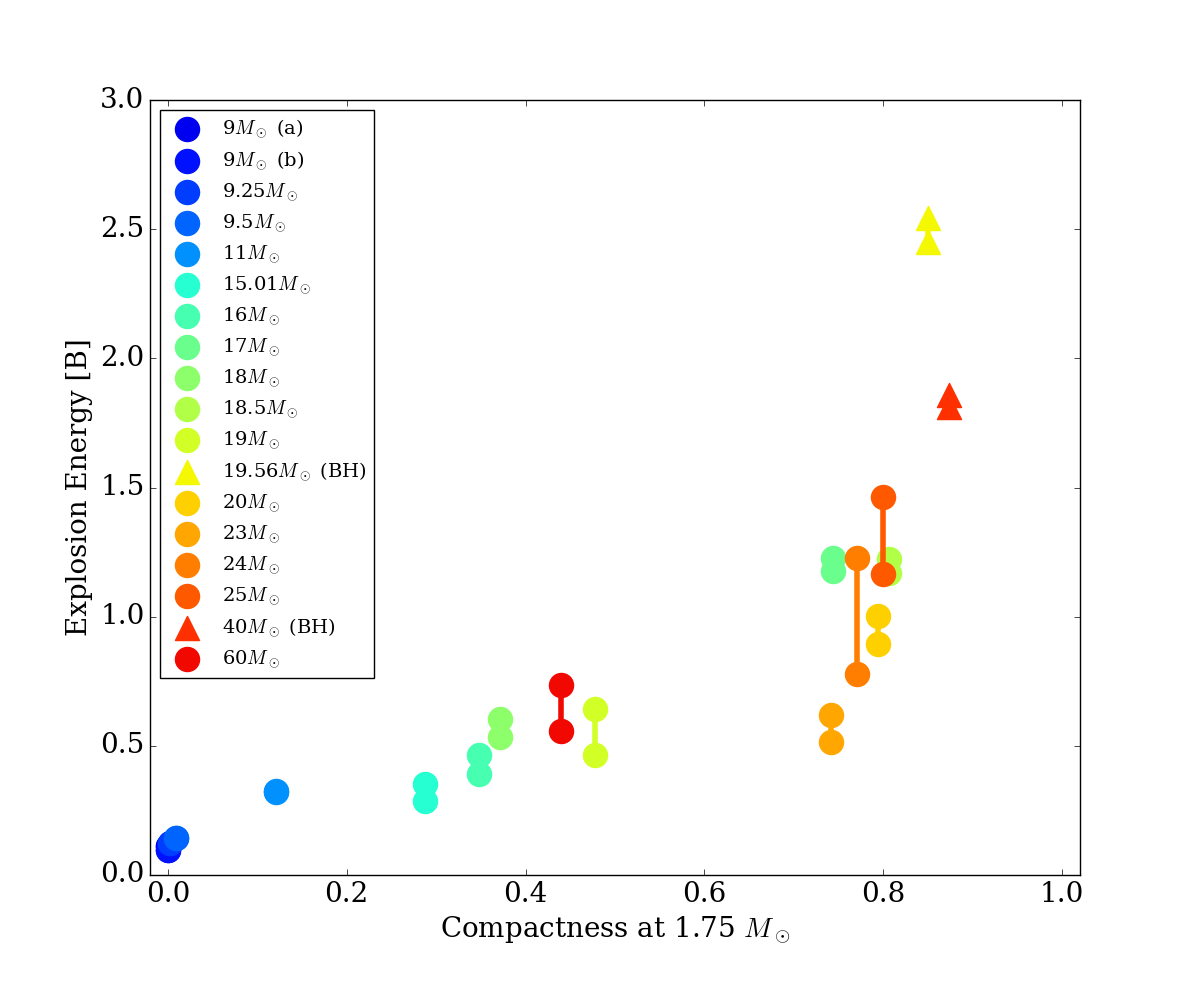}
    \caption{The relation between explosion energy and binding energy exterior to 2.0 $M_{\odot}$ (left) and to the compactness (right). The connection between binding energy and explosion energy is physically one of the most important findings from the theoretical perspective. The only major outlier is the 19.56 $M_{\odot}$ exploding black hole former. See text in \S\ref{energy} for a discussion.}
    \label{fig:E-intrinsic}
\end{figure}

Figure \ref{fig:E-intrinsic} plots the connection between binding energy and explosion energy (left) and compactness and explosion energy (right). Both relationships are roughly monotonic and robust, but the explosion energy/binding energy correlation is physically one of the most important findings from the theoretical perspective {\citep{Janka_2017b}}. It implies that the supernova energy is a ``self-regulating" phenomenon, akin in broad outline to winds. Such self-regulation, mediated through the shared implicit dependence upon compactness, may become one of the unifying principles of CCSN theory going forward.

\subsection{$^{56}$Ni Yields and Nucleosynthesis}
\label{nickel_nucleo}

Supernova nucleosynthesis is a vast subject that spans almost eight decades. Components are pre-supernova progenitor yield calculations, explosive nucleosynthesis, $\alpha$-rich freeze-out, and various sub-processes of differing import. Reviews can be found in \citet{woosley2002,nomoto2013,fischer2023,arcones2023}. The decades-long tradition of 1D models \citet{thielemann1996,rauscher2002,woosley2002,heger2010,nomoto2013,Sukhbold2016,limongi2018,curtis2019} is now being challenged by 3D models \citep{Lentz2015,wongwathanarat2015,harris2017,wanajo2018,sieverding2020,sieverding2023,wang2023,wang2023b}. Winds play an important role \citep{duncan1986,burrows1987,Burrows1995,qian1996,wang2023}, as do neutrinos \citep{woosley1990,frohlich2006,Pruet2006} and $\gamma$-ray emissions of the ejecta can play a diagnostic role \citep{diehl1998,diehl2021,diehl2023}.



\begin{figure}
    \centering
    \includegraphics[width=0.48\textwidth]{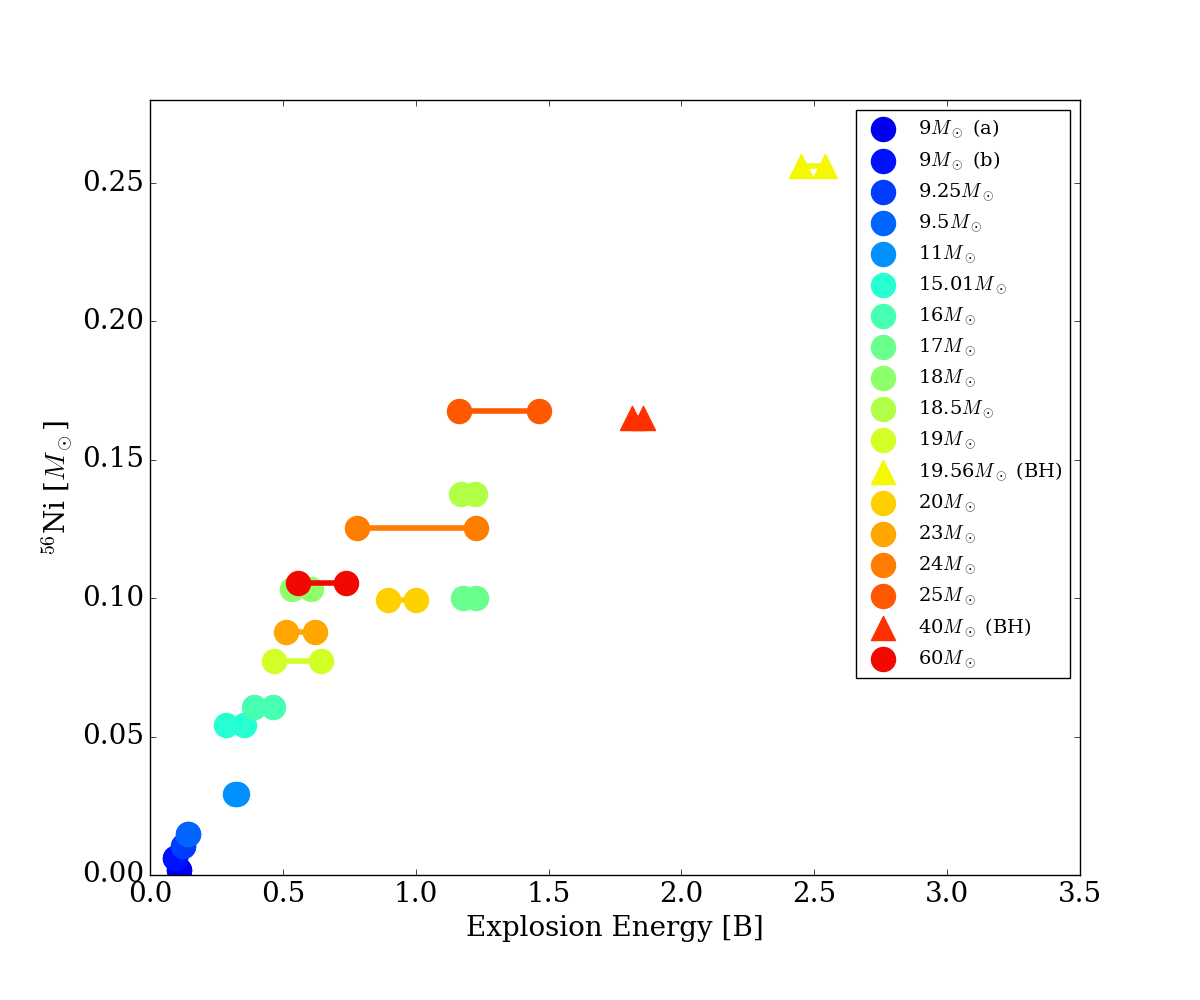}
    \includegraphics[width=0.48\textwidth]{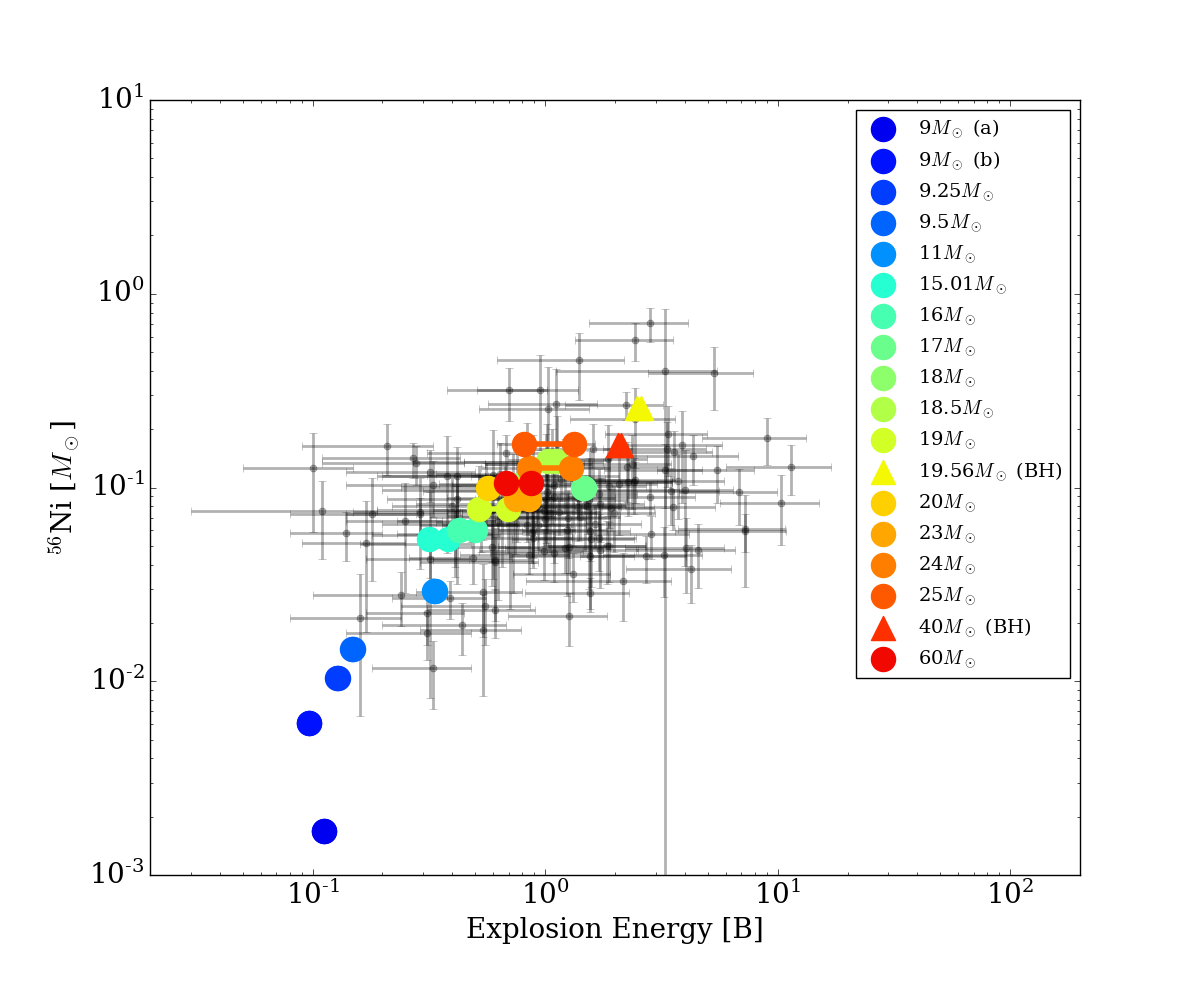}
    \caption{Relations between $^{56}$Ni (in $M_{\odot}$) and explosion energy (in Bethes), with the range in the former indicated for some models that have yet to completely asymptote. The triangles are for the exploding black hole formers, yet they still follow the nearly monotonic curve.
    The right hand side includes the data of \citet{rodriguez2023} and suggests by the overlap of the two a reasonable concordance. {Similar observational correlations are found in \citet{prieto2017} and \citet{Pejcha_2020}.}}
    \label{fig:E-ni56}
\end{figure}

One of the central observables of core-collapse supernova explosions is the $^{56}$Ni yield (in solar masses), since the $^{56}$Ni decay chain to $^{56}$Fe powers some of the supernova light curve and can easily be measured.  Hence, the predicted $^{56}$Ni yields are central physical quantities of the CCSN phenomenon. Figure \ref{fig:E-ni56} provides the relation that emerges in our 3D simulation suite between explosion energy (in Bethes $\equiv$10$^{51}$ ergs) and $^{56}$Ni production (in $M_{\odot}$). {$^{56}$Ni production, and in general the ejecta isotopic abundances, are calculated using SkyNet \citep{lippuner2017} with a 1530-isotope network including elements up to $A=100$, together with about 340,000 post-processed backwardly-integrated tracers in each simulation. More details concerning the formalism can be found in \citet{wang2023b}.}  This too is one of the central results of this paper, but the correlation between the two has already been anticipated theoretically (e.g. \citet{Sukhbold2016}) and observed astronomically \citep{rodriguez2023}. The left hand side provides the $^{56}$Ni correlation with explosion energy, with the range of predicted energies due to the fact that not all 3D models have yet asymptoted to their final values indicated. The right hand side shows the same numerical data, but superposed on the \citet{rodriguez2023} measurements. The overlap between the two suggests a reasonable concordance, though both trends are subject to natural variation. Note that the strong theoretical trend  of $^{56}$Ni mass with explosion energy includes the exploding black hole formers (triangles). 

\begin{figure}
    \centering
    \includegraphics[width=0.48\textwidth]{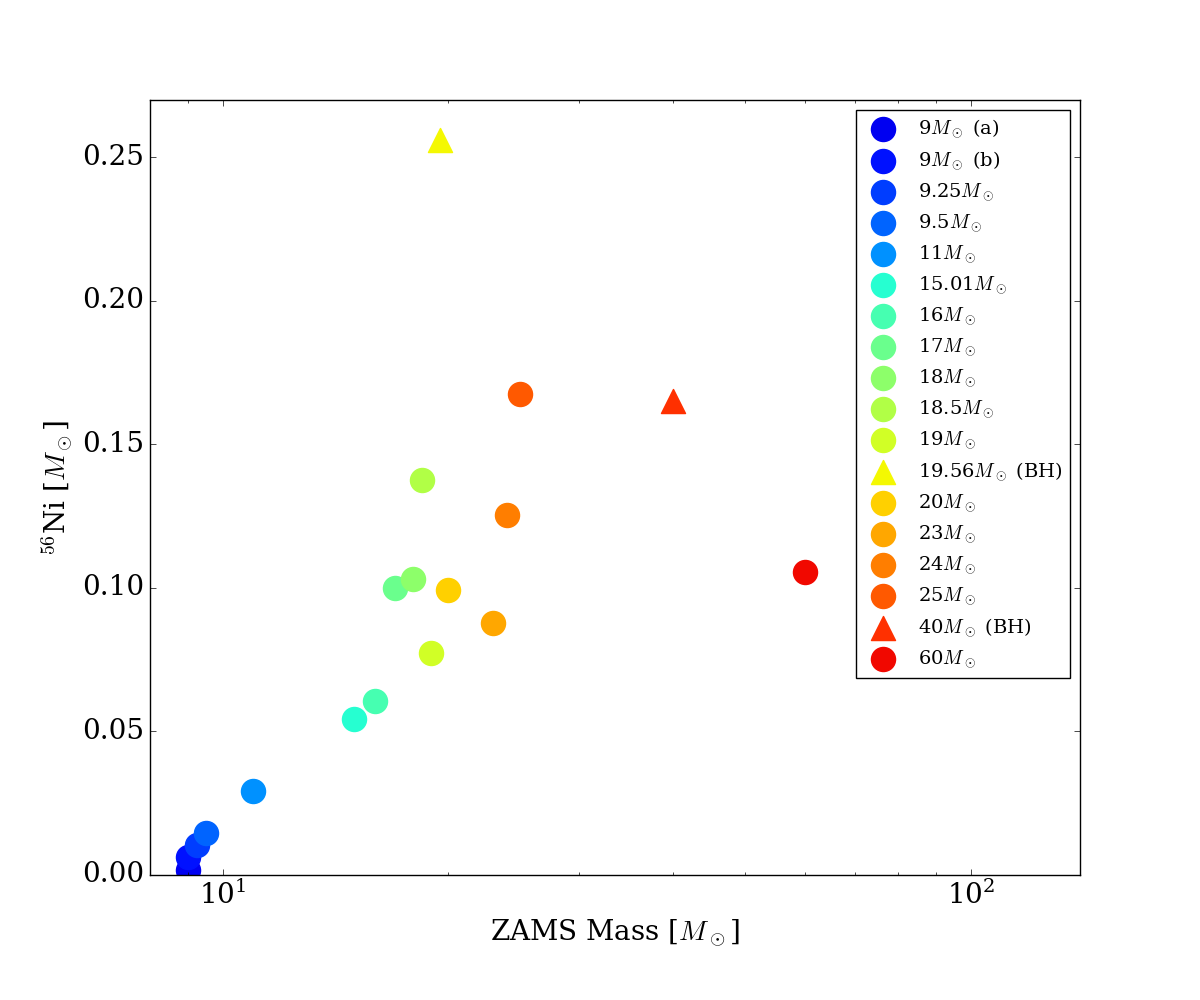}
    \includegraphics[width=0.48\textwidth]{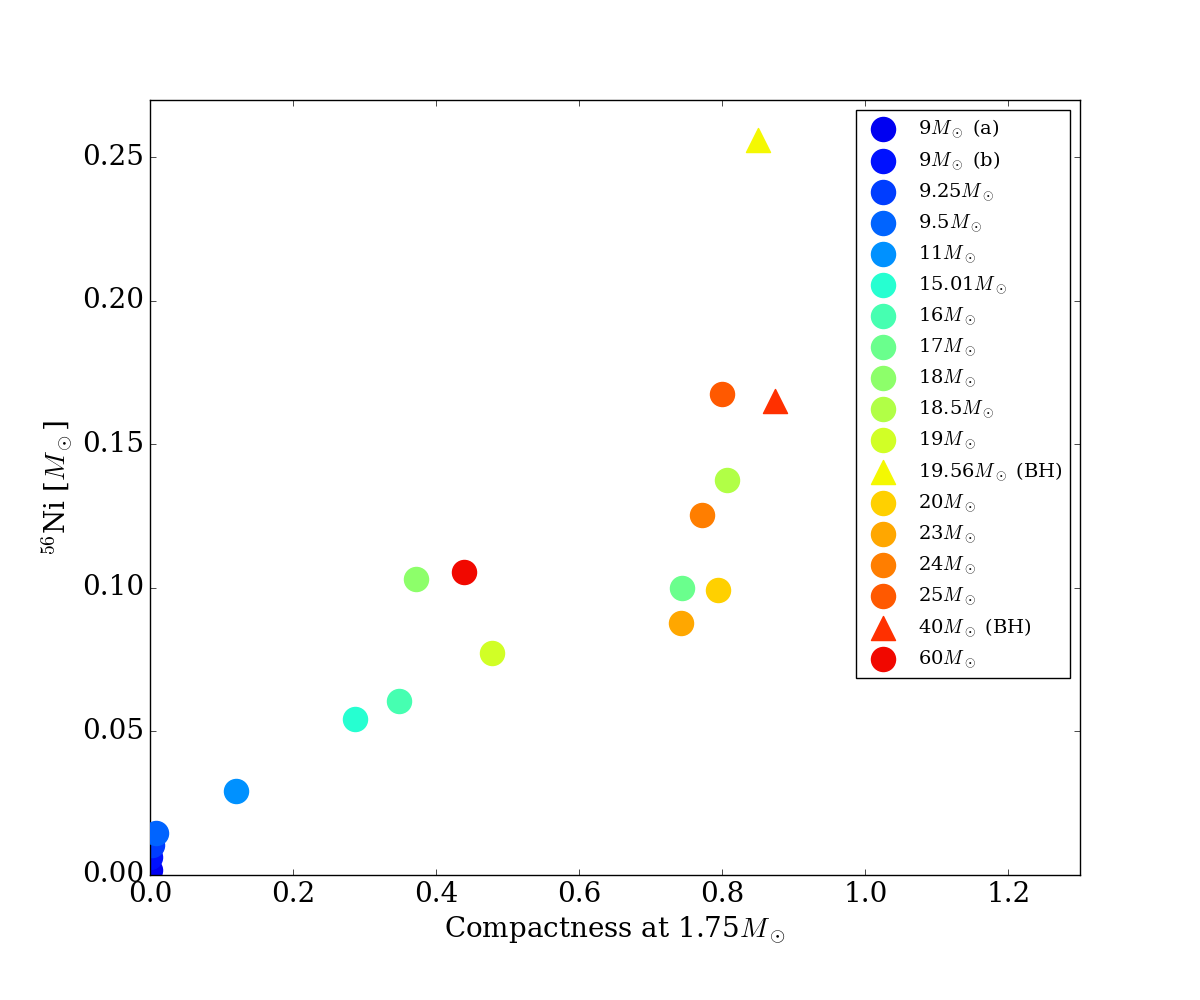}
    \caption{The relationship between the ejected $^{56}$Ni mass and 1) the {log$_{10}$ of the} ZAMS mass (left) and 2) the compactness parameter (right). See text in \S\ref{nickel_nucleo} for a discussion.}
    \label{fig:ni56-zams-cpt}
\end{figure}

Figure \ref{fig:ni56-zams-cpt} shows the relationship between $^{56}$Ni production, ZAMS masses, and compactness.  While there is some scatter, the general monotonic trend is good, with two exceptions versus ZAMS mass. The first in the 19.56 $M_{\odot}$ exploding black hole former, which produces a lot of $^{56}$Ni and the other is the 60 $M_{\odot}$ model, which is underenergetic and leaves behind a neutron star. It also experienced pronounced mass loss prior to explosion, which may not comport with the emerging understanding of stellar mass loss \citep{smith2014}. The deviations for both models from the trend is less pronounced versus compactness. Though we still witness some degree of stochasticity, the better (though not perfect) correlation with compactness than with ZAMS mass reflects again the non-monotonicity of the two, one with the other in the \citet{Sukhbold2016} and \citet{Sukhbold2018} progenitor collection. What other progenitor collections would provide is yet to be determined and is an important future question in supernova and stellar evolution theory.

\begin{figure}
    \centering
    \includegraphics[width=0.80\textwidth]{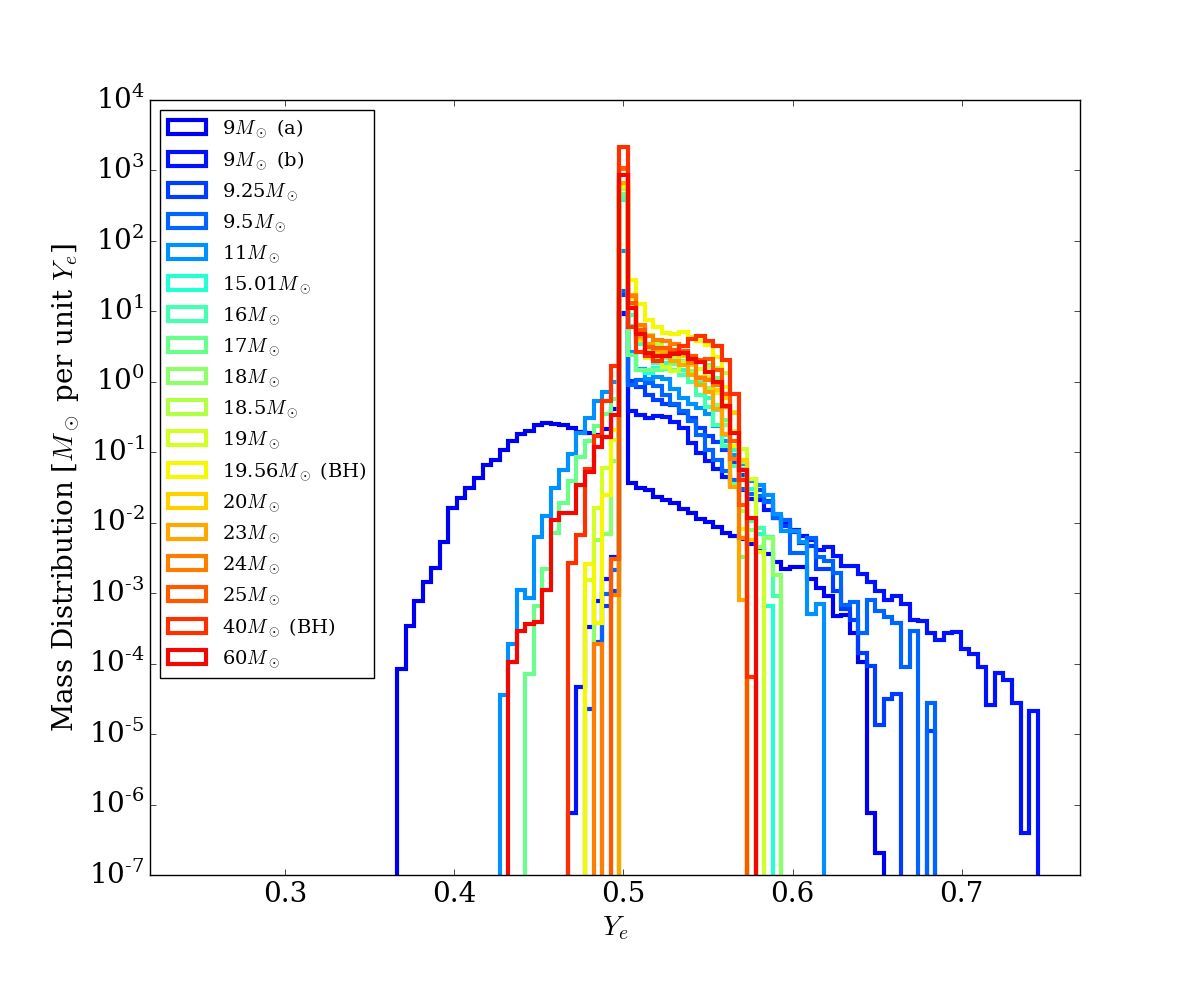}
    \caption{Ejecta $Y_e$ distributions for our exploding models.  Most of the ejecta are at $Y_e =$ 0.5 or higher, in the proton-rich regime.  However, a few of the models, in particular the 9(a) $M_{\odot}$ model that boasted imposed initial perturbations in the progenitor, have an interesting neutron-rich tail. Such a tail can give rise to the first peak on the r-process (see \citet{wang2023,wang2023b}).}
    \label{fig:all-ye}  
\end{figure}

Figure \ref{fig:all-ye} shows the final ejecta electron fraction ($Y_e$) distributions of all our exploding models. The competition between $\nu_e$ and $\bar{\nu}_e$ absorption in the ejecta when near the inner core pushes much of the $Y_e$ above 0.5 into the proton-rich domain.  Most of the mass, however, is near 0.5.  However, for some models (such as model 9(a)), which experience faster ejecta speeds during the wind and $\alpha$-rich freeze-out phases, some neutron-rich starting material does not have enough time to be pushed above $Y_e = 0.5$ before being ejected.  The result is a component of the explosion debris that emerges neutron-rich, with consequences for the nucleosynthesis. In particular, isotopes near the first peak of the r-process can be produced.  There are many other consequences of this \citep{Pruet2006,wang2023,wang2023b}, but we prefer here to highlight only a few. 

\begin{figure}
    \centering
    \includegraphics[width=0.48\textwidth]{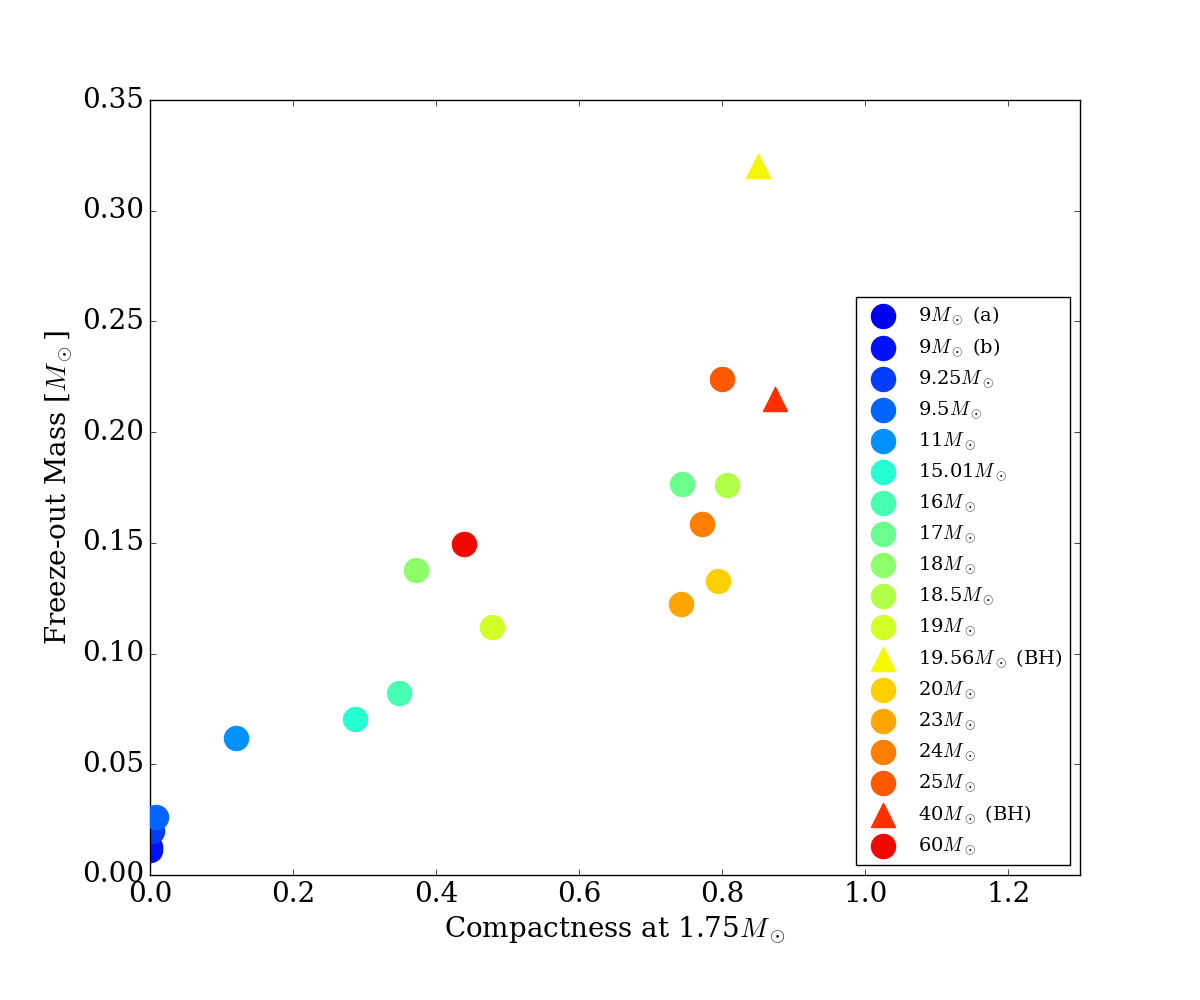}
    \includegraphics[width=0.48\textwidth]{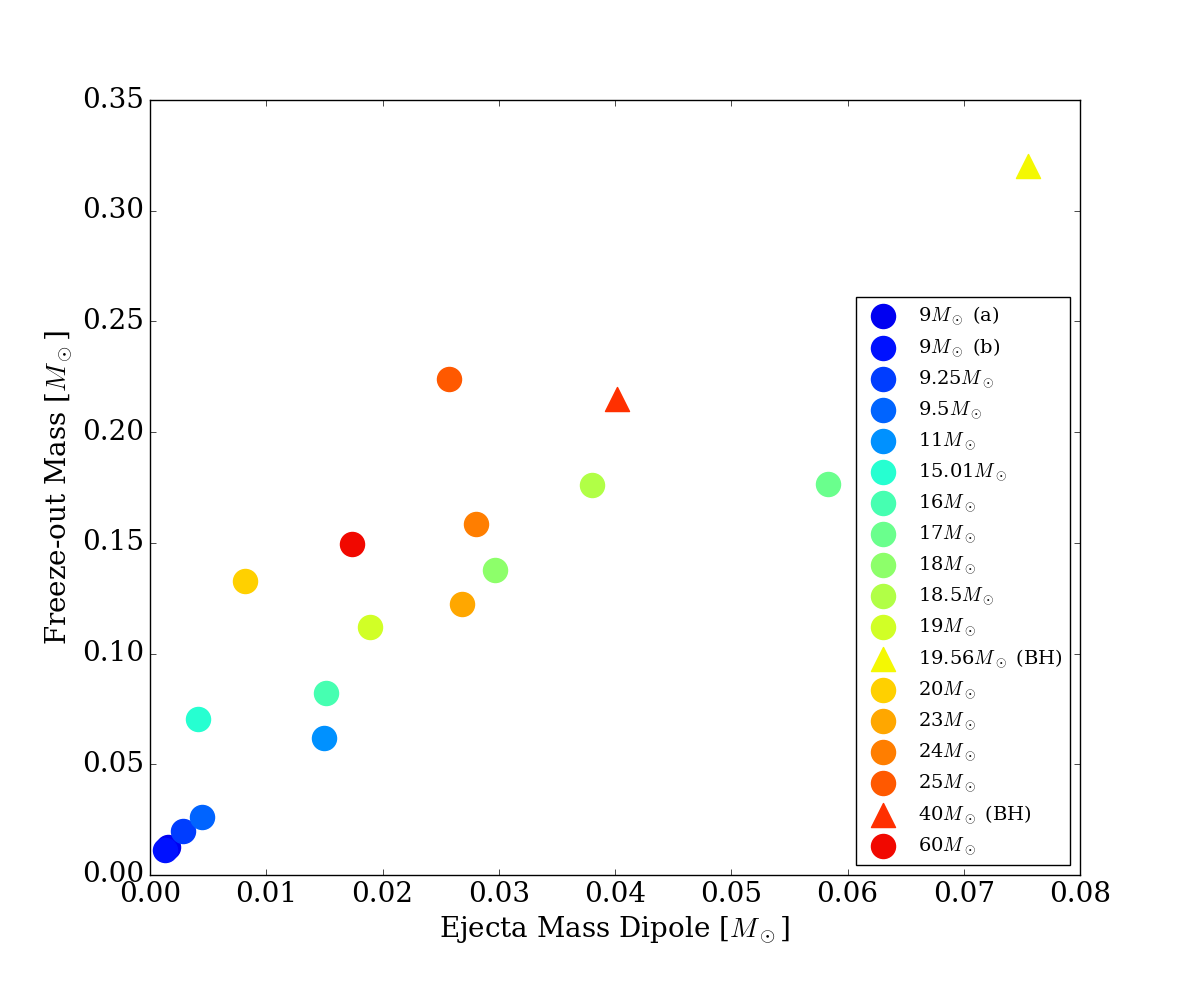}
    \includegraphics[width=0.48\textwidth]{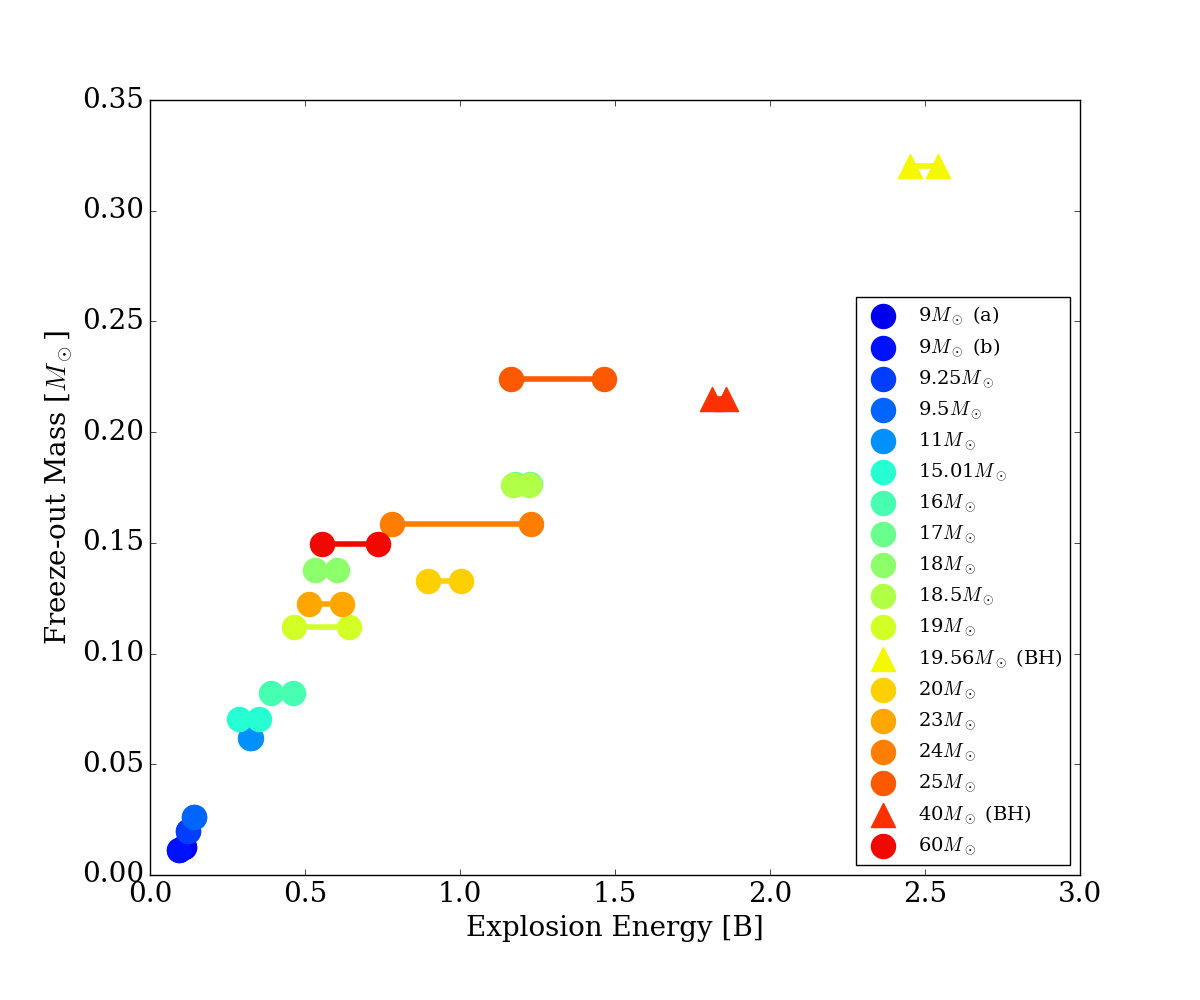}
    \caption{Relations between the freeze-out ejecta mass and the compactness parameter (top left), ejecta mass dipole (top right), and explosion energy (bottom).  The reasonably robust correlations between the freeze-out mass and both the explosion energy and the ejecta mass dipole are predictions of the theory that can be tested.  See the text at \S\ref{nickel_nucleo} for a discussion.}
    \label{fig:freezeout-cpt-dipole}
\end{figure}

Figure \ref{fig:freezeout-cpt-dipole} shows the dependence of the freeze-out mass\footnote{{The NSE temperature is set at 7 GK, and the freeze-out mass} represents the total amount of matter that has ever achieved this temperature \citep{wang2023b}.} upon compactness (top left), ejecta mass dipole (top right), and explosion energy (bottom). We see here too the rough monotonicity, in particular versus the explosion energy, of the mass in the freeze-out component.
This follows the corresponding correlation of the explosion energy with $^{56}$Ni mass depicted in Figure \ref{fig:E-ni56}
and is not surprising. It is nevertheless a strong prediction of the theory of CCSN explosions, as we see it, and is testable.

\begin{figure}
    \centering
    \includegraphics[width=0.48\textwidth]{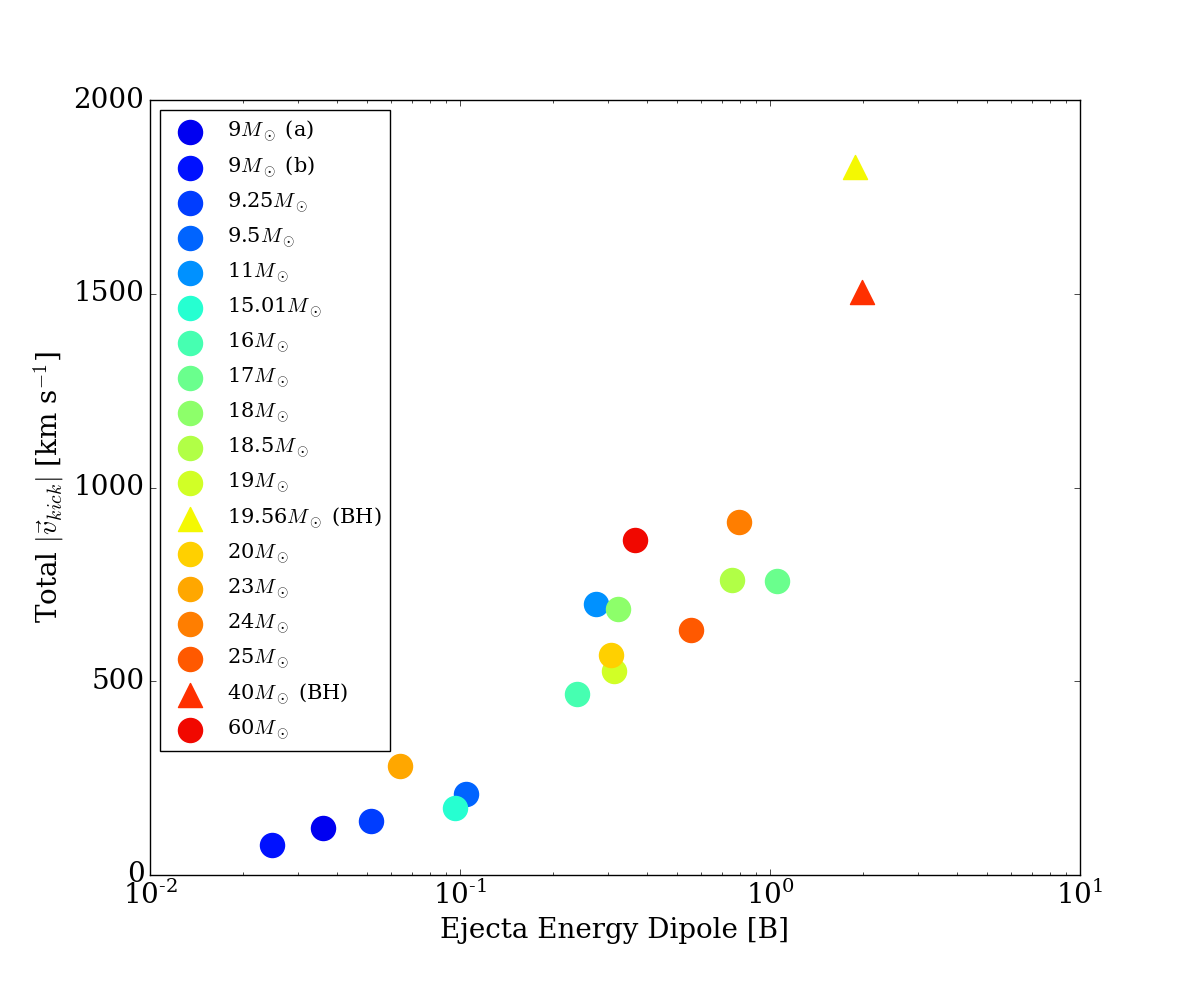}
    \includegraphics[width=0.48\textwidth]{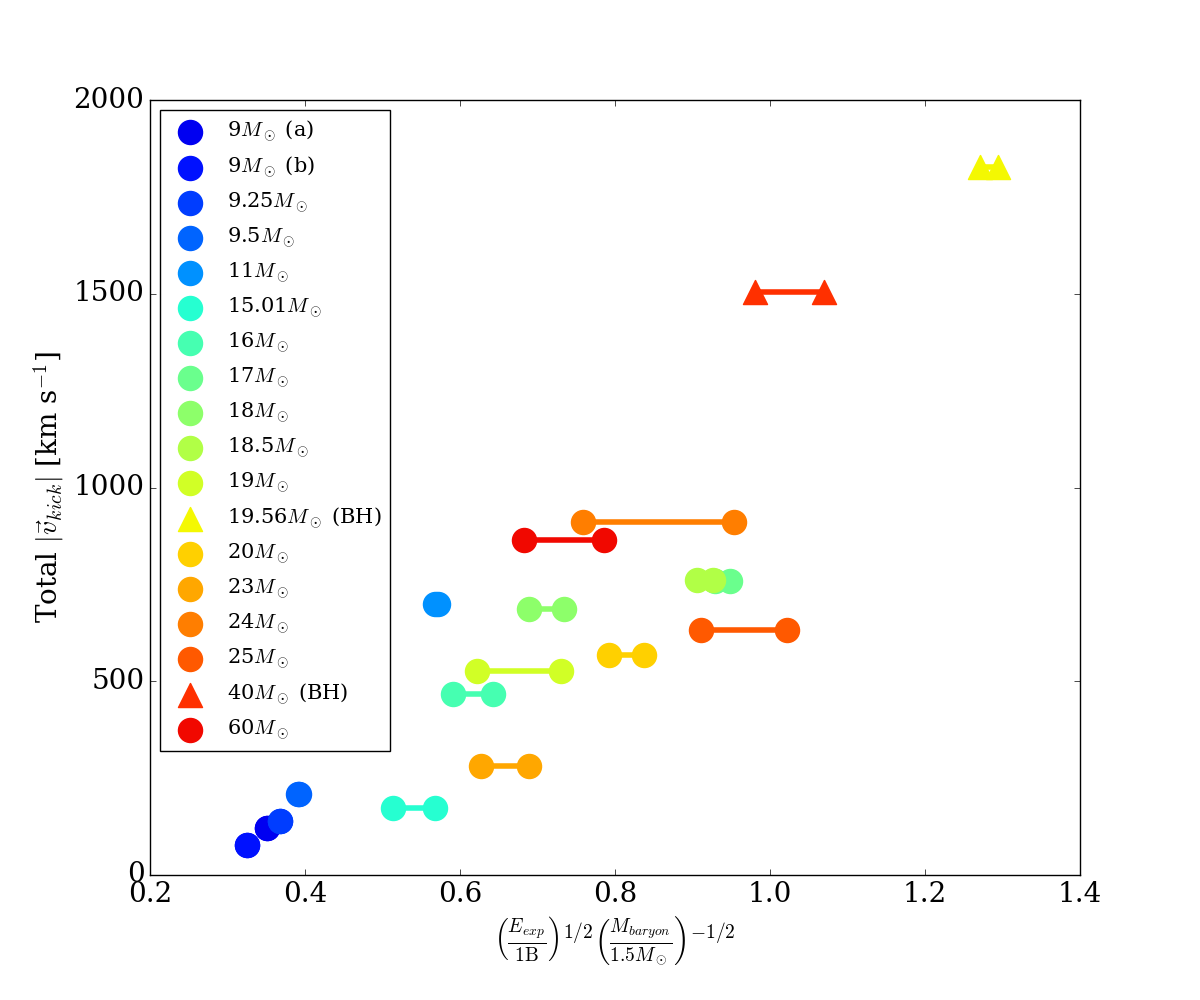}
    \caption{Relations between the neutron star kick speed and the ejecta energy dipole (left) and the quantity ($\frac{E_{exp}}{M_{bar}})^{1/2}$ (right), where E$_{exp}$ is the explosion energy and $M_{bar}$ is the baryon mass of the neutron star.  See \citet{burrows_kick_2023} for a more thorough discussion.}
    \label{fig:kick-zams-dipoleE}
\end{figure}

\subsection{Kicks}
\label{kick}

\citet{burrows_kick_2023} have recently crafted and {had accepted} a paper on the physics and systematics of the recoil kicks and induced spins, and their potential correlations, for our long-term 3D simulations. Though that paper contains the same models, the time between this paper and that has allowed us to carry those models even further.  However, the general conclusions in \citet{burrows_kick_2023} are unaltered and we won't repeat its discussion and conclusions in detail here. We found an important correlation between the kick speed and both the energy of explosion and the ejecta mass dipole.  We identified a weakly inverse relation between induced spin and kick speed, with a lot of scatter, and found a weak induced-spin/kick correlation. If the initial rotational vector of the collapsing core correlates somewhat with the direction of explosion, a spin/kick correlation would naturally emerge \citep{Holland-Ashford2017,2018ApJ...856...18K,Ng2007}. We also found a compelling explanation for the lower average gravitational masses of binary neutron stars in the neutron star mass/kick connection that emerges from the simulations, viewed collectively. The lowest mass progenitors give birth to the lowest mass neutron stars with the slowest kicks, and this would make binary disruption harder.

Therefore, in this paper we add only two more plots that reveal two additional correlations.  Figure \ref{fig:kick-zams-dipoleE} indicates that the kick speed and the energy dipole of the ejecta (left) and the combination $(\frac{E_{\rm exp}}{M_{\rm baryon}})^{1/2}$ (right) (where E$_{\rm exp}$ is the explosion energy and $M_{\rm baryon}$ is the baryon mass of the neutron star) are tightly correlated. The more asymmetrical the explosion the greater the kick speed.  Moreover, the kick speed is related to the explosion energy itself, divided by the residual neutron star mass.  In \citet{burrows_kick_2023}, we found similar relations between kick speed and explosion energy and ejecta mass dipole, as did \citet{Janka2017}\footnote{{However, we note that \citet{Janka2017} ``calibrates" his alpha parameter on the \citet{Scheck2006} and \citet{Wongwathanarat2013} papers.  The former is 2D and the latter used a light bulb to artificially explode models. Moreover, \citet{Wongwathanarat2013} found a ``reduced ejecta asymmetry with higher explosion energies," something we do not see.}}.  However, the correlation between kick speed and ejecta energy dipole is actually a bit tighter. In any case, the connections between kick speed and various measures of anisotropy, while not unexpected, are testable.

\section{Conclusions}
\label{conclusion}   

In this paper, we have derived correlations between core-collapse supernova observables and progenitor core structures that naturally emerge from our extensive suite of twenty state-of-the-art 3D CCSN simulations carried to late times.  This is the largest such collection of 3D supernova models ever generated and allows one to witness and derive patterns that might otherwise be obscured when studying one or a few models in isolation. Moreover, such long-term simulations are necessary to capture the asymptotic values of many of the most important observable quantities, and such simulations have until recently been too expensive to generate. However, with the efficiency of our new supernova code F{\sc{ornax}} and recent generous allocations of supercomputer time, this limitation has now been overcome. Furthermore, a large collection of such models allows one partially to mitigate against the fact that the physical turbulence and chaos that attends supernova dynamics will naturally translate into a distribution of outcomes and explosion parameters in deriving trends.\footnote{The associated scatter and spread in observables is unknown and is properly a subject of future research.}.  Looking at single models one at a time tends to obscure the overall correlations, with both progenitor structure and between observables.

From this new panoramic perspective, we have {identified} 1) observationally useful correlations between explosion energy, neutron star gravitational birth masses, $^{56}$Ni and $\alpha$-rich freeze-out yields, and pulsar kicks \citep{burrows_kick_2023} and 2) physically and theoretically important correlations with the compactness parameter of progenitor structure. The theoretical correlation between $^{56}$Ni and explosion energy has been seen before (e.g. \citet{Sukhbold2016}), but not in the context of such a large 3D model suite carried to, or near to, the asymptotic state. Such a correlation is also strongly suggested by observations \citep{rodriguez2023}. The intriguing correlation between explosion energy and neutron star gravitational mass is new {in the detailed 3D modeling context, but was anticipated in the 2D results of \citet{nakamura2015} and in the semi-analytic modeling of \citet{Muller2016}}. In addition, we find a correlation between explosion energy and progenitor mantle binding energy, suggesting that neutrino-driven explosions are self-regulating. We also find a testable correlation between explosion energy and measures of explosion asymmetry, such as the ejecta energy and mass dipoles. 

While the correlations between two observables are roughly independent of the progenitor ZAMS mass, the many correlations we have derived with compactness can not unambiguously be tied to a particular progenitor ZAMS mass. The compactness/ZAMS mass mapping depends upon the massive star progenitor models employed and the mapping between compactness and/or initial density structure and progenitor mass is still in flux. Therefore, our derived correlations between compactness and observables may be more robust than with ZAMS mass. There remain may important issues, such as binarity, shell mixing, overshoot, wind mass loss, semi-convection, the $^{12}$C($\alpha$,$\gamma$) nuclear rate, and Roche-lobe overflow, that remain to be resolved related to the endpoint structures of massive stars.  Hence, it is likely that the mapping between compactness and ZAMS mass found in the \citet{Sukhbold2016} and \citet{Sukhbold2018}
compendia is not the final word. Using a simple analytic model for core-collapse supernovae and estimates of some explosion observables, this point is clearly made in \citet{temaj2023}.

In particular, we find that there is a range of progenitors in the Sukhbold et al. compilation that don't explode, but this gap in progenitor masses may be displaced from Nature's and from those in other progenitor model sets.  Hence, a gap may indeed exist to explain one channel of black hole formation \citep{Burrows2023} and an associated measured luminosity gap \citep{smartt2009,smartt2015,adams2017}, but it may reside at a different range of progenitor masses than we have found.  Nevertheless, the theoretical derivation of a range of ZAMS masses in which explosions are difficult may survive, but what that range is is likely still in play.   

Though F{\sc{ornax}} is a sophisticated, multi-physics multi-group radiation/hydrodynamics code that has been matured over the last eight years to address the complicated multi-dimensional phenomena encountered in core collapse, there are numerous upgrades and improvements that suggest themselves.  We are bundling the $\nu_{\mu}$, $\bar{\nu}_{\mu}$ 
$\nu_{\tau}$, $\bar{\nu}_{\tau}$ neutrinos into one pseudo-species.  This is done because their neutrino-matter interactions are quite similar and to trim the computational load. However, their interactions are not exactly the same \citep{Bollig2017}. Also, we are using 12 ($\times$3) energy groups and though we have in the past determined that this number is adequate, more would be better. In the same spirit, an augmentation of the number of zones and enhanced spatial resolution are always good, if affordable.  Currently, we are using a 1024($r$)$\times$128($\theta$)$\times$256($\phi$) grid and this is better than most 3D production simulations.  However, we have shown that resolution can make a difference, perhaps qualitatively \citep{Nagakura2019}. Importantly, we are using approximate general relativity \citep{Marek2006}, which captures the basic differences in the strength of gravity between the Newtonian and Einsteinian realizations, provides an accurate lapse function, and allows for the incorporation of gravitational redshifts. However, full general relativity is the correct approach.  Also, we have not attempted to capture the effects of neutrino oscillations and we are wedded in this simulation suite to the \citet{Sukhbold2016} and \citet{Sukhbold2018} initial models.  The latter are certainly not the last word, are not 3D, and do not include the many effects of binarity. Furthermore, as we have demonstrated, the initial structures are crucial to determining the mapping between star and outcome and we have not incorporated the effects of magnetic fields \citep{Muller2020,varma21,Obergaulinger2020,Kuroda2020,Obergaulinger2021}, nor addressed the effects of initial core rotation.  Finally, the chaos of the turbulent context of post-bounce evolution introduces a degree of randomness and stochasticity that will translate into distribution functions for the observables, even for the same progenitor, whose widths are currently unknown.  Hence, one can envision many potential upgrades to current computational platforms, and many necessary and more detailed investigations to pursue, in order to further improve the 3D theory of core-collapse supernova explosions. Nevertheless, our current collection of twenty state-of-the-art 3D models provides a synoptic view unavailable in the past upon which future work can profitably build.

\section*{Acknowledgments}

We thank Chris White, Hiroki Nagakura, and Matt Coleman for previous insights, collaboration, and conversations. We also thank Joe Insley for very helpful graphics support. We acknowledge support from the U.~S.\ Department of Energy Office of Science and the Office of Advanced Scientific Computing Research via the Scientific Discovery through Advanced Computing (SciDAC4) program and Grant DE-SC0018297 (subaward 00009650) and support from the U.~S.\ National Science Foundation (NSF) under Grants AST-1714267 and PHY-1804048 (the latter via the Max-Planck/Princeton Center (MPPC) for Plasma Physics). Some of the models were simulated on the Frontera cluster (under awards AST20020 and AST21003), and this research is part of the Frontera computing project at the Texas Advanced Computing Center \citep{Stanzione2020}. Frontera is made possible by NSF award OAC-1818253. Additionally, a generous award of computer time was provided by the INCITE program, enabling this research to use resources of the Argonne Leadership Computing Facility, a DOE Office of Science User Facility supported under Contract DE-AC02-06CH11357. Finally, the authors acknowledge computational resources provided by the high-performance computer center at Princeton University, which is jointly supported by the Princeton Institute for Computational Science and Engineering (PICSciE) and the Princeton University Office of Information Technology, and our continuing allocation at the National Energy Research Scientific Computing Center (NERSC), which is supported by the Office of Science of the U.~S.\ Department of Energy under contract DE-AC03-76SF00098.

\appendix

The code that we employ to perform state-of-the-art 3D, 2D, and 1D multi-group radiation/hydrodynamic simulations
of core collapse and explosion is F{\sc{ornax}} \citep{Skinner2019}. When created, F{\sc{ornax}} was already $\sim$5 times faster than competitive codes, due in part to its explicit transport solver and
its ``dendritic" grid.  We have ported to, and efficiently run F{\sc{ornax}} on, 1) the CPU machines Blue Waters, XSEDE/Stampede 2, NERSC/Cori II, TACC/Frontera, and ALCF/Theta and 2) the GPU machines ALCF/ThetaGPU, ALCF/Polaris, and NERSC/Perlmutter. To this list of machines and architectures can be added local (though smaller) CPU and GPU clusters at Princeton. Both the CPU and GPU variants scale well on all machines out to $\sim$150,000 CPU cores and $\sim$480 GPU nodes.

F{\sc{ornax}} is written in C. We use an MPI programming model for the CPU variant and an MPI/OpenMP-4.5/5.0 hybrid paralellism model for the GPU variant.  The code employs spherical coordinates in one, two, and three spatial dimensions, solves the comoving-frame, multi-group, two-moment, velocity-dependent transport equations, and uses the M1 tensor closure for the second and third moments of the radiation fields \cite{vaytet:11}\footnote{Most, though not all, supernova groups in the world employ the so-called ``ray-by-ray+" dimensional reduction approach to transport \citep{Bruenn2020,2020ApJ...891...27M}.  This is not true multi-D transport, but multiple spherical (1D) solves, and is done for speed. However, the results for multi-D evolutions that are not spherical can be in qualitative error \citep[in 2D]{skinner2016}. {\citet{Glas2019} explored the same issues in 3D and showed that when the explosion is roughly spherical (as in their 9 $M_{\odot}$ model) the ray-by-ray+ approach is adequate. However, they make a less convincing argument when the explosion is aspherical (as in their 20 $M_{\odot}$ model).}.} Three species of neutrino ($\nu_e$, $\bar{\nu}_e$, and ``$\nu_{\mu}$" [$\nu_{\mu}$, $\bar{\nu}_{\mu}$, $\nu_{\tau}$, and $\bar{\nu}_{\tau}$ lumped together]) are followed using an explicit Godunov characteristic method applied to the radiation transport operators, but an implicit solver for the radiation source terms. This solver for the transport operators can be explicit since the speed of light and the speed of sound in the inner core are close and we are already timestep-limited by the hydro Courant condition \citep{2015MNRAS.453.3386J,2015ApJS..219...24O,Skinner2019}.
By addressing the transport operator with an explicit method, we significantly reduce the computational complexity and communication overhead of traditional multi-dimensional radiative transfer solutions by bypassing the need for global iterative solvers. Radiation quantities are reconstructed with linear profiles and the calculated edge states are used to determine fluxes via an HLLE solver.  In the non-hyperbolic regime, the HLLE fluxes are corrected to reduce numerical diffusion \citep{oconnor_ott:2013}. The momentum and energy transfer between the radiation and the gas are operator-split and addressed implicitly.

The hydrodynamics in F{\sc{ornax}} is based on a directionally unsplit Godunov-type finite-volume method.  Fluxes at cell faces are computed with a fast and accurate HLLC approximate Riemann solver based on left and right states reconstructed from the underlying volume-averaged
states.
Without gravity, the coupled set of radiation/hydrodynamic equations conserves energy and momentum to machine accuracy.  With gravity, energy and momentum conservation, handled with source terms in the energy and momentum equations, is excellent before and after core bounce.

Gravity is handled in 2D and 3D with a multipole solver \citep{mueller:95}, where we generally set the maximum spherical harmonic order necessary equal to twelve. The monopole gravitational term is altered to approximately accommodate general-relativistic gravity \citep{Marek2006},
and we employ the metric terms, $g_{rr}$ and $g_{tt}$, derived from this potential in the neutrino transport equations to incorporate general relativistic redshift effects.

In the interior, to circumvent Courant limits due to converging angular zones, the code deresolves in both angles ($\theta$ and $\phi$) independently with decreasing radius, conserving hydrodynamic and radiative fluxes in a manner similar to the method employed in AMR codes at refinement boundaries. The use of such static-mesh refinement, or ``dendritic grid," allows us to avoid angular Courant limits at the center, while maintaining accuracy and enabling us to employ the useful spherical coordinate system natural for the supernova problem. We have set up eleven nuclear equations of state in F{\sc{ornax}} form, but focused on
the SFHo EOS \citep{Steiner2013} for this study.

A comprehensive set of neutrino-matter interactions are followed in F{\sc{ornax}} and these are described in \citet{Burrows2006}. They include weak magnetism and recoil corrections to neutrino-nucleon scattering and absorption \citep{Horowitz2002} and ion-ion-correlations, weak screening, and form-factor corrections for neutrino-nucleus scattering. For inelastic neutrino-electron scattering, we use the scheme of \citet{2003ApJ...592..434T} and the relativistic formalism summarized in \citet{reddy1999}. Inelastic neutrino-nucleon scattering is now handled using the much more efficient, faster, and more accurate Kompaneets approach we pioneered \citep{Tianshu}. We note that most other supernova codes, with the exceptions of those fielded by the Garching \citep{Buras2006} and Oak Ridge \citep{Bruenn2020} groups, do not generally include inelastic redistribution. Neutrino sources and sinks due to nucleon-nucleon bremsstrahlung and electron-positron annihilation are included, as described in \citet{thomp_bur_horvath}. We currently include a many-body structure factor ($S_A$) correction to the axial-vector term in the neutrino-nucleon scattering rate due to the neutrino response to nuclear matter at (low) densities below $\sim$10$^{13}$ gm cm$^{-3}$ derived by \citet{Horowitz2017} using a virial approach. \citet{Horowitz2017} attach their fit to $S_A$ to that of \citet{1998PhRvC..58..554B} at higher densities.

The code checkpoints periodically, and the I/O is fully parallel. In fact, the code uses parallel HDF5 writes to a single file with underlying MPIIO coordination.  Each processor writes at a different offset into a single HDF5 file, which we have found useful for restarting checkpoints
with different processor numbers and topologies. We have also optimized the striping on simulation runs to get further speed in the I/O and can dump data on local SSDs (if available) using 48 stripes.
Twelve energy groups arrayed logarithmically per neutrino species works well for 3D simulations. As stated, the current main advantages of the F{\sc{ornax}} code are its
efficiency due to its explicit nature, its truly multi dimensional transport solution, and the interior static mesh derefinement in the core. For the long-term 3D simulations 
employed for this paper, we use 12 ($\times$3) energy groups and a 1024($r$)$\times$128($\theta$)$\times$256($\phi$) spatial grid.

We realized early in 2022 a factor of $\sim$2 speed-up in the CPU version of F{\sc{ornax}}.  This was the result of more aggressive inlining and vectorization, and of the incorporation of the Kompaneets scheme to handle inelastic scattering on nucleons \citep{Tianshu}.  Also in 2022, we created, tested, and fielded the MPI/OpenMP-4.5 GPU version of F{\sc{ornax}}. To achieve this, we converted the entire code to C (it was originally a hybrid of C and Fortran). Our dendritic grid is now mapped to an 1D array which is linearly allocated in memory.  The whole grid is now divided into three parts: a bulk segment (not receiving MPI messages) and a sending and a receiving segment, now all continuous in memory. First focussing on the hydro only, we added GPU pragmas, writing some functions to fit the GPU format and redefining variables to minimize local memory usage.  All calculations are done on the GPUs, while communication is done on the CPUs.  Only the sending and receiving components are mapped between the CPUs and GPUs, greatly reducing data transfer.  The grid bulk is transferred to the CPU only to write dump and restart/checkpoint files. We executed the same porting steps for the radiation transport modules.  This entailed a lot more work.  Noting that statically allocated arrays can not be used as pointers on GPUs using OpenMP, we redefined them dynamically. We also rewrote the flux calculation. Previously, this was based on a 1D pencil layout and parallelization could only be done along the perpendicular two dimensions (in 3D). This has been altered so that all three dimensions can be parallelized.  Though this entails a bit more calculation, the overall speed-up on the GPUs is improved. We also decreased the memory footprint of the gravitational potential calculation, and substituted ``atomic add" for ``array reduction." This enabled the use of more threads and, hence, is much faster. We changed the index order of all physical variables so that the fastest changing index is the grid cell index.  This reduced the uncoalesced memory access on the GPUs.

Finally, in 2022 we optimized the grid-to-node decomposition
mapping and the load balance figure-of-merit in anticipation of the much higher node counts anticipated in the future. Importantly, we have already achieved very good load balancing efficiency for all decompositions for both the GPU and CPU variants of F{\sc{ornax}}.

\bigskip

\section*{Data Availability}

The numerical data underlying this article will be shared upon reasonable request to the corresponding author.



\bibliography{citations}{}

\begin{thebibliography}{}
\expandafter\ifx\csname natexlab\endcsname\relax\def\natexlab#1{#1}\fi
\providecommand{\url}[1]{\href{#1}{#1}}
\providecommand{\dodoi}[1]{doi:~\href{http://doi.org/#1}{\nolinkurl{#1}}}
\providecommand{\doeprint}[1]{\href{http://ascl.net/#1}{\nolinkurl{http://ascl.net/#1}}}
\providecommand{\doarXiv}[1]{\href{https://arxiv.org/abs/#1}{\nolinkurl{https://arxiv.org/abs/#1}}}

\bibitem[{{Adams} {et~al.}(2017){Adams}, {Kochanek}, {Gerke}, {Stanek}, \&
  {Dai}}]{adams2017}
{Adams}, S.~M., {Kochanek}, C.~S., {Gerke}, J.~R., {Stanek}, K.~Z., \& {Dai},
  X. 2017, \mnras, 468, 4968, \dodoi{10.1093/mnras/stx816}

\bibitem[{Andresen {et~al.}(2019)Andresen, Muller, Janka, Summa, Gill, \&
  Zanolin}]{Andresen2019}
Andresen, H., Muller, E., Janka, H.~T., {et~al.} 2019, Monthly Notices of the
  Royal Astronomical Society, 486, 2238–2253, \dodoi{10.1093/mnras/stz990}

\bibitem[{{Arcones} \& {Thielemann}(2023)}]{arcones2023}
{Arcones}, A., \& {Thielemann}, F.-K. 2023, \aapr, 31, 1,
  \dodoi{10.1007/s00159-022-00146-x}

\bibitem[{{Arnett}(1967)}]{arnett1967}
{Arnett}, D. 1967, Canadian Journal of Physics, 45, 1621,
  \dodoi{10.1139/p67-126}

\bibitem[{{Bethe} \& {Wilson}(1985)}]{Bethe1985}
{Bethe}, H.~A., \& {Wilson}, J.~R. 1985, \apj, 295, 14, \dodoi{10.1086/163343}

\bibitem[{{Blondin} {et~al.}(2003){Blondin}, {Mezzacappa}, \&
  {DeMarino}}]{blondin2003}
{Blondin}, J.~M., {Mezzacappa}, A., \& {DeMarino}, C. 2003, \apj, 584, 971,
  \dodoi{10.1086/345812}

\bibitem[{{Blondin} \& {Shaw}(2007)}]{Blondin2007b}
{Blondin}, J.~M., \& {Shaw}, S. 2007, \apj, 656, 366, \dodoi{10.1086/510614}

\bibitem[{Bollig {et~al.}(2017)Bollig, Janka, Lohs, Martinez-Pinedo, Horowitz,
  \& Melson}]{Bollig2017}
Bollig, R., Janka, H.-T., Lohs, A., {et~al.} 2017, Physical Review Letters,
  119, 242702, \dodoi{10.1103/PhysRevLett.119.242702}

\bibitem[{Bollig {et~al.}(2021)Bollig, Yadav, Kresse, Janka, Mueller, \&
  Heger}]{Bollig2021}
Bollig, R., Yadav, N., Kresse, D., {et~al.} 2021, The Astrophysical Journal,
  915, 28, \dodoi{10.3847/1538-4357/abf82e}

\bibitem[{{Bowers} \& {Wilson}(1982)}]{bowers1982}
{Bowers}, R.~L., \& {Wilson}, J.~R. 1982, \apjs, 50, 115,
  \dodoi{10.1086/190822}

\bibitem[{{Bruenn} {et~al.}(2020){Bruenn}, {Blondin}, {Hix}, {Lentz}, {Messer},
  {Mezzacappa}, {Endeve}, {Harris}, {Marronetti}, {Budiardja}, {Chertkow}, \&
  {Lee}}]{Bruenn2020}
{Bruenn}, S.~W., {Blondin}, J.~M., {Hix}, W.~R., {et~al.} 2020, \apjs, 248, 11,
  \dodoi{10.3847/1538-4365/ab7aff}

\bibitem[{{Buras} {et~al.}(2006){Buras}, {Rampp}, {Janka}, \&
  {Kifonidis}}]{Buras2006}
{Buras}, R., {Rampp}, M., {Janka}, H.~T., \& {Kifonidis}, K. 2006, \aap, 447,
  1049, \dodoi{10.1051/0004-6361:20053783}

\bibitem[{{Burrows}(1987{\natexlab{a}})}]{burrows1987}
{Burrows}, A. 1987{\natexlab{a}}, \apjl, 318, L57, \dodoi{10.1086/184937}

\bibitem[{{Burrows}(1987{\natexlab{b}})}]{Burrows_PT}
---. 1987{\natexlab{b}}, Physics Today, 40, 28, \dodoi{10.1063/1.881086}

\bibitem[{{Burrows}(2013)}]{Burrows2013}
---. 2013, Reviews of Modern Physics, 85, 245,
  \dodoi{10.1103/RevModPhys.85.245}

\bibitem[{{Burrows}(2019)}]{burrows_paris_2019}
{Burrows}, A. 2019, in Proceedings of the International Conference on History
  of the Neutrino, ed. M.~{Cribier}, J.~{Dumarchez}, \& D.~{Vignaud}, 1774457,
  \dodoi{10.48550/arXiv.1812.05612}

\bibitem[{{Burrows} \& {Goshy}(1993)}]{goshy}
{Burrows}, A., \& {Goshy}, J. 1993, \apjl, 416, L75, \dodoi{10.1086/187074}

\bibitem[{{Burrows} {et~al.}(1995){Burrows}, {Hayes}, \&
  {Fryxell}}]{Burrows1995}
{Burrows}, A., {Hayes}, J., \& {Fryxell}, B.~A. 1995, \apj, 450, 830,
  \dodoi{10.1086/176188}

\bibitem[{{Burrows} {et~al.}(2019){Burrows}, {Radice}, \&
  {Vartanyan}}]{Burrows2019}
{Burrows}, A., {Radice}, D., \& {Vartanyan}, D. 2019, \mnras, 485, 3153,
  \dodoi{10.1093/mnras/stz543}

\bibitem[{Burrows {et~al.}(2020)Burrows, Radice, Vartanyan, Nagakura, Skinner,
  \& Dolence}]{Burrows2020}
Burrows, A., Radice, D., Vartanyan, D., {et~al.} 2020, Monthly Notices of the
  Royal Astronomical Society, 491, 2715–2735, \dodoi{10.1093/mnras/stz3223}

\bibitem[{Burrows {et~al.}(2006)Burrows, Reddy, \& Thompson}]{Burrows2006}
Burrows, A., Reddy, S., \& Thompson, T.~A. 2006, Nuclear Physics A, 777,
  356–394, \dodoi{10.1016/j.nuclphysa.2004.06.012}

\bibitem[{{Burrows} \& {Sawyer}(1998)}]{1998PhRvC..58..554B}
{Burrows}, A., \& {Sawyer}, R.~F. 1998, \prc, 58, 554,
  \dodoi{10.1103/PhysRevC.58.554}

\bibitem[{{Burrows} \& {Vartanyan}(2021)}]{Burrows2021}
{Burrows}, A., \& {Vartanyan}, D. 2021, \nat, 589, 29,
  \dodoi{10.1038/s41586-020-03059-w}

\bibitem[{Burrows {et~al.}(2018)Burrows, Vartanyan, Dolence, Skinner, \&
  Radice}]{Burrows2018}
Burrows, A., Vartanyan, D., Dolence, J.~C., Skinner, M.~A., \& Radice, D. 2018,
  Space Science Reviews, 214, 33, \dodoi{10.1007/s11214-017-0450-9}

\bibitem[{{Burrows} {et~al.}(2023{\natexlab{a}}){Burrows}, {Vartanyan}, \&
  {Wang}}]{Burrows2023}
{Burrows}, A., {Vartanyan}, D., \& {Wang}, T. 2023{\natexlab{a}}, arXiv
  e-prints, arXiv:2308.05798, \dodoi{10.48550/arXiv.2308.05798}

\bibitem[{{Burrows} {et~al.}(2023{\natexlab{b}}){Burrows}, {Wang}, {Vartanyan},
  \& {Coleman}}]{burrows_kick_2023}
{Burrows}, A., {Wang}, T., {Vartanyan}, D., \& {Coleman}, M. S.~B.
  2023{\natexlab{b}}, arXiv e-prints, arXiv:2311.12109,
  \dodoi{10.48550/arXiv.2311.12109}

\bibitem[{Chatzopoulos {et~al.}(2016)Chatzopoulos, Couch, Arnett, \&
  Timmes}]{Chatzopoulos2016}
Chatzopoulos, E., Couch, S.~M., Arnett, W.~D., \& Timmes, F.~X. 2016, The
  Astrophysical Journal, 822, 61, \dodoi{10.3847/0004-637X/822/2/61}

\bibitem[{{Chevalier} \& {Emmering}(1986)}]{Chevalier1986}
{Chevalier}, R.~A., \& {Emmering}, R.~T. 1986, \apj, 304, 140,
  \dodoi{10.1086/164150}

\bibitem[{{Coleman} \& {Burrows}(2022)}]{coleman}
{Coleman}, M. S.~B., \& {Burrows}, A. 2022, \mnras, 517, 3938,
  \dodoi{10.1093/mnras/stac2573}

\bibitem[{{Colgate} \& {White}(1966)}]{colgate1966}
{Colgate}, S.~A., \& {White}, R.~H. 1966, \apj, 143, 626,
  \dodoi{10.1086/148549}

\bibitem[{{Couch} {et~al.}(2015){Couch}, {Chatzopoulos}, {Arnett}, \&
  {Timmes}}]{2015ApJ...808L..21C}
{Couch}, S.~M., {Chatzopoulos}, E., {Arnett}, W.~D., \& {Timmes}, F.~X. 2015,
  \apjl, 808, L21, \dodoi{10.1088/2041-8205/808/1/L21}

\bibitem[{{Curtis} {et~al.}(2019){Curtis}, {Ebinger}, {Fr{\"o}hlich}, {Hempel},
  {Perego}, {Liebend{\"o}rfer}, \& {Thielemann}}]{curtis2019}
{Curtis}, S., {Ebinger}, K., {Fr{\"o}hlich}, C., {et~al.} 2019, \apj, 870, 2,
  \dodoi{10.3847/1538-4357/aae7d2}

\bibitem[{{Diehl} \& {Prantzos}(2023)}]{diehl2023}
{Diehl}, R., \& {Prantzos}, N. 2023, arXiv e-prints, arXiv:2303.01825,
  \dodoi{10.48550/arXiv.2303.01825}

\bibitem[{{Diehl} \& {Timmes}(1998)}]{diehl1998}
{Diehl}, R., \& {Timmes}, F.~X. 1998, \pasp, 110, 637, \dodoi{10.1086/316169}

\bibitem[{{Diehl} {et~al.}(2021){Diehl}, {Lugaro}, {Heger}, {Sieverding},
  {Tang}, {Li}, {Li}, {Doherty}, {Krause}, {Wallner}, {Prantzos}, {Brinkman},
  {den Hartogh}, {Wehmeyer}, {Yag{\"u}e L{\'o}pez}, {Pleintinger}, {Banerjee},
  \& {Wang}}]{diehl2021}
{Diehl}, R., {Lugaro}, M., {Heger}, A., {et~al.} 2021, \pasa, 38, e062,
  \dodoi{10.1017/pasa.2021.48}

\bibitem[{{Duncan} {et~al.}(1986){Duncan}, {Shapiro}, \&
  {Wasserman}}]{duncan1986}
{Duncan}, R.~C., {Shapiro}, S.~L., \& {Wasserman}, I. 1986, \apj, 309, 141,
  \dodoi{10.1086/164587}

\bibitem[{Ertl {et~al.}(2016)Ertl, Janka, Woosley, Sukhbold, \&
  Ugliano}]{Ertl2016}
Ertl, T., Janka, H.-T., Woosley, S.~E., Sukhbold, T., \& Ugliano, M. 2016, The
  Astrophysical Journal, 818, 124, \dodoi{10.3847/0004-637X/818/2/124}

\bibitem[{{Fan} {et~al.}(2023){Fan}, {Han}, {Jiang}, {Shao}, \&
  {Tang}}]{fan2023}
{Fan}, Y.-Z., {Han}, M.-Z., {Jiang}, J.-L., {Shao}, D.-S., \& {Tang}, S.-P.
  2023, arXiv e-prints, arXiv:2309.12644, \dodoi{10.48550/arXiv.2309.12644}

\bibitem[{{Fang} {et~al.}(2023){Fang}, {Maeda}, {Kuncarayakti}, \&
  {Nagao}}]{qiliang2023}
{Fang}, Q., {Maeda}, K., {Kuncarayakti}, H., \& {Nagao}, T. 2023, Nature
  Astronomy, \dodoi{10.1038/s41550-023-02120-8}

\bibitem[{{Faucher-Gigu{\`e}re} \& {Kaspi}(2006)}]{Faucher2006}
{Faucher-Gigu{\`e}re}, C.-A., \& {Kaspi}, V.~M. 2006, \apj, 643, 332,
  \dodoi{10.1086/501516}

\bibitem[{{Fields} \& {Couch}(2020)}]{Fields2020}
{Fields}, C.~E., \& {Couch}, S.~M. 2020, \apj, 901, 33,
  \dodoi{10.3847/1538-4357/abada7}

\bibitem[{{Fields} \& {Couch}(2021)}]{Fields2021}
---. 2021, \apj, 921, 28, \dodoi{10.3847/1538-4357/ac24fb}

\bibitem[{{Fischer} {et~al.}(2023){Fischer}, {Guo}, {Langanke},
  {Martinez-Pinedo}, {Qian}, \& {Wu}}]{fischer2023}
{Fischer}, T., {Guo}, G., {Langanke}, K., {et~al.} 2023, arXiv e-prints,
  arXiv:2308.03962, \dodoi{10.48550/arXiv.2308.03962}

\bibitem[{{Foglizzo} {et~al.}(2007){Foglizzo}, {Galletti}, {Scheck}, \&
  {Janka}}]{foglizzo2007}
{Foglizzo}, T., {Galletti}, P., {Scheck}, L., \& {Janka}, H.~T. 2007, \apj,
  654, 1006, \dodoi{10.1086/509612}

\bibitem[{{Foglizzo} {et~al.}(2006){Foglizzo}, {Scheck}, \&
  {Janka}}]{foglizzo2006}
{Foglizzo}, T., {Scheck}, L., \& {Janka}, H.~T. 2006, \apj, 652, 1436,
  \dodoi{10.1086/508443}

\bibitem[{{Fr{\"o}hlich} {et~al.}(2006){Fr{\"o}hlich}, {Mart{\'\i}nez-Pinedo},
  {Liebend{\"o}rfer}, {Thielemann}, {Bravo}, {Hix}, {Langanke}, \&
  {Zinner}}]{frohlich2006}
{Fr{\"o}hlich}, C., {Mart{\'\i}nez-Pinedo}, G., {Liebend{\"o}rfer}, M.,
  {et~al.} 2006, \prl, 96, 142502, \dodoi{10.1103/PhysRevLett.96.142502}

\bibitem[{{Glas} {et~al.}(2019){Glas}, {Just}, {Janka}, \&
  {Obergaulinger}}]{Glas2019}
{Glas}, R., {Just}, O., {Janka}, H.~T., \& {Obergaulinger}, M. 2019, \apj, 873,
  45, \dodoi{10.3847/1538-4357/ab0423}

\bibitem[{{Harris} {et~al.}(2017){Harris}, {Hix}, {Chertkow}, {Lee}, {Lentz},
  \& {Messer}}]{harris2017}
{Harris}, J.~A., {Hix}, W.~R., {Chertkow}, M.~A., {et~al.} 2017, \apj, 843, 2,
  \dodoi{10.3847/1538-4357/aa76de}

\bibitem[{{Heger} {et~al.}(2003){Heger}, {Fryer}, {Woosley}, {Langer}, \&
  {Hartmann}}]{heger2003b}
{Heger}, A., {Fryer}, C.~L., {Woosley}, S.~E., {Langer}, N., \& {Hartmann},
  D.~H. 2003, \apj, 591, 288, \dodoi{10.1086/375341}

\bibitem[{{Heger} \& {Woosley}(2010)}]{heger2010}
{Heger}, A., \& {Woosley}, S.~E. 2010, \apj, 724, 341,
  \dodoi{10.1088/0004-637X/724/1/341}

\bibitem[{{Herant} {et~al.}(1994){Herant}, {Benz}, {Hix}, {Fryer}, \&
  {Colgate}}]{Herant1994}
{Herant}, M., {Benz}, W., {Hix}, W.~R., {Fryer}, C.~L., \& {Colgate}, S.~A.
  1994, \apj, 435, 339, \dodoi{10.1086/174817}

\bibitem[{Holland-Ashford {et~al.}(2017)Holland-Ashford, Lopez, Auchettl,
  Temim, \& Ramirez-Ruiz}]{Holland-Ashford2017}
Holland-Ashford, T., Lopez, L.~A., Auchettl, K., Temim, T., \& Ramirez-Ruiz, E.
  2017, The Astrophysical Journal, 844, 84, \dodoi{10.3847/1538-4357/aa7a5c}

\bibitem[{{Horowitz}(2002)}]{Horowitz2002}
{Horowitz}, C.~J. 2002, \prd, 65, 043001, \dodoi{10.1103/PhysRevD.65.043001}

\bibitem[{{Horowitz} {et~al.}(2017){Horowitz}, {Caballero}, {Lin}, {O'Connor},
  \& {Schwenk}}]{Horowitz2017}
{Horowitz}, C.~J., {Caballero}, O.~L., {Lin}, Z., {O'Connor}, E., \& {Schwenk},
  A. 2017, \prc, 95, 025801, \dodoi{10.1103/PhysRevC.95.025801}

\bibitem[{Igoshev \& Popov(2013)}]{Igoshev2013}
Igoshev, A.~P., \& Popov, S.~B. 2013, Monthly Notices of the Royal Astronomical
  Society, 432, 967–972, \dodoi{10.1093/mnras/stt519}

\bibitem[{{Janka}(2012)}]{Janka2012}
{Janka}, H.-T. 2012, Annual Review of Nuclear and Particle Science, 62, 407,
  \dodoi{10.1146/annurev-nucl-102711-094901}

\bibitem[{Janka(2017{\natexlab{a}})}]{Janka_2017b}
Janka, H.-T. 2017{\natexlab{a}}, Neutrino-Driven Explosions (Springer
  International Publishing), 1095–1150, \dodoi{10.1007/978-3-319-21846-5_109}

\bibitem[{Janka(2017{\natexlab{b}})}]{Janka2017}
---. 2017{\natexlab{b}}, The Astrophysical Journal, 837, 84,
  \dodoi{10.3847/1538-4357/aa618e}

\bibitem[{{Janka} \& {Mueller}(1996)}]{janka1996}
{Janka}, H.~T., \& {Mueller}, E. 1996, \aap, 306, 167

\bibitem[{{Jones} {et~al.}(2017){Jones}, {Andrassy}, {Sandalski}, {Davis},
  {Woodward}, \& {Herwig}}]{jones2017}
{Jones}, S., {Andrassy}, R., {Sandalski}, S., {et~al.} 2017, \mnras, 465, 2991,
  \dodoi{10.1093/mnras/stw2783}

\bibitem[{{Jones} {et~al.}(2016){Jones}, {R{\"o}pke}, {Pakmor}, {Seitenzahl},
  {Ohlmann}, \& {Edelmann}}]{jones2016}
{Jones}, S., {R{\"o}pke}, F.~K., {Pakmor}, R., {et~al.} 2016, \aap, 593, A72,
  \dodoi{10.1051/0004-6361/201628321}

\bibitem[{{Just} {et~al.}(2015){Just}, {Obergaulinger}, \&
  {Janka}}]{2015MNRAS.453.3386J}
{Just}, O., {Obergaulinger}, M., \& {Janka}, H.~T. 2015, \mnras, 453, 3386,
  \dodoi{10.1093/mnras/stv1892}

\bibitem[{{Katsuda} {et~al.}(2018){Katsuda}, {Morii}, {Janka},
  {Wongwathanarat}, {Nakamura}, {Kotake}, {Mori}, {M{\"u}ller}, {Takiwaki},
  {Tanaka}, {Tominaga}, \& {Tsunemi}}]{2018ApJ...856...18K}
{Katsuda}, S., {Morii}, M., {Janka}, H.-T., {et~al.} 2018, \apj, 856, 18,
  \dodoi{10.3847/1538-4357/aab092}

\bibitem[{Kuroda {et~al.}(2020)Kuroda, Arcones, Takiwaki, \&
  Kotake}]{Kuroda2020}
Kuroda, T., Arcones, A., Takiwaki, T., \& Kotake, K. 2020, The Astrophysical
  Journal, 896, 102, \dodoi{10.3847/1538-4357/ab9308}

\bibitem[{{Lentz} {et~al.}(2015){Lentz}, {Bruenn}, {Hix}, {Mezzacappa},
  {Messer}, {Endeve}, {Blondin}, {Harris}, {Marronetti}, \&
  {Yakunin}}]{Lentz2015}
{Lentz}, E.~J., {Bruenn}, S.~W., {Hix}, W.~R., {et~al.} 2015, \apjl, 807, L31,
  \dodoi{10.1088/2041-8205/807/2/L31}

\bibitem[{{Leonard} {et~al.}(2000){Leonard}, {Filippenko}, {Barth}, \&
  {Matheson}}]{leonard_polar_2000}
{Leonard}, D.~C., {Filippenko}, A.~V., {Barth}, A.~J., \& {Matheson}, T. 2000,
  \apj, 536, 239, \dodoi{10.1086/308910}

\bibitem[{{Leonard} {et~al.}(2006){Leonard}, {Filippenko}, {Ganeshalingam},
  {Serduke}, {Li}, {Swift}, {Gal-Yam}, {Foley}, {Fox}, {Park}, {Hoffman}, \&
  {Wong}}]{leonard2006}
{Leonard}, D.~C., {Filippenko}, A.~V., {Ganeshalingam}, M., {et~al.} 2006,
  \nat, 440, 505, \dodoi{10.1038/nature04558}

\bibitem[{{Limongi} \& {Chieffi}(2018)}]{limongi2018}
{Limongi}, M., \& {Chieffi}, A. 2018, \apjs, 237, 13,
  \dodoi{10.3847/1538-4365/aacb24}

\bibitem[{{Lippuner} \& {Roberts}(2017)}]{lippuner2017}
{Lippuner}, J., \& {Roberts}, L.~F. 2017, \apjs, 233, 18,
  \dodoi{10.3847/1538-4365/aa94cb}

\bibitem[{Lyne \& Lorimer(1994)}]{Lyne1994}
Lyne, A.~G., \& Lorimer, D.~R. 1994, Nature, 369, 127–129,
  \dodoi{10.1038/369127a0}

\bibitem[{{Manchester} {et~al.}(2005){Manchester}, {Hobbs}, {Teoh}, \&
  {Hobbs}}]{Manchester2005}
{Manchester}, R.~N., {Hobbs}, G.~B., {Teoh}, A., \& {Hobbs}, M. 2005, \aj, 129,
  1993, \dodoi{10.1086/428488}

\bibitem[{{Mandel} \& {M{\"u}ller}(2020)}]{Mandel2020}
{Mandel}, I., \& {M{\"u}ller}, B. 2020, \mnras, 499, 3214,
  \dodoi{10.1093/mnras/staa3043}

\bibitem[{{Marek} {et~al.}(2006){Marek}, {Dimmelmeier}, {Janka}, {M{\"u}ller},
  \& {Buras}}]{Marek2006}
{Marek}, A., {Dimmelmeier}, H., {Janka}, H.~T., {M{\"u}ller}, E., \& {Buras},
  R. 2006, \aap, 445, 273, \dodoi{10.1051/0004-6361:20052840}

\bibitem[{McNeill \& M\"uller(2021)}]{McNeill2021}
McNeill, L.~O., \& M\"uller, B. 2021, arXiv e-prints, arXiv:2107.00173

\bibitem[{{Melson} {et~al.}(2015){Melson}, {Janka}, {Bollig}, {Hanke}, {Marek},
  \& {M{\"u}ller}}]{Melson2015}
{Melson}, T., {Janka}, H.-T., {Bollig}, R., {et~al.} 2015, \apjl, 808, L42,
  \dodoi{10.1088/2041-8205/808/2/L42}

\bibitem[{{Melson} {et~al.}(2020){Melson}, {Kresse}, \&
  {Janka}}]{2020ApJ...891...27M}
{Melson}, T., {Kresse}, D., \& {Janka}, H.-T. 2020, \apj, 891, 27,
  \dodoi{10.3847/1538-4357/ab72a7}

\bibitem[{{M{\"u}ller} {et~al.}(2018){M{\"u}ller}, {Gay}, {Heger}, {Tauris}, \&
  {Sim}}]{Muller2018_ultra}
{M{\"u}ller}, B., {Gay}, D.~W., {Heger}, A., {Tauris}, T.~M., \& {Sim}, S.~A.
  2018, \mnras, 479, 3675, \dodoi{10.1093/mnras/sty1683}

\bibitem[{M\"uller {et~al.}(2016{\natexlab{a}})M\"uller, Heger, Liptai, \&
  Cameron}]{Muller2016}
M\"uller, B., Heger, A., Liptai, D., \& Cameron, J.~B. 2016{\natexlab{a}},
  Monthly Notices of the Royal Astronomical Society, 460, 742–764,
  \dodoi{10.1093/mnras/stw1083}

\bibitem[{M\"uller {et~al.}(2017)M\"uller, Melson, Heger, \&
  Janka}]{Muller2017}
M\"uller, B., Melson, T., Heger, A., \& Janka, H.-T. 2017, Monthly Notices of
  the Royal Astronomical Society, 472, 491–513, \dodoi{10.1093/mnras/stx1962}

\bibitem[{M\"uller \& Varma(2020)}]{Muller2020}
M\"uller, B., \& Varma, V. 2020, arXiv:2007.04775 [astro-ph].
\newblock \url{http://arxiv.org/abs/2007.04775}

\bibitem[{M\"uller {et~al.}(2016{\natexlab{b}})M\"uller, Viallet, Heger, \&
  Janka}]{Muller2016b}
M\"uller, B., Viallet, M., Heger, A., \& Janka, H.-T. 2016{\natexlab{b}}, The
  Astrophysical Journal, 833, 124, \dodoi{10.3847/1538-4357/833/1/124}

\bibitem[{{M{\"u}ller} {et~al.}(2019){M{\"u}ller}, {Tauris}, {Heger},
  {Banerjee}, {Qian}, {Powell}, {Chan}, {Gay}, \&
  {Langer}}]{muller_low_kick2019}
{M{\"u}ller}, B., {Tauris}, T.~M., {Heger}, A., {et~al.} 2019, \mnras, 484,
  3307, \dodoi{10.1093/mnras/stz216}

\bibitem[{{M{\"u}ller} \& {Steinmetz}(1995)}]{mueller:95}
{M{\"u}ller}, E., \& {Steinmetz}, M. 1995, Computer Physics Communications, 89,
  45, \dodoi{10.1016/0010-4655(94)00185-5}

\bibitem[{{M{\"u}ller} {et~al.}(2017){M{\"u}ller}, {Prieto}, {Pejcha}, \&
  {Clocchiatti}}]{prieto2017}
{M{\"u}ller}, T., {Prieto}, J.~L., {Pejcha}, O., \& {Clocchiatti}, A. 2017,
  \apj, 841, 127, \dodoi{10.3847/1538-4357/aa72f1}

\bibitem[{Nagakura {et~al.}(2019)Nagakura, Burrows, Radice, \&
  Vartanyan}]{Nagakura2019}
Nagakura, H., Burrows, A., Radice, D., \& Vartanyan, D. 2019, Monthly Notices
  of the Royal Astronomical Society, 490, 4622–4637,
  \dodoi{10.1093/mnras/stz2730}

\bibitem[{{Nagao} {et~al.}(2023){Nagao}, {Patat}, {Cikota}, {Baade}, {Mattila},
  {Kotak}, {Kuncarayakti}, {Bulla}, \& {Ayala}}]{nagao_2023_polar}
{Nagao}, T., {Patat}, F., {Cikota}, A., {et~al.} 2023, arXiv e-prints,
  arXiv:2308.00996, \dodoi{10.48550/arXiv.2308.00996}

\bibitem[{{Nakamura} {et~al.}(2015){Nakamura}, {Takiwaki}, {Kuroda}, \&
  {Kotake}}]{nakamura2015}
{Nakamura}, K., {Takiwaki}, T., {Kuroda}, T., \& {Kotake}, K. 2015, \pasj, 67,
  107, \dodoi{10.1093/pasj/psv073}

\bibitem[{Ng \& Romani(2007)}]{Ng2007}
Ng, C.-Y., \& Romani, R.~W. 2007, The Astrophysical Journal, 660, 1357–1374,
  \dodoi{10.1086/513597}

\bibitem[{{Nomoto} {et~al.}(2013){Nomoto}, {Kobayashi}, \&
  {Tominaga}}]{nomoto2013}
{Nomoto}, K., {Kobayashi}, C., \& {Tominaga}, N. 2013, \araa, 51, 457,
  \dodoi{10.1146/annurev-astro-082812-140956}

\bibitem[{Noutsos {et~al.}(2013)Noutsos, Schnitzeler, Keane, Kramer, \&
  Johnston}]{Noutsos2013}
Noutsos, A., Schnitzeler, D. H. F.~M., Keane, E.~F., Kramer, M., \& Johnston,
  S. 2013, Monthly Notices of the Royal Astronomical Society, 430, 2281–2301,
  \dodoi{10.1093/mnras/stt047}

\bibitem[{{Obergaulinger} \& {Aloy}(2020)}]{Obergaulinger2020}
{Obergaulinger}, M., \& {Aloy}, M.~{\'A}. 2020, \mnras, 492, 4613,
  \dodoi{10.1093/mnras/staa096}

\bibitem[{Obergaulinger \& Aloy(2021)}]{Obergaulinger2021}
Obergaulinger, M., \& Aloy, M.~A. 2021, Monthly Notices of the Royal
  Astronomical Society, 503, 4942–4963, \dodoi{10.1093/mnras/stab295}

\bibitem[{{Obergaulinger} {et~al.}(2018){Obergaulinger}, {Just}, \&
  {Aloy}}]{Obergaulinger2018}
{Obergaulinger}, M., {Just}, O., \& {Aloy}, M.~{\'A}. 2018, Journal of Physics
  G Nuclear Physics, 45, 084001, \dodoi{10.1088/1361-6471/aac982}

\bibitem[{{O'Connor}(2015)}]{2015ApJS..219...24O}
{O'Connor}, E. 2015, \apjs, 219, 24, \dodoi{10.1088/0067-0049/219/2/24}

\bibitem[{{O'Connor} \& {Ott}(2011)}]{oconnor2011}
{O'Connor}, E., \& {Ott}, C.~D. 2011, \apj, 730, 70,
  \dodoi{10.1088/0004-637X/730/2/70}

\bibitem[{{O'Connor} \& {Ott}(2013)}]{oconnor_ott:2013}
---. 2013, \apj, 762, 126, \dodoi{10.1088/0004-637X/762/2/126}

\bibitem[{{Ott} {et~al.}(2018){Ott}, {Roberts}, {da Silva Schneider}, {Fedrow},
  {Haas}, \& {Schnetter}}]{ott2018_rel}
{Ott}, C.~D., {Roberts}, L.~F., {da Silva Schneider}, A., {et~al.} 2018, \apjl,
  855, L3, \dodoi{10.3847/2041-8213/aaa967}

\bibitem[{{{\"O}zel} \& {Freire}(2016)}]{feryal_masses_2016}
{{\"O}zel}, F., \& {Freire}, P. 2016, \araa, 54, 401,
  \dodoi{10.1146/annurev-astro-081915-023322}

\bibitem[{Pejcha(2020)}]{Pejcha_2020}
Pejcha, O. 2020, The Explosion Mechanism of Core-Collapse Supernovae and Its
  Observational Signatures (Springer International Publishing), 189–211,
  \dodoi{10.1007/978-3-030-38509-5_7}

\bibitem[{Popov \& Turolla(2012)}]{Popov2012}
Popov, S.~B., \& Turolla, R. 2012, Astrophysics and Space Science, 341,
  457–464, \dodoi{10.1007/s10509-012-1100-z}

\bibitem[{Powell \& M\"uller(2020)}]{Powell2020}
Powell, J., \& M\"uller, B. 2020, Monthly Notices of the Royal Astronomical
  Society, 494, 4665–4675, \dodoi{10.1093/mnras/staa1048}

\bibitem[{{Pruet} {et~al.}(2006){Pruet}, {Hoffman}, {Woosley}, {Janka}, \&
  {Buras}}]{Pruet2006}
{Pruet}, J., {Hoffman}, R.~D., {Woosley}, S.~E., {Janka}, H.~T., \& {Buras}, R.
  2006, \apj, 644, 1028, \dodoi{10.1086/503891}

\bibitem[{{Qian} \& {Woosley}(1996)}]{qian1996}
{Qian}, Y.~Z., \& {Woosley}, S.~E. 1996, \apj, 471, 331, \dodoi{10.1086/177973}

\bibitem[{{Rauscher} {et~al.}(2002){Rauscher}, {Heger}, {Hoffman}, \&
  {Woosley}}]{rauscher2002}
{Rauscher}, T., {Heger}, A., {Hoffman}, R.~D., \& {Woosley}, S.~E. 2002, \apj,
  576, 323, \dodoi{10.1086/341728}

\bibitem[{{Reddy} {et~al.}(1999){Reddy}, {Prakash}, {Lattimer}, \&
  {Pons}}]{reddy1999}
{Reddy}, S., {Prakash}, M., {Lattimer}, J.~M., \& {Pons}, J.~A. 1999, \prc, 59,
  2888, \dodoi{10.1103/PhysRevC.59.2888}

\bibitem[{{Roberts} {et~al.}(2016){Roberts}, {Ott}, {Haas}, {O'Connor},
  {Diener}, \& {Schnetter}}]{roberts2016}
{Roberts}, L.~F., {Ott}, C.~D., {Haas}, R., {et~al.} 2016, \apj, 831, 98,
  \dodoi{10.3847/0004-637X/831/1/98}

\bibitem[{{Rodr{\'\i}guez} {et~al.}(2023){Rodr{\'\i}guez}, {Maoz}, \&
  {Nakar}}]{rodriguez2023}
{Rodr{\'\i}guez}, {\'O}., {Maoz}, D., \& {Nakar}, E. 2023, \apj, 955, 71,
  \dodoi{10.3847/1538-4357/ace2bd}

\bibitem[{{Sato} {et~al.}(2021){Sato}, {Maeda}, {Nagataki}, {Yoshida},
  {Grefenstette}, {Williams}, {Umeda}, {Ono}, \&
  {Hughes}}]{sato_hughes_bubbles_2021}
{Sato}, T., {Maeda}, K., {Nagataki}, S., {et~al.} 2021, \nat, 592, 537,
  \dodoi{10.1038/s41586-021-03391-9}

\bibitem[{Scheck {et~al.}(2006)Scheck, Kifonidis, Janka, \&
  Müller}]{Scheck2006}
Scheck, L., Kifonidis, K., Janka, H.-T., \& Müller, E. 2006, Astronomy and
  Astrophysics, 457, 963–986, \dodoi{bb}

\bibitem[{{Sieverding} {et~al.}(2023){Sieverding}, {Kresse}, \&
  {Janka}}]{sieverding2023}
{Sieverding}, A., {Kresse}, D., \& {Janka}, H.-T. 2023, arXiv e-prints,
  arXiv:2308.09659, \dodoi{10.48550/arXiv.2308.09659}

\bibitem[{{Sieverding} {et~al.}(2020){Sieverding}, {M{\"u}ller}, \&
  {Qian}}]{sieverding2020}
{Sieverding}, A., {M{\"u}ller}, B., \& {Qian}, Y.~Z. 2020, \apj, 904, 163,
  \dodoi{10.3847/1538-4357/abc61b}

\bibitem[{{Skinner} {et~al.}(2016){Skinner}, {Burrows}, \&
  {Dolence}}]{skinner2016}
{Skinner}, M.~A., {Burrows}, A., \& {Dolence}, J.~C. 2016, \apj, 831, 81,
  \dodoi{10.3847/0004-637X/831/1/81}

\bibitem[{Skinner {et~al.}(2019)Skinner, Dolence, Burrows, Radice, \&
  Vartanyan}]{Skinner2019}
Skinner, M.~A., Dolence, J.~C., Burrows, A., Radice, D., \& Vartanyan, D. 2019,
  The Astrophysical Journal Supplement Series, 241, 7,
  \dodoi{10.3847/1538-4365/ab007f}

\bibitem[{{Smartt}(2015)}]{smartt2015}
{Smartt}, S.~J. 2015, \pasa, 32, e016, \dodoi{10.1017/pasa.2015.17}

\bibitem[{{Smartt} {et~al.}(2009){Smartt}, {Eldridge}, {Crockett}, \&
  {Maund}}]{smartt2009}
{Smartt}, S.~J., {Eldridge}, J.~J., {Crockett}, R.~M., \& {Maund}, J.~R. 2009,
  \mnras, 395, 1409, \dodoi{10.1111/j.1365-2966.2009.14506.x}

\bibitem[{{Smith}(2014)}]{smith2014}
{Smith}, N. 2014, \araa, 52, 487, \dodoi{10.1146/annurev-astro-081913-040025}

\bibitem[{Stanzione {et~al.}(2020)Stanzione, West, Evans, Minyard, Ghattas, \&
  Panda}]{Stanzione2020}
Stanzione, D., West, J., Evans, R.~T., {et~al.} 2020, in Practice and
  Experience in Advanced Research Computing, PEARC '20 (New York, NY, USA:
  Association for Computing Machinery), 106–111,
  \dodoi{10.1145/3311790.3396656}

\bibitem[{{Steiner} {et~al.}(2013){Steiner}, {Hempel}, \&
  {Fischer}}]{Steiner2013}
{Steiner}, A.~W., {Hempel}, M., \& {Fischer}, T. 2013, \apj, 774, 17,
  \dodoi{10.1088/0004-637X/774/1/17}

\bibitem[{{Stockinger} {et~al.}(2020){Stockinger}, {Janka}, {Kresse}, {Melson},
  {Ertl}, {Gabler}, {Gessner}, {Wongwathanarat}, {Tolstov}, {Leung}, {Nomoto},
  \& {Heger}}]{Stockinger2020}
{Stockinger}, G., {Janka}, H.~T., {Kresse}, D., {et~al.} 2020, \mnras, 496,
  2039, \dodoi{10.1093/mnras/staa1691}

\bibitem[{{Sukhbold} {et~al.}(2016){Sukhbold}, {Ertl}, {Woosley}, {Brown}, \&
  {Janka}}]{Sukhbold2016}
{Sukhbold}, T., {Ertl}, T., {Woosley}, S.~E., {Brown}, J.~M., \& {Janka}, H.~T.
  2016, \apj, 821, 38, \dodoi{10.3847/0004-637X/821/1/38}

\bibitem[{{Sukhbold} {et~al.}(2018){Sukhbold}, {Woosley}, \&
  {Heger}}]{Sukhbold2018}
{Sukhbold}, T., {Woosley}, S.~E., \& {Heger}, A. 2018, \apj, 860, 93,
  \dodoi{10.3847/1538-4357/aac2da}

\bibitem[{{Takiwaki} {et~al.}(2016){Takiwaki}, {Kotake}, \&
  {Suwa}}]{Takiwaki2016}
{Takiwaki}, T., {Kotake}, K., \& {Suwa}, Y. 2016, \mnras, 461, L112,
  \dodoi{10.1093/mnrasl/slw105}

\bibitem[{{Tanaka} {et~al.}(2017){Tanaka}, {Maeda}, {Mazzali}, {Kawabata}, \&
  {Nomoto}}]{tanaka_polar_2017}
{Tanaka}, M., {Maeda}, K., {Mazzali}, P.~A., {Kawabata}, K.~S., \& {Nomoto}, K.
  2017, \apj, 837, 105, \dodoi{10.3847/1538-4357/aa6035}

\bibitem[{{Taubenberger} {et~al.}(2009){Taubenberger}, {Valenti}, {Benetti},
  {Cappellaro}, {Della Valle}, {Elias-Rosa}, {Hachinger}, {Hillebrandt},
  {Maeda}, {Mazzali}, {Pastorello}, {Patat}, {Sim}, \&
  {Turatto}}]{taubenberger_profiles_2009}
{Taubenberger}, S., {Valenti}, S., {Benetti}, S., {et~al.} 2009, \mnras, 397,
  677, \dodoi{10.1111/j.1365-2966.2009.15003.x}

\bibitem[{{Temaj} {et~al.}(2023){Temaj}, {Schneider}, {Laplace}, {Wei}, \&
  {Podsiadlowski}}]{temaj2023}
{Temaj}, D., {Schneider}, F.~R.~N., {Laplace}, E., {Wei}, D., \&
  {Podsiadlowski}, P. 2023, arXiv e-prints, arXiv:2311.05701,
  \dodoi{10.48550/arXiv.2311.05701}

\bibitem[{{Thielemann} {et~al.}(1996){Thielemann}, {Nomoto}, \&
  {Hashimoto}}]{thielemann1996}
{Thielemann}, F.-K., {Nomoto}, K., \& {Hashimoto}, M.-A. 1996, \apj, 460, 408,
  \dodoi{10.1086/176980}

\bibitem[{{Thompson} {et~al.}(2000){Thompson}, {Burrows}, \&
  {Horvath}}]{thomp_bur_horvath}
{Thompson}, T.~A., {Burrows}, A., \& {Horvath}, J.~E. 2000, \prc, 62, 035802,
  \dodoi{10.1103/PhysRevC.62.035802}

\bibitem[{{Thompson} {et~al.}(2003){Thompson}, {Burrows}, \&
  {Pinto}}]{2003ApJ...592..434T}
{Thompson}, T.~A., {Burrows}, A., \& {Pinto}, P.~A. 2003, \apj, 592, 434,
  \dodoi{10.1086/375701}

\bibitem[{{Varma} \& {M{\"u}ller}(2021)}]{varma21}
{Varma}, V., \& {M{\"u}ller}, B. 2021, arXiv e-prints, arXiv:2101.00213.
\newblock \doarXiv{2101.00213}

\bibitem[{{Vartanyan} {et~al.}(2018){Vartanyan}, {Burrows}, {Radice},
  {Skinner}, \& {Dolence}}]{vartanyan2018}
{Vartanyan}, D., {Burrows}, A., {Radice}, D., {Skinner}, M.~A., \& {Dolence},
  J. 2018, \mnras, 477, 3091, \dodoi{10.1093/mnras/sty809}

\bibitem[{{Vartanyan} {et~al.}(2023){Vartanyan}, {Burrows}, {Wang}, {Coleman},
  \& {White}}]{Vartanyan2023}
{Vartanyan}, D., {Burrows}, A., {Wang}, T., {Coleman}, M. S.~B., \& {White},
  C.~J. 2023, \prd, 107, 103015, \dodoi{10.1103/PhysRevD.107.103015}

\bibitem[{{Vartanyan} {et~al.}(2022){Vartanyan}, {Coleman}, \&
  {Burrows}}]{Vartanyan2021}
{Vartanyan}, D., {Coleman}, M. S.~B., \& {Burrows}, A. 2022, \mnras, 510, 4689,
  \dodoi{10.1093/mnras/stab3702}

\bibitem[{{Vaytet} {et~al.}(2011){Vaytet}, {Audit}, {Dubroca}, \&
  {Delahaye}}]{vaytet:11}
{Vaytet}, N.~M.~H., {Audit}, E., {Dubroca}, B., \& {Delahaye}, F. 2011, \jqsrt,
  112, 1323, \dodoi{10.1016/j.jqsrt.2011.01.027}

\bibitem[{{Wanajo} {et~al.}(2018){Wanajo}, {M{\"u}ller}, {Janka}, \&
  {Heger}}]{wanajo2018}
{Wanajo}, S., {M{\"u}ller}, B., {Janka}, H.-T., \& {Heger}, A. 2018, \apj, 852,
  40, \dodoi{10.3847/1538-4357/aa9d97}

\bibitem[{{Wang} \& {Burrows}(2020)}]{Tianshu}
{Wang}, T., \& {Burrows}, A. 2020, \prd, 102, 023017,
  \dodoi{10.1103/PhysRevD.102.023017}

\bibitem[{{Wang} \& {Burrows}(2023{\natexlab{a}})}]{wang2023}
---. 2023{\natexlab{a}}, \apj, 954, 114, \dodoi{10.3847/1538-4357/ace7b2}

\bibitem[{{Wang} \& {Burrows}(2023{\natexlab{b}})}]{wang2023b}
---. 2023{\natexlab{b}}, arXiv e-prints, arXiv:2311.03446,
  \dodoi{10.48550/arXiv.2311.03446}

\bibitem[{{Wang} {et~al.}(2022){Wang}, {Vartanyan}, {Burrows}, \&
  {Coleman}}]{wang2022}
{Wang}, T., {Vartanyan}, D., {Burrows}, A., \& {Coleman}, M. S.~B. 2022,
  \mnras, 517, 543, \dodoi{10.1093/mnras/stac2691}

\bibitem[{{Wilson}(1971)}]{wilson1971}
{Wilson}, J.~R. 1971, \apj, 163, 209, \dodoi{10.1086/150759}

\bibitem[{Wongwathanarat {et~al.}(2013)Wongwathanarat, Janka, \&
  Müller}]{Wongwathanarat2013}
Wongwathanarat, A., Janka, H.-T., \& Müller, E. 2013, Astronomy and
  Astrophysics, 552, A126, \dodoi{10.1051/0004-6361/201220636}

\bibitem[{{Wongwathanarat} {et~al.}(2015){Wongwathanarat}, {M{\"u}ller}, \&
  {Janka}}]{wongwathanarat2015}
{Wongwathanarat}, A., {M{\"u}ller}, E., \& {Janka}, H.~T. 2015, \aap, 577, A48,
  \dodoi{10.1051/0004-6361/201425025}

\bibitem[{{Woosley} {et~al.}(1990){Woosley}, {Hartmann}, {Hoffman}, \&
  {Haxton}}]{woosley1990}
{Woosley}, S.~E., {Hartmann}, D.~H., {Hoffman}, R.~D., \& {Haxton}, W.~C. 1990,
  \apj, 356, 272, \dodoi{10.1086/168839}

\bibitem[{{Woosley} {et~al.}(2002){Woosley}, {Heger}, \&
  {Weaver}}]{woosley2002}
{Woosley}, S.~E., {Heger}, A., \& {Weaver}, T.~A. 2002, Reviews of Modern
  Physics, 74, 1015, \dodoi{10.1103/RevModPhys.74.1015}

\bibitem[{Yoshida {et~al.}(2021)Yoshida, Takiwaki, Aguilera-Dena, Kotake,
  Takahashi, Nakamura, Umeda, \& Langer}]{Yoshida2021}
Yoshida, T., Takiwaki, T., Aguilera-Dena, D.~R., {et~al.} 2021, Monthly Notices
  of the Royal Astronomical Society, 506, L20–L25,
  \dodoi{10.1093/mnrasl/slab067}

\bibitem[{{Yoshida} {et~al.}(2019){Yoshida}, {Takiwaki}, {Kotake}, {Takahashi},
  {Nakamura}, \& {Umeda}}]{yoshida2019}
{Yoshida}, T., {Takiwaki}, T., {Kotake}, K., {et~al.} 2019, \apj, 881, 16,
  \dodoi{10.3847/1538-4357/ab2b9d}

\end{thebibliography}
\bibliographystyle{aasjournal}




\end{document}